\documentclass[11pt]{article}

\usepackage{graphicx}
\usepackage{epsfig,cite}
\usepackage{amssymb}
\usepackage{amsmath}
\usepackage{dsfont}

\usepackage{url}

\def\smallfrac#1#2{\hbox{${{#1}\over {#2}}$}}
\newcommand{\be}{\begin{equation}}
\newcommand{\ee}{\end{equation}}
\newcommand{\bea}{\begin{eqnarray}}
\newcommand{\eea}{\end{eqnarray}}
\newcommand{\bi}{\begin{itemize}}
\newcommand{\ei}{\end{itemize}}
\newcommand{\ben}{\begin{enumerate}}
\newcommand{\een}{\end{enumerate}}
\newcommand{\la}{\left\langle}
\newcommand{\ra}{\right\rangle}
\newcommand{\lc}{\left[}
\newcommand{\rc}{\right]}
\newcommand{\lp}{\left(}
\newcommand{\rp}{\right)}
\newcommand{\as}{\alpha_s}

\def\frac#1#2{{{#1}\over {#2}}}
\def\gsim{\mathrel{\rlap{\lower4pt\hbox{\hskip1pt$\sim$}}
    \raise1pt\hbox{$>$}}}         %greater than or approx. symbol
\def\lsim{\mathrel{\rlap{\lower4pt\hbox{\hskip1pt$\sim$}}
    \raise1pt\hbox{$<$}}}         %less than or approx. symbol
\newcommand{\mrexp}{\mathrm{exp}}
\newcommand{\dat}{\mathrm{dat}}

\newcommand{\art}{\mathrm{art}} 
\newcommand{\rep}{\mathrm{rep}}
\newcommand{\net}{\mathrm{net}}

\newcommand{\tot}{\mathrm{tot}}

\newcommand{\draft}[1]{}

\begin{document}
\begin{flushright}
Edinburgh 2008/25\\
IFUM-923-FT\\
Freiburg 2008/08
\end{flushright}
\begin{center}
{\Large \bf
A determination of parton distributions\\ with faithful
uncertainty estimation \\}
\vspace{0.8cm}

{\bf  The NNPDF Collaboration:}\\
Richard~D.~Ball$^{1,2}$,
 Luigi~Del~Debbio$^1$, Stefano~Forte$^3$, Alberto~Guffanti$^4$, 
Jos\'e~I.~Latorre$^5$, Andrea~Piccione$^{3}$, 
Juan~Rojo$^6$ and Maria~Ubiali$^1$.

\vspace{1.cm}
{\it ~$^1$ School of Physics and Astronomy, University of Edinburgh,\\
JCMB, KB, Mayfield Rd, Edinburgh EH9 3JZ, Scotland\\
~$^2$ Niels Bohr International Academy, Niels Bohr Institute,\\ 
Blegdamsvej 17, 2100 K{\o}benhavn {\O}, Danmark\\
~$^3$ Dipartimento di Fisica, Universit\`a di Milano and
INFN, Sezione di Milano,\\ Via Celoria 16, I-20133 Milano, Italy\\
~$^4$  Physikalisches Institut, Albert-Ludwigs-Universit\"at Freiburg
\\ Hermann-Herder-Stra\ss e 3, D-79104 Freiburg i. B., Germany  \\
~$^5$ Departament d'Estructura i Constituents de la Mat\`eria, 
Universitat de Barcelona,\\ Diagonal 647, E-08028 Barcelona, Spain\\
~$^6$ LPTHE, CNRS UMR 7589, Universit\'es Paris VI-Paris VII,\\
F-75252, Paris Cedex 05, France}
\end{center}

\begin{center}
{\bf \large Abstract:}
\end{center}

We present the determination of a set of parton distributions of
the nucleon, at next-to-leading order, from a global set of 
deep-inelastic scattering data: NNPDF1.0. 
The determination is based on a
Monte Carlo approach, with neural networks used as unbiased
interpolants. This method, previously discussed by us
and applied to a determination of the
nonsinglet quark distribution, is designed to provide a faithful and
statistically sound representation of the uncertainty on parton
distributions.  We discuss our
dataset, its statistical features, and its Monte Carlo
representation. We summarize the technique used
to solve the evolution equations and its
benchmarking, and the method used to compute physical observables. 
We discuss the parametrization and
fitting of neural networks, and the algorithm used to
determine the optimal fit. We finally present our set of parton
distributions. We discuss its statistical properties, test for its
stability upon various modifications of the fitting procedure,
and compare it to other recent parton sets. We use it to compute 
the benchmark W and Z cross sections at the LHC. We discuss issues
of delivery and interfacing to commonly used packages such as LHAPDF.

\clearpage

\tableofcontents

\clearpage

%----------------------------
%
% \section{Introduction}
%
%---------------------------------
%------------------------------------------------

\section{Introduction}
\label{sec-intro}
\subsection{Determination of parton distributions}
\label{sec-det}
The determination of parton distributions has gone through various
phases which mirror the evolution  of theoretical and
phenomenological  understanding of the theory of strong
interactions. At a very early
stage~\cite{McElhaney:1973nj,Kawaguchi:1976wm,DeRujula:1976tz,
Johnson:1976xc,Gluck:1976iz,Hinchliffe:1977jy}, parton distributions 
were determined through a
combination of general physical principles (as embodied in sum rules),
model assumptions and the first crude experimental information coming
from Bjorken scaling and its violation. These determinations were
semi-quantitative at best, and they were aimed at showing the
compatibility of the data with the partonic interpretation of hard
processes. The parton sets were used to compare 
the observed scaling violations with those
predicted by perturbative QCD~\cite{Kawaguchi:1976wm,DeRujula:1976tz,Johnson:1976xc,Gluck:1976iz,Hinchliffe:1977jy},
thereby leading to  first tests of the theory of strong interactions. 
These early investigations
met with such
success that the parton set of Buras and
Gaemers~\cite{Buras:1977yj}  is sometimes still
used today~\cite{Zeller:2001hh}. 

As the accuracy of the data and the confidence in perturbative QCD
improved, the gluon distribution was extracted from scaling
violations~\cite{Gluck:1980cp}, 
and first parton sets based on consistent global fits were
performed~\cite{Duke:1983gd,Eichten:1984eu}. Despite the availability
of next-to-leading order evolution tools~\cite{Devoto:1983sh}, these
analyses were performed at leading order, which was accurate enough
for these sets to be widely used for phenomenology in the ensuing
decade.

However, thanks to a second generation of
 high--precision deep-inelastic scattering~\cite{Roberts:1990ww}
and hadron collider~\cite{Ellis:1991qj} experiments, QCD gradually
evolved towards being viewed as precision physics --- an integral part
of the standard model. This required an approach to parton
determination based on next-to-leading order theory (in order to have
perturbative uncertainties under control), and also based on fairly
wide ``global'' sets of data of a varied nature, in order to minimize
as much as possible the role of theoretical prejudice in the
determination of the shape of the parton distributions at the initial
scale~\cite{Martin:1987vw,dflm,Gluck:1989ze,Morfin:1990ck}. 

Next-to-leading order parton sets evolved into standard analysis
tools and were constantly updated throughout the ensuing
decade~\cite{HEPDATAurl}.  In
particular, the wealth of data from the HERA
collider~\cite{Klein:2008di} led to a considerable increase in the size of 
the kinematic region over which parton distributions could be
determined, along with a substantial improvement in accuracy,
especially in the determination of quantities which are sensitive to 
scaling violations. Accumulated knowledge eventually led to parton
sets (such as as the  
CTEQ5~\cite{Lai:1999wy} and MRST2001~\cite{Martin:2001es} sets)
very likely to have an accuracy comparable to that of
next-to-leading order QCD computation,
adequate for the determination of most hard processes at
collider energies. These parton sets differ in many
technical details, but are rooted in a similar approach: a 
parton parametrization is assumed,
based on the functional form $f(x)\sim
x^\alpha(1-x)^\beta$ (used since the earliest
investigations~\cite{McElhaney:1973nj}), and its
parameters are then tuned so that the various computed 
observables fit the experimental data. 

With parton distributions now a tool for precision physics, it 
becomes important
to be able to assess accurately the uncertainty on any given parton
set. This need was recognized at a
relatively early stage, and in fact the parton set of Ref.~\cite{dflm}
included error parton sets along with average ones. However, providing
error estimates which can be relied upon raises many subtle issues,
the most obvious  of which is the need for a full treatment of correlated
uncertainties of the underlying data. In the absence of a full
understanding of the problem, the only way of estimating the
uncertainty related to the parton distribution was to compare results
obtained with several parton sets, an especially unsatisfactory
procedure given that many possible sources of systematic bias are
likely to be common to several parton determinations. 

Some first determinations of parton distributions with
uncertainties were obtained by only fitting to restricted data sets (typically
from a subset of deep-inelastic experiments), but retaining
all the information on the correlated uncertainties in the underlying data, and
propagating it through the fitting
procedure~\cite{Alekhin:1996za,Barone:1999yv,Botje:1999dj}. The need
for a systematic approach which could lead to parton distributions
with reliable uncertainty estimation was stressed in the seminal papers
Ref.~\cite{Giele:1998gw,Giele:2001mr}, where an entirely different approach to
parton determination was suggested, based on Bayesian inference
combined with a Monte Carlo approach. While the approach of
Ref.~\cite{Giele:2001mr} was never fully implemented, the need for parton sets
with uncertainties is now generally recognized, and there are currently 
at least three sets of parton distributions with uncertainties available, 
maintained by the
CTEQ~\cite{Pumplin:2002vw,Huston:2005jm,Tung:2006tb,Owens:2007kp,Lai:2007dq}, MRST-MSTW~\cite{Martin:2002aw,Martin:2003sk,Martin:2004ir,Martin:2007bv} and
Alekhin~\cite{Alekhin:2000ch,Alekhin:2002fv,Alekhin:2005gq,Alekhin:2006zm}
groups. 

Parton distributions with
uncertainties~\cite{Pumplin:2002vw,Martin:2002aw,Alekhin:2006zm} have
now become standard. Nevertheless, many of the problems raised in
Refs.~\cite{Giele:1998gw,Giele:2001mr} are still only partly solved. In particular, 
benchmark comparisons performed between some of these
sets~\cite{heralhc} have shown that the uncertainties that come with them
are not
easily interpreted in a statistical sense, in that they are to a
significant amount determined or 
constrained by theoretical or phenomenological expectations. Indeed,
whereas uncertainty bands for parton determinations based on
restricted data
sets~\cite{Alekhin:2006zm}
are obtained by using standard error propagation of one--sigma
contours, those for global fits which include a large variety of
data~\cite{Martin:2002aw,Pumplin:2002vw} are obtained on the basis
of  
a tolerance~\cite{Pumplin:2002vw}, determined
by studying  the
compatibility of the data with each other and with the underlying
theory. The effect of this tolerance is equivalent to multiplying
experimental errors by a   factor between four and six.

This state of affairs might be the inevitable consequences of incompatibilities
between data and, possibly, of inadequacy of the theory used to
describe them. Be that as it may, the standard parton determination method
based on fitting a particular functional form does not seem to be sufficiently
flexible to ascertain whether this is the case: in the absence of a
term of comparison,  it is hard to tell to which extent the current
difficulties are due to an intrinsic limitation of the methodology.

An altogether new approach was proposed in
Ref.~\cite{f2ns}. The general aim of this approach is to determine
objectively both the value and the uncertainty of a function 
(or set of functions)
from a discrete set of many independent (and possibly
incompatible) experimental measurements. 
Its viability was originally demonstrated by using it to provide a 
determination of the structure function 
$F_2(x,Q^2)$ of the proton and neutron from its direct measurement 
at around 600 points,
each by two independent experiments~\cite{f2ns}. The method was then
used  in Ref.~\cite{f2p} to provide a state-of-the art 
determination of the same structure function for the proton, by 
combining almost 2000 different
measurements in 13 different data sets, thereby addressing
issues of data incompatibility. Finally, in
Ref.~\cite{DelDebbio:2007ee} it was used to provide the determination
of a single parton distribution (the nonsinglet quark distribution),
thereby addressing the issue of determining a quantity which is not
measured directly, but rather related through theory to an
experimental observable. In the present paper, we use this method for the
construction of a first parton set from deep-inelastic data: we determine 
five parton distributions from around 3000 measurements in 25 different 
data sets.

\subsection{The NNPDF approach}
\label{sec-nnpdfapp}

%------------------------------------------------------------
\begin{figure}[t!]
\begin{center}
\epsfig{width=0.95\textwidth,figure=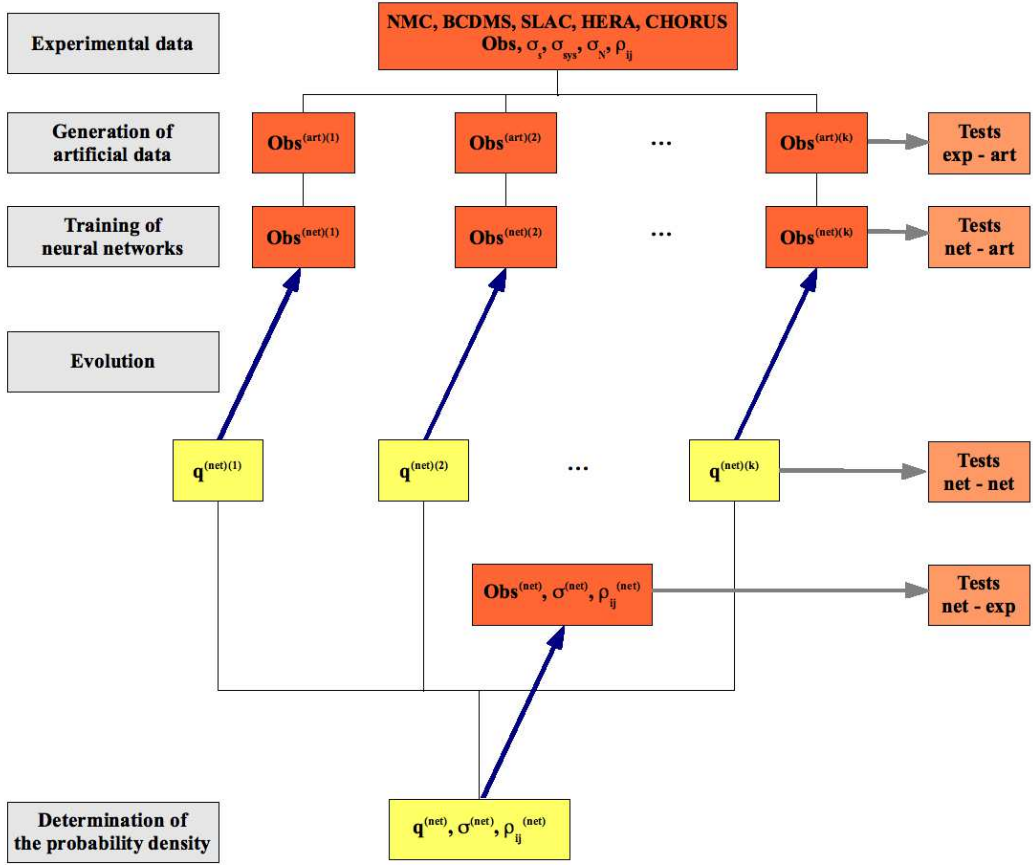} 
\caption{\small Schematic representation of the NNPDF approach.}
\label{fig:scheme}
\end{center}
\end{figure}
%------------------------------------------------------------------
The approach adopted here for the determination of parton
distributions is based on a combination of a Monte Carlo method with
the use of neural networks as basic interpolating functions. The
general idea is twofold: first, problems related to the
possibility of non-gaussian errors and nontrivial error propagation are
best addressed through
the use of a representation
whereby central values are obtained from a Monte Carlo 
sample as averages, uncertainties as
standard deviations, and so forth. 
Second, problems which require the reconstruction of a function 
through its discrete sampling,
without making assumptions on its functional form, are best addressed
using neural networks as unbiased interpolants. The combination of
these two techniques works well in situations where data
are partly inconsistent, in that neural networks are
well suited to the separation of a smooth signal from background
fluctuations, while the Monte Carlo handles the
fluctuations themselves.

The strategy is summarized
in Figure~\ref{fig:scheme}, and it involves two stages. In the first
stage, one generates a Monte Carlo ensemble of replicas of the original
data. This ensemble is generated with the probability
distribution of the data, and it is large enough that the statistical
properties of the data are reproduced to the desired accuracy. In
practice, most data are given with multigaussian probability
distributions of statistical and systematic errors,
described by a covariance matrix and a normalization error, and in
such cases
this is the distribution that will be used to generate the pseudodata. However,
any other probability distribution can be used if and
when required by the experimental
data. Each element in the Monte Carlo
set is a replica of the experimental data: each replica contains as
many data points as are originally available. 
The ensemble
contains all the available experimental information, which can be reproduced by
performing statistical operations on the replicas which form the
ensemble.  That indeed the given ensemble has the desired
statistical features can be verified by means of standard tests, such
as comparison of quantities calculated from it with the original
properties of the data: this is denoted in
Fig.~\ref{fig:scheme} as ``tests exp-art'', namely, the comparison of
experimental and artificial data.
  
In the second stage,  
a set of parton distributions is constructed from each replica of the
data. Each parton distribution function (PDF) at a given scale is
parametrized by an individual neural network: the neural network is
just an especially convenient functional form of parton parametrization, 
used in place of the usual functional forms. Physical observables are
computed from parton distributions in the usual way. One first chooses
a basis set of initial parton distributions, 
typically smaller than the maximal set of twelve quarks and 
antiquarks plus one gluon --- here we
use a set of five independent parton distributions. One
then evolves from the initial scale to the scale at which data are 
available by using standard QCD 
evolution equations, and physical observables are 
computed by convoluting the evolved parton distributions with hard
partonic cross sections.  The best fit set of parton distribution is
finally determined by comparing
the theoretical computation
of the observable for a given PDF set with their replica experimental
values. The experimental values will of course be different in each
replica --- they will fluctuate  according to their distribution in
the Monte Carlo ensemble --- and the best fit PDFs will be
accordingly different for each replica. The ensemble of these best fit
PDFs, which contains as many elements as the set of replicas of
the data that were generated, is the final result of the parton
determination. 

The way in which the best fit set of PDFs is determined from
each data replica is especially important. A first obvious
requirement is that the best fit be independent of
any assumptions made about the parton parametrization. 
This requirement is met by
adopting a redundant parametrization: the size of the neural networks
used, i.e. the number of parameters used to parametrize them, is much
larger than the minimum required in order to reproduce the data. This
redundancy may be checked a posteriori, by verifying that results are
independent of the size and architecture of the neural network. 

A more subtle issue is
that of establishing how the best fit is to be determined. A first possible
answer might be to determine the best fit as the absolute minimum of
the $\chi^2$ (i.e. absolute maximum of the likelihood) of the
comparison between theory and data for a given replica. As already
pointed out in Ref.~\cite{f2ns}, however, this procedure  does not
produce the 
optimal fit for quantities with some built--in smoothness, such as
physical cross sections. Indeed, even for fully compatible data,
independent measurements of the same quantity at the same point will
fluctuate within the uncertainty of the
measurement. If fitted by maximum likelihood, such independent
measurements will automatically be combined into their weighted
average~\cite{cowan}. However, 
assume now that two independent measurements are performed of the same
observable, but measured at very close
values of the underlying kinematic variables: for example the
structure function $F(x,Q^2)$ at the same $Q^2$ and two different but 
close values
$x$. Then a
fit which goes through the central values of both measurements might
be possible, but in the limit in which the two measurements are
performed at infinitesimally close points this would correspond to a
discontinuous behaviour of the observable, which is surely
unphysical. This problem is exacerbated in the case of incompatible
measurements. 

It was thus suggested already in Ref.~\cite{f2ns} that
the best fit should be characterized by a  value of the $\chi^2$ which
is not as small as possible, but rather equal to the value expected on
the basis of the fluctuations of the data.
In order to determine this value, a strategy was developed in
Ref.~\cite{DelDebbio:2007ee}, based on the cross-validation method
used quite generally in neural network studies~\cite{Bishop:1995}. 
Namely, for each replica, the data are divided
randomly into a training set and a validation set. The fit is then performed
on the data in the training set, and the $\chi^2$ computed
from data in both sets is monitored. Minimization is stopped when the
$\chi^2$ in the validation set (not used for fitting) stops
decreasing. The method is made possible by the availability of a very
large and mostly compatible set of data, and it guarantees that the
best fit does not attempt to reproduce random fluctuations of the data. The
method also handles incompatible data, by automatically
tolerating fluctuations in the data even when they are larger than the
nominal uncertainty, whenever fitting these fluctuations would not
lead to an improvement of the global quality of the fit. This analysis
is denoted as ``tests net-art'' 
in Fig.~\ref{fig:scheme}, namely, the comparison of neural net to the
previously generated artificial data.

An important feature of this approach is that many issues of parton
determination can be now addressed using standard statistical
tools. For example, the stability of results upon a change of
parametrization can be verified by computing the distance 
between results in units of
their standard deviation. An
advantage of the Monte Carlo approach is that it is no more
difficult to do this for uncertainties, correlation
coefficients or even more indirect quantities, than it is for 
values of physical observables. Likewise, it is possible to verify
that fits performed by removing data from the set have wider error
bands but remain compatible within these enlarged uncertainties, and
so forth.
The reliability of the results, in particular for the 
uncertainties on the PDFs, can thus be assessed directly.
This assessment
is denoted as ``tests net-net'' in Fig.~\ref{fig:scheme}.

Of course, it is also possible to perform the 
standard tests based on the comparison of the final fit
prediction with the original input data set. The most fundamental of
these is the comparison  to the
data (and the computation of the corresponding $\chi^2$) for the
best fit obtained by averaging results over all neural nets in the
final sample. These tests are denoted as ``tests net-exp'' in
Fig.~\ref{fig:scheme}: the comparison of the final set of neural nest
to the starting experimental data set. 

The implementation of this approach for the parton fit presented here
follows the principles and techniques presented in
Ref.~\cite{DelDebbio:2007ee}. In this paper
we present the adaptations which are needed in order to go 
from the determination of a single
parton distribution to that of a full set. We also present in
detail the features and results of this specific PDF determination, 
NNPDF1.0, and in particular the results of the tests discussed above.

The paper is organized as follows. In Section~\ref{sec:expdata} we
present the data which will be used in our determination, along with their
main statistical features, and the kinematic cuts that we have applied
to them, and the results of tests to verify that the Monte Carlo sample
provides a faithful representation of these data.
In Section~\ref{evolution} we summarize the method we use to solve
evolution equations (already introduced in
Ref.~\cite{DelDebbio:2007ee} in the nonsinglet case), provide
results from its benchmarking, describe 
the specific basis of PDFs we use, and discuss  
the computation of physical observables with the inclusion
target-mass corrections. Full details of the hard kernels used to 
construct the physical observables are given in
Appendix~\ref{sec:kernels}.
In Section~\ref{sec:minim} we turn to the neural network
parametrization and minimization: we summarize the structure of neural
networks used, the weighted genetic algorithm which has been
employed to train them, and the
stopping algorithm used to determine the best fit along the lines
discussed above (already introduced  in
Ref.~\cite{DelDebbio:2007ee}) and present the 
parameter settings and specific features used in the current fits. 
In Section~\ref{sec:results} we discuss in detail our results: we
discuss the general statistical features of our reference parton set,
and present the individual PDFs and their correlations; 
we discuss the
results of various stability tests related to the architecture of
neural network (size and choice of the preprocessing function) and the
dataset (kinematic cuts and reduced datasets); we study theoretical
uncertainties, specifically higher order corrections and the choice of
value of the strong coupling; finally we present results computed
using our dataset both for some of the physical observables entering
the fit (such as structure functions and reduced cross sections) and
for benchmark LHC observables (the W and Z total cross sections).
Useful formulae for the determination of uncertainties on physical
observables using various different PDF sets are summarized in
Appendix~\ref{sec:app-pdferr}.
An outlook on future developments is provided in
Section~\ref{sec:conclusions}.

%------------------------------
%
%\section{Experimental data and Monte Carlo generation}
%
%------------------------------
%----------------------------------------------------------

\section{Experimental data and Monte Carlo generation}
\label{sec:expdata}

The determination of PDFs presented in this work is based on a
comprehensive set of  experimental data from deep-inelastic scattering
(DIS) 
with various lepton beams and nucleon targets. We choose a purely 
DIS data set for this first fit because of the known general
consistency of DIS data. We include  both proton data and neutron
data from nuclear targets 
in order to be able to disentangle isospin triplet and isospin
singlet contribution. We also include charged current scattering data
from charged lepton beams and neutrino scattering data in order to be
able to disentangle the quark and antiquark distributions. 
Because of the limitations due to only fitting DIS data, we will
take a basis of five independent parton distributions, namely the two
light flavours and antiflavours and the gluon, as we discuss below in
Sect.~\ref{sec:pdfbas}.
Further constraints on PDFs could be obtained from various other
experiments, in particular from hadron-hadron scattering.
The inclusion of these data in our fit is
conceptually straightforward, and is left to forthcoming publications.
We shall first present the general features of the data we use, then
discuss the construction of the covariance matrix, provide definitions
of relevant observables and finally present the generation and testing
of the Monte Carlo sample of pseudodata.

\subsection{The data set}
\label{sec:data}

The data sets used in 
this study are listed in Table~\ref{tab:exps-sets}, and their 
kinematic coverage is shown in Fig.~\ref{fig:dataplot}.  They can be
summarized as follows. 

We use the data for proton and deuteron structure functions
$F_2^{p,d}$ determined in fixed-target experiments 
by the 
BCDMS~\cite{Benvenuti:1989rh,Benvenuti:1989fm} and 
NMC~\cite{Arneodo:1996kd,Arneodo:1996qe} collaborations, which
were already included in our previous analysis of the nonsinglet quark
distribution in Ref.~\cite{DelDebbio:2007ee}. They provide the most
accurate and up-to-date information on the valence region of parton
distributions.  They are supplemented
with data on the structure functions from
SLAC~\cite{Whitlow:1991uw} which, though rather older and less
precise, improve the kinematic coverage in the large $x$ region. 
Compared to previous studies by our
collaboration, we now use the ratio $F_2^d/F_2^p$ whenever data for
this observable are available, thereby benefitting from cancellations
in the correlated systematic uncertainties. Altogether these data 
cover the middle- to large-$x$ and
smaller $Q^2$ region of the kinematical range, corresponding to the
lower-right corner in Fig.~\ref{fig:dataplot}.  

Collider experiments have explored a larger kinematical range in great
detail. Neutral and charged current reduced cross sections from the
H1~\cite{Adloff:2000qk,Adloff:1999ah,Adloff:2000qj,Adloff:2003uh} and
ZEUS~\cite{Chekanov:2001qu,Breitweg:1999aa,Chekanov:2002ej,Chekanov:2002zs,Chekanov:2003yv,Chekanov:2003vw}
collaborations are used in the current fit. As shown in
Fig.~\ref{fig:dataplot} these data sets yield informations in a much
wider region of the $(x,Q^2)$ plane, in both
the small-$x$ and the large-$Q^2$ directions. We also include
the data for $F_L$ that have recently appeared in
Ref.~\cite{h1fl}. This is a rather small data set, but it provides the
only direct measurement of $F_L$. We refer to
Refs.~\cite{f2ns,f2p,DelDebbio:2007ee} for additional informations on
all the data sets that were used in our earlier studies.

In order to be able to control the valence--sea (or quark--antiquark)
separation, in this fit we also include neutrino DIS data.
Specifically, we use the large, up-to-date, and consistent set of 
neutrino and antineutrino scattering data by the
CHORUS collaboration~\cite{Onengut:2005kv}. These data have a similar
kinematic coverage to the fixed target charged lepton DIS data.
 
The main features of our data sets are summarized in
Table~\ref{tab:exps-sets}, where we show the beam, target and observable,
 the number of data points,
the kinematic range, the size of uncertainties averaged over the data points.
The observable chosen  is generally that which is closest to the
experimental measurement and minimizes the pre-analysis by the
experimental collaboration: in particular we have used  
the reduced cross section for all collider and neutrino data sets.
The various systematics and their correlations are treated according
to the information provided by the experimental collaborations themselves 
(see Section~2 in Ref.~\cite{f2ns} for a detailed
description of NMC and BCDMS data, Table~1 in Ref.~\cite{Chekanov:2005nn}
for ZEUS data, Table~2 in Ref.~\cite{Adloff:2003uh} for H1
data,Ref.~\cite{CHORUSurl} for CHORUS data). 

In Table~\ref{tab:exps-sets} 
we distinguish between ``Experiments'', defined as groups
of data that are not correlated to each other, and ``Sets'' within an
experiment, which are correlated with each other. They
correspond to measurements of different observables in the same
experiment, or measurements of the same observables in different years
which retain some correlated systematics. This distinction will be
important in the minimization strategy, discussed in
Section~\ref{sec:nn_ga_minim} below.
\begin{table}[t!]
\begin{center}

 \tiny
 \begin{tabular}{|ll|c|c|c|c|c|c|c|c|}
  \hline
  Experiment & Set & $N_{\rm dat}$ & $x_{\rm min}$ &  $x_{\rm max}$ 
 &  $Q^2_{\rm min}$ &  $Q^2_{\rm max}$ & $\sigma_{\rm tot}$ (\%) & $F$ & Ref.\\ \hline
 \hline
 \multicolumn{2}{|l|}{SLAC} & & & & & & & &\\
 &  SLACp    &  211 (47) & .07000 & .85000 &    0.6 &    29. &   3.6 & $F_2^p$ &\cite{Whitlow:1991uw}\\
 &  SLACd    &  211 (47) & .07000 & .85000 &    0.6 &    29. &   3.2 & $F_2^d$ & \cite{Whitlow:1991uw}\\
 \hline
 \multicolumn{2}{|l|}{BCDMS} & & & & & & & &\\
 &  BCDMSp   &  351 (333) & .07000 & .75000 &    7.5 &   230. &   5.5 & $F_2^p$ & \cite{Benvenuti:1989rh}\\
 &  BCDMSd   &  254 (248) & .07000 & .75000 &    8.8 &   230. &   6.6 & $F_2^d$ &\cite{Benvenuti:1989fm}\\
  \hline
NMC & 
  &  288 (245) & .00350 & .47450 &    0.8 &    61. &   5.0 & $F_2^p$
 & \cite{Arneodo:1996qe}\\
\hline
NMC-pd &  &  260 (153) & .00150 & .67500 &    0.2 &    99. &   2.1 & $F_2^d/F_2^p$ &\cite{Arneodo:1996kd}\\
 \hline
 \multicolumn{2}{|l|}{ZEUS} & & & & & & & &\\
 &  Z97lowQ2 &   80 & .00006 & .03200 &    2.7 &    27. &   4.9 & $\tilde{\sigma}^{NC,e^+}$ &\cite{Chekanov:2001qu}\\
 &  Z97NC    &  160 & .00080 & .65000 &   35.0 & 20000. &   7.7 & $\tilde{\sigma}^{NC,e^+}$ &\cite{Chekanov:2001qu}\\
 &  Z97CC    &   29 & .01500 & .42000 &  280.0 & 17000. &  34.2 & $\tilde{\sigma}^{CC,e^+}$ &\cite{Breitweg:1999aa}\\
 &  Z02NC    &   92 & .00500 & .65000 &  200.0 & 30000. &  13.2 & $\tilde{\sigma}^{NC,e^-}$ &\cite{Chekanov:2002ej}\\
 &  Z02CC    &   26 & .01500 & .42000 &  280.0 & 30000. &  40.2 & $\tilde{\sigma}^{CC,e^-}$ &\cite{Chekanov:2002zs}\\
 &  Z03NC    &   90 & .00500 & .65000 &  200.0 & 30000. &   9.1 & $\tilde{\sigma}^{NC,e^+}$ &\cite{Chekanov:2003yv}\\
 &  Z03CC    &   30 & .00800 & .42000 &  280.0 & 17000. &  31.0 & $\tilde{\sigma}^{CC,e^+}$ &\cite{Chekanov:2003vw}\\
 \hline
 \multicolumn{2}{|l|}{H1} & & & & & & & &\\
 &  H197mb   &   67 (55) & .00003 & .02000 &    1.5 &    12. &   4.9 & $\tilde{\sigma}^{NC,e^+}$ &\cite{Adloff:2000qk}\\
 &  H197lowQ2 &   80 & .00016 & .20000 &   12.0 &   150. &   4.2 & $\tilde{\sigma}^{NC,e^+}$ &\cite{Adloff:2000qk}\\
 &  H197NC   &  130 & .00320 & .65000 &  150.0 & 30000. &  13.3 & $\tilde{\sigma}^{NC,e^+}$ &\cite{Adloff:1999ah}\\
 &  H197CC   &   25 & .01300 & .40000 &  300.0 & 15000. &  29.8 & $\tilde{\sigma}^{CC,e^+}$ &\cite{Adloff:1999ah}\\
 &  H199NC   &  126 & .00320 & .65000 &  150.0 & 30000. &  15.5 & $\tilde{\sigma}^{NC,e^-}$ &\cite{Adloff:2000qj}\\
 &  H199CC   &   28 & .01300 & .40000 &  300.0 & 15000. &  27.6 & $\tilde{\sigma}^{CC,e^-}$ &\cite{Adloff:2000qj}\\
 &  H199NChy &   13 & .00130 & .01050 &  100.0 &   800. &   9.2 & $\tilde{\sigma}^{NC,e^-}$ &\cite{Adloff:2003uh}\\
 &  H100NC   &  147 & .00131 & .65000 &  100.0 & 30000. &   10.4 & $\tilde{\sigma}^{NC,e^+}$ &\cite{Adloff:2003uh}\\
 &  H100CC   &   28 & .01300 & .40000 &  300.0 & 15000. &  21.8 & $\tilde{\sigma}^{CC,e^+}$ &\cite{Adloff:2003uh}\\
 \hline
 \multicolumn{2}{|l|}{CHORUS} & & & & & & & &\\
 &  CHORUS$\nu$ &  607 (471) & .02000 & .65000 &    0.3 &    95. &   11.2 & $\tilde{\sigma}^{\nu}$ &\cite{Onengut:2005kv}\\
 &  CHORUS$\bar{\nu}$ &  607 (471) & .02000 & .65000 &    0.3 &    95. &  18.7 & $\tilde{\sigma}^{\bar{\nu}}$ &\cite{Onengut:2005kv}\\
 \hline
\multicolumn{2}{|l|}{FLH108} &    8 & .00028 & .00360 &   12.0 &    90. &  69.2 & $F_L$ & \cite{h1fl}\\
\hline
%Total  &   &  3948 &  &  &    &    &  & & &  &\\
%\cline{1-3}
\multicolumn{2}{|l|}{Total}     &  3948 (3161) & \multicolumn{7}{|c}{}\\
\cline{1-3}
 \end{tabular}

\end{center}
\caption{\small The experiments included in the
present analysis divided in the respective data sets.
We show the number of points before (after) applying kinematic cuts, 
the kinematic range, the average total
uncertainty and the measured observable.
Different sets within an experiment are correlated
with each other, but data from different experiments are not.
\label{tab:exps-sets}}
\end{table}

\subsection{Uncertainties and correlations}
\label{sec:errcorr}

The covariance matrix for each experiment can be computed from
knowledge of statistical, systematic and normalization
uncertainties:
\bea
\label{eq:covmat}
{\rm cov}_{pq}=
\lp\sum_{l=1}^{N_c}\sigma_{p,l}\sigma_{q,l}+
\sum_{n=1}^{N_a} \sigma_{p,n}\sigma_{q,n}+
\sum_{n=1}^{N_r} \sigma_{p,n}\sigma_{q,n}
+\delta_{pq}\sigma_{p,s}^2\rp F_{p,I} F_{q,J}\ ,
\eea
where $p$ and $q$ run over the experimental points,
$F_{I,p}=F_I(x_p,Q_p^2)$ and $F_{J,q}=F_J(x_q,Q_q^2)$ are the measured
central values for the observables $I$ and
 $J$, and the various uncertainties, given as relative values, are:
$\sigma_{p,l}$, the
$N_c$ correlated systematic uncertainties; $\sigma_{p,n}$,  the $N_a$
($N_r$) absolute (relative) normalization uncertainties;
$\sigma_{p,s}$ the statistical uncertainty. 

The correlation matrix is  defined as 
\bea
\rho_{pq}=\frac{ {\rm cov}_{pq}}{\sigma_{p,\tot}\sigma_{q,\tot}F_{p,I}
  F_{q,J}} \ ,
\label{eq:cormat}
\eea
where the total uncertainty
 $\sigma_{p,{\rm tot}}$ for the $p$-th point is given by
\bea
\sigma_{p,\tot}=\sqrt{\sigma_{p,s}^2+\sigma_{p,c}^2+ \sigma_{p,N}^2}\,,
\label{eq:toterr}
\eea
the total correlated uncertainty $\sigma_{p,c}$ is the sum of all 
correlated systematics
\bea
\sigma_{p,c}^2=\sum_{l=1}^{N_c} \sigma_{p,l}^2\,,
\label{eq:totsyst}
\eea
and the total normalization uncertainty is
\bea
\sigma_{p,N}^2 = 
\sum_{n=1}^{N_a} \sigma_{p,n}^2+
\sum_{n=1}^{N_r} \left( \smallfrac{1}{2} \sigma_{p,n}\right)^2 .
\label{eq:totnorm}
\eea
The factor of one half in the relative normalization uncertainties comes
from the first order expansion of Eq.~(\ref{eq:totalnorm}) below.

The $N_u$ uncorrelated systematic uncertainties quoted for HERA data sets are
combined with the statistical uncertainty according to
\bea
\sigma_{p,s}^2=\sigma_{p,stat}^2 + \sum_{k=1}^{N_u}\sigma_{p,k}^2 .
\label{eq:unctot}
\eea 
Asymmetric uncertainties  quoted for some ZEUS data sets in
Refs.~\cite{Chekanov:2002ej,Chekanov:2002zs,Chekanov:2003yv,Chekanov:2003vw}
are symmetrized as described in Section 2 of Ref.~\cite{f2p} and references
therein. For the case of SLAC data the single systematic uncertainty is
taken to be fully correlated for all the data points.

\subsection{Observables and kinematic cuts}
\label{sec:obscuts}

The deep-inelastic observables used in our fit are either structure
functions or reduced cross sections.
The neutral current deep-inelastic scattering cross section 
involving a generic charged lepton $\ell^\pm$ is defined as
\be
\frac{d^2\sigma^{\rm NC,\ell^{\pm}}}{dxdQ^2} (x,y,Q^2)=\frac{2\pi \alpha^2}{ x Q^4}
 \lc
Y_+ F_2^{NC}(x,Q^2) \mp Y_- x F_3^{NC}(x,Q^2)-y^2 F_L^{NC}(x,Q^2)\rc \ ,
\label{eq:ncxsect}
\ee
where 
\begin{equation}
 Y_{\pm}=1\pm (1-y)^2.
\label{eq:ypmdef}
\end{equation}
In the case of NMC, BCDMS, and SLAC data,
we use the quoted value for the structure function $F_2 (x, Q^2)$. For
ZEUS and H1 data, we use the quoted reduced cross section defined as
\begin{equation}
  \label{eq:rednc}
  \widetilde{\sigma}^{\rm NC,e^{\pm}}(x,y,Q^2)=\lc
  \frac{2\pi \alpha^2}{ x Q^4} Y_+\rc^{-1}\frac{d^2\sigma^{\rm NC,e^{\pm}}}{dxdQ^2}(x,y,Q^2) \,.
\end{equation}

For charged current deep-inelastic scattering, the measured double differential
cross section in the case of unpolarized beams is given by
\begin{eqnarray}
  \frac{d^2\sigma^{\rm CC,e^{\pm}}}{dxdQ^2}(x,y,Q^2) &=&\frac{G_F^2}{4\pi x}
  \lp \frac{M_W^2}{M_W^2+Q^2}\rp^2 \\\nonumber
&\times&\frac{1}{2}\lc
  Y_+ F_2^{CC,e^{\pm}}(x,Q^2)\mp Y_- x F_3^{CC,e^{\pm}}(x,Q^2)
  -y^2 F_L^{CC,e^{\pm}}(x,Q^2)\rc\,.
\end{eqnarray}
As in the neutral current case we use the reduced cross section defined as
\begin{eqnarray}
  \label{eq:redcc}
  \widetilde{\sigma}^{\rm CC,e^{\pm}}(x,y,Q^2)=\lc
  \frac{G_F^2}{4\pi x}
  \lp \frac{M_W^2}{M_W^2+Q^2}\rp^2 \rc^{-1}
  %\frac{1}{2}
  \frac{d^2\sigma^{\rm CC,e^{\pm}}}{dxdQ^2}(x,y,Q^2)\,.
\end{eqnarray}
In the case of CHORUS data, we use the neutrino-nucleon reduced cross section,
which in the single $W$-exchange approximation, can be written as
\begin{eqnarray}
\label{eq:nuxsec}
\tilde{\sigma}^{\nu (\bar{\nu})}(x,y,Q^2)&=&\frac{1}{E_{\nu}}\frac{d^2\sigma^{\nu(\bar{\nu})}}{dx\,dy}
(x,y,Q^2)\\\nonumber
&=&\frac{G_F^2M_N}{2\pi(1+Q^2/M_W^2)^2}
\left[\left(Y_+ - \frac{2M^2_Nx^2y^2}{Q^2}\right)F_2^{\nu(\bar{\nu})}\pm \,Y_-\,xF_3^{\nu(\bar{\nu})}
-y^2F_L^{\nu(\bar{\nu})}\right].
\end{eqnarray}

For all nuclear targets, 
namely NMC, BCDMS and SLAC deuteron data and
CHORUS heavy nuclei (mostly lead with a small admixture of iron and
other materials), no nuclear corrections are applied.

\begin{figure}[t!]
\begin{center}
\epsfig{width=0.95\textwidth,figure=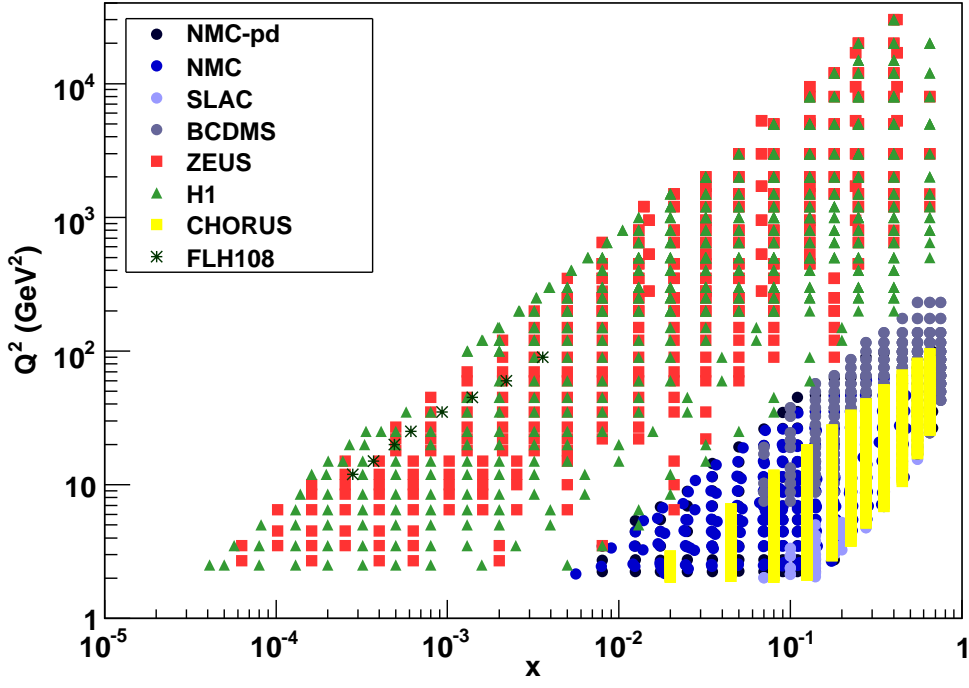} 
\caption{\small Experimental data in the $(x,Q^2)$ plane used
    in the present analysis after kinematic cuts.}
\label{fig:dataplot}
\end{center}
\end{figure}

In order to keep higher--twist corrections under control, only data
with $Q^2 > 2\,{\rm GeV^2}$ and $W^2 > 12.5\,{\rm GeV^2}$ are
retained. The changes, if any, in the number of data points after
kinematic cuts for each set are reported in Table
\ref{tab:exps-sets} between parenthesis. The experimental data
actually used in the present analysis are summarized in
Fig.~\ref{fig:dataplot}. Since the kinematic cuts we use
are not too conservative, we will supplement our fit
with target mass corrections, as discussed in Section \ref{sec:tmc}.

%------------------------------------------------------------
%------------------------------------------------------------------

\subsection{Generation of the pseudo-data sample}
\label{sec:mcgen}

Error propagation from experimental data to the fit is handled
by a Monte Carlo sampling of the probability distribution defined by
data. The statistical sample is obtained by generating $N_{\rm rep}$
artificial replicas of data points following a multi-gaussian
distribution centered on each data point with the variance given by
the experimental uncertainty. More precisely, given a data point
$F_{I,p}^{\rm (exp)}\equiv F_I (x_p,Q^2_p)$ we generate
$k=1,\ldots,N_\rep$ artificial points $F_{I,p}^{(\art)(k)}$ as follows
\be
\label{eq:replicas}
F_{I,p}^{(\art)(k)}=S_{p,N}^{(k)} F_{I,p}^{\rm (\mrexp)}\lp 1+
 \sum_{l=1}^{N_c}r_{p,l}^{(k)}\sigma_{p,l}+r_{p}^{(k)}\sigma_{p,s}\rp
 \ , \ k=1,\ldots,N_{\rep} \ ,
\ee
where
\be
\label{eq:totalnorm}
S_{p,N}^{(k)}=\prod_{n=1}^{N_a}\lp1+r_{p,n}^{(k)}\sigma_{p,n}\rp
\prod_{n=1}^{N_r}\sqrt{1+r_{p,n}^{(k)}\sigma_{p,n}}.
\ee 
The variables $r_{p,l}^{(k)},r_{p}^{(k)},r_{p,n}^{(k)}$ are
all  univariate gaussian random numbers that generate
fluctuations of the artificial data around the central value given by
the experiments. For each replica $k$, if two experimental
points $p$ and $p^\prime$ have correlated systematic uncertainties, 
then $r^{(k)}_{p,l}=r^{(k)}_{p^\prime,l}$, i.e. the
fluctuations due to the correlated systematic uncertainties are the same for
both points. A similar condition on $r^{(k)}_{p,n}$ ensures that
correlations between normalization uncertainties are properly taken into
account. 

\begin{table}[t!]
\begin{center}

%\begin{flushleft}
{\tiny
\begin{tabular}{|c|c|c|c|c|c}
\cline{1-5} 
Experiment & NMC & NMC-pd & SLAC & BCDMS \\
\cline{1-5} 
$\la PE\left [\la F^{\rm (art)} \ra_{\rm rep}\right]\ra_{\rm dat}$ & 9.0 $\cdot 10^{-5}$ & 1.8 $\cdot 10^{-5}$ & 3.1 $\cdot 10^{-4}$& 
1.3 $\cdot 10^{-3}$\\
$ r\left[ F^{\rm (art)} \right] $ & 1.000 & 1.000 & 1.000& 1.000\\
\cline{1-5} 
$\la PE\left [\la \sigma^{\rm (art)} \ra_{\rm rep}\right]\ra_{\rm dat}$ & 1.5 $\cdot 10^{-3}$ & 4.2 $\cdot 10^{-3}$& 3.1 $\cdot 10^{-3}$&
 4.0 $\cdot 10^{-3}$\\
$ \la \sigma^{\rm (exp)} \ra_{\rm dat} $ & 0.0147& 0.0170& 0.0104& 0.0698\\
$ \la \sigma^{\rm (art)} \ra_{\rm dat} $ & 0.0146& 0.0171& 0.0104& 0.0692\\
$ r\left[ \sigma^{\rm (art)} \right] $ & 1.000& 0.998& 0.998& 0.999\\
\cline{1-5} 
$ \la \rho^{\rm (exp)} \ra_{\rm dat} $ & 0.033& 0.165& 0.312& 0.470\\
$ \la \rho^{\rm (art)} \ra_{\rm dat} $ & 0.033& 0.176& 0.311& 0.463\\
$ r\left[ \rho^{\rm (art)} \right] $ & 0.963& 0.988& 0.987& 0.994\\
\cline{1-5} 
$ \la {\rm cov}^{\rm (exp)} \ra_{\rm dat} $ & 6.52 $\cdot 10^{-6}$& 4.39 $\cdot 10^{-5}$& 3.07 $\cdot 10^{-5}$& 
2.90 $\cdot 10^{-5}$\\
$ \la {\rm cov}^{\rm (art)} \ra_{\rm dat} $ & 6.78 $\cdot 10^{-6}$& 4.73 $\cdot 10^{-5}$& 3.03 $\cdot 10^{-5}$& 
2.82 $\cdot 10^{-5}$\\
$ r\left[ {\rm cov}^{\rm (art)} \right] $ & 0.989& 0.984& 0.988& 0.999\\
\hline
%\end{tabular}
%}
%
%{\tiny
%\begin{tabular}{|c|c|c|c|c|c|}
%\hline 
Experiment & ZEUS & H1 & CHORUS & FLH108 & \multicolumn{1}{|c|}{Total}\\
\hline
$\la PE\left [\la F^{\rm (art)} \ra_{\rm rep}\right]\ra_{\rm dat}$ & 8.5 $\cdot 10^{-4}$ & 1.1 $\cdot 10^{-4}$ & 1.8 $\cdot 10^{-3}$& 
1.3 $\cdot 10^{-2}$ & \multicolumn{1}{|c|}{7.1 $\cdot 10^{-5}$}\\
$ r\left[ F^{\rm (art)} \right] $ & 1.000 & 1.000 & 1.000&  1.000& \multicolumn{1}{|c|}{0.980}\\
\hline
$\la PE\left [\la \sigma^{\rm (art)} \ra_{\rm rep}\right]\ra_{\rm dat}$ & 9.6 $\cdot 10^{-3}$ & 4.2 $\cdot 10^{-3}$& 1.8 $\cdot 10^{-2}$&
6.1 $\cdot 10^{-4}$ &  \multicolumn{1}{|c|}{3.0 $\cdot 10^{-3}$}\\
$ \la \sigma^{\rm (exp)} \ra_{\rm dat} $ & 0.0607& 0.0472& 0.1088& 0.1744& \multicolumn{1}{|c|}{0.0556}\\
$ \la \sigma^{\rm (art)} \ra_{\rm dat} $ & 0.0603& 0.0472& 0.1109& 0.1756& \multicolumn{1}{|c|}{0.0562}\\
$ r\left[ \sigma^{\rm (art)} \right] $ & 1.000& 1.000& 0.998& 0.999& \multicolumn{1}{|c|}{0.980}\\
\hline
$ \la \rho^{\rm (exp)} \ra_{\rm dat} $ & 0.079& 0.027& 0.094& 0.650& \multicolumn{1}{|c|}{0.145}\\
$ \la \rho^{\rm (art)} \ra_{\rm dat} $ & 0.082& 0.028& 0.096& 0.657& \multicolumn{1}{|c|}{0.146}\\
$ r\left[ \rho^{\rm (art)} \right] $ & 0.982& 0.952& 0.998& 0.996& \multicolumn{1}{|c|}{0.996}\\
\hline
$ \la {\rm cov}^{\rm (exp)} \ra_{\rm dat} $ & 1.53 $\cdot 10^{-4}$& 4.93 $\cdot 10^{-5}$& 2.16 $\cdot 10^{-3}$& 
2.03 $\cdot 10^{-2}$ & \multicolumn{1}{|c|}{1.07 $\cdot 10^{-3}$}\\
$ \la {\rm cov}^{\rm (art)} \ra_{\rm dat} $ & 1.57 $\cdot 10^{-4}$& 5.03 $\cdot 10^{-5}$& 2.31 $\cdot 10^{-3}$& 
2.11 $\cdot 10^{-2}$ & \multicolumn{1}{|c|}{1.01 $\cdot 10^{-3}$}\\
$ r\left[ {\rm cov}^{\rm (art)} \right] $ & 0.996& 0.987& 0.998& 0.998& \multicolumn{1}{|c|}{0.997}\\
\hline
\end{tabular}
}
 
%\end{flushleft}
\end{center}
\caption{\small Statistical estimators for the
Monte Carlo artificial data
generation with $N_{\rep}=1000$. 
The definition of the statistical estimators 
is given in Appendix B of~\cite{DelDebbio:2007ee}.
}
\label{tab:mcest}
\end{table}

The treatment of normalization uncertainties needs some care:
as is well known, including normalization uncertainties in the
covariance matrix would lead to a fit that is systematically biased
to lie below the data~\cite{dagos}. 
Rather, normalization uncertainties are included by rescaling all
uncertainties, i.e.  by constructing 
for each replica a modified covariance matrix:
\be
\label{eq:covmatnn}
{\overline{\rm cov}^{(k)}}_{pq}=
\lp\sum_{l=1}^{N_c}{\overline\sigma^{(k)}}_{p,l}
{\overline\sigma^{(k)}}_{q,l}+\delta_{pq}\lp{\overline\sigma^{(k)}}_{p,s}\rp^2\rp 
F_{I,p}F_{J,q} \ ,
\ee
with the statistical uncertainties $\sigma_{p,s}$ and each
systematic uncertainty $\sigma_{p,l}$ being rescaled according to
\be
\overline \sigma_{p,s}^{(k)}= S_{p,N}^{(k)} \sigma_{p,s}\,,
\qquad \qquad
\overline \sigma_{p,l}^{(k)}= S_{p,N}^{(k)} \sigma_{p,l}\,,
\qquad l=1,\ldots,N_c\,.
\label{eq:errrescal}
\ee
It can be readily seen that:
\be
{\overline{\rm cov}^{(k)}}_{pq}={\overline{\rm cov}^{(\mrexp)}}_{pq}
S_{p,N}^{(k)}S_{q,N}^{(k)}\, ,
\label{eq:rescalcov}
\ee 
and therefore the experimental correlation matrix without
normalization uncertainties needs to be evaluated only once, while
${\overline{\rm cov}^{(k)}}_{pq}$ is obtained by multiplying by the
normalization factors $S_{p,N}^{(k)}$ and $S_{q,N}^{(k)}$ for each
replica. If within an experiment all the sets have only a common
global normalization uncertainty, the rescaling is an overall multiplicative
factor. The covariance matrix Eq.~(\ref{eq:covmatnn}) is that which is
used in order to perform a fit to the $k$-th data replica.

Appropriate statistical estimators have been devised in
Ref.~\cite{DelDebbio:2007ee} in order to quantify the accuracy of the
statistical sampling obtained from a given ensemble of replicas. We
refer the reader to Appendix~B of Ref.~\cite{DelDebbio:2007ee} for a
detailed explanation of the meaning of these statistical
estimators. Using these estimators, 
we have verified that a Monte Carlo sample of
pseudo-data with $N_{\rep}=1000$ is sufficient to reproduce the mean
values, the 
variances, and the correlations of experimental data with a 1\%
accuracy for all the experiments.  Results for the estimators computed
from a sample of $N_{\rep}=1000$ replicas are shown 
in Table~\ref{tab:mcest}. This set of Monte Carlo replicas will be
used in the rest of this paper.

%--------------------------------------------------------

% ------------------------------------------------------
%
%\section{From parton distributions to physical observables}
%
% ----------------------------------------------------
%---------------------------------------------------------

\section{From parton distributions to physical observables}
\label{evolution}

In this section we provide all the technical details for the calculation
in perturbation theory of deep--inelastic observables from a set of 
initial PDFs. First, we briefly review the strategy for the solution
of QCD evolution equations in terms of pre-computable perturbative
hard kernels $K_{Ij}((x,\as(Q^2),\as(Q_0^2))$, originally introduced in
Ref.~\cite{DelDebbio:2007ee}. Then
 we review the calculation of the DGLAP 
evolution factors $\Gamma_{ij}(x,\as(Q^2),\as(Q_0^2))$ using Mellin space 
techniques, including our prescription for heavy quarks, and give details of
their benchmarking. Next, we turn to the particular choice of basis for 
the input PDFs. Finally, we describe the calculation of physical
observables by combining evolved PDFs with the hard 
coefficient functions (including their target mass corrections) and  
the procedure for obtaining the hard kernels
$K_{Ij}((x,\as(Q^2),\as(Q_0^2))$
for  deep inelastic observables.

\subsection{Leading-twist factorization and evolution}
\label{ltev}

The perturbative computation
of physical observables involves first evolving the PDFs up to the scale of 
the measurement, and then their convolution with a hard cross-section to 
give the observable.  Here this is done following the strategy of
Ref.~\cite{DelDebbio:2007ee}, whereby evolution kernels are
pre-computed, and then convoluted with parton distributions. This 
separates the numerical computation of
the solutions to evolution equations from the computation of input
parton distributions. The advantage of this is that each of the two
computations can be optimized separately from a numerical point of
view: in particular, we can thus use a Mellin-space approach to solve
evolution equations, but adopt $x$-space parametrization of
PDFs. Also, evolution kernels can thus be pre-computed, benchmarked,
and stored for future use during the fitting procedure.

The basic ideas behind this technique were discussed in
\cite{DelDebbio:2007ee}. The extension from the nonsinglet structure 
function, with only one PDF, to a number of different deep inelastic 
structure functions and reduced cross-sections, expressed in terms of 
several singlet and nonsinglet PDFs, is in principle 
straightforward, but in practice complicated by a number of subtleties which 
will be discussed as they arise.

Deep inelastic observables $F_I(x,Q^2)$ (which may be structure functions 
or reduced cross-sections) may always be expressed at leading twist as
a convolution of parton distributions $f_j(x,Q^2)$ and hard 
coefficient functions 
$C_{Ij}(x,\as(Q^2))$, computed in perturbation theory:
\be
  \label{eq:fact}
  F_I(x,Q^2) =  \sum_j C_{Ij}(x,\as(Q^2))\otimes f_j(x,Q^2),\,
\ee
where  $\otimes$ denotes the convolution
\be
\label{eq:conv}
f(x)\otimes g(x) \equiv \int_x^1 \frac{dy}{y} f(y) g\left(\frac{x}{y}\right),\,
\ee
and the indices $I$ and $j$ run over observables and parton distribution 
functions respectively. The scale dependence of the parton distribution 
functions is in turn given by the renormalisation group, or DGLAP equations
\be
  \label{eq:dglap}
  Q^2\frac{\partial}{\partial Q^2}f_i(x,Q^2) = \sum_j 
                                   P_{ij}(x,\as(Q^2))\otimes f_j(x,Q^2),\,
\ee
where $P_{ij}$ are the Altarelli-Parisi splitting functions, also calculable 
in perturbation theory. 

The solution of these coupled integro-differential
equations may be written as
\be
  \label{eq:evfact}
  f_i(x,Q^2) =  \sum_j \Gamma_{ij}(x,\as,\as^0)
\otimes f_j(x,Q_0^2),\,
\ee
where $f_j(x,Q_0^2)$ are the input PDFs, to be determined empirically,  
$\Gamma_{ij}(x,\as,\as^0)$ are the evolution factors, and we use the shorthand
notation
\be
\label{asnotation}
\as \equiv \as(Q^2),\qquad \as^0\equiv \as(Q_0^2). 
\ee
The evolution factors also 
satisfy evolution equations:
\be
  \label{evolx}
  Q^2\frac{\partial}{\partial Q^2}\Gamma_{ij}(x,\as,\as^0) 
           = \sum_k P_{ik}(x,\as)\otimes 
                \Gamma_{kj}(x,\as,\as^0),\,
\ee
with boundary conditions $\Gamma_{ij}(x,\as^0,\as^0)
=\delta_{ij}\delta(1-x)$.

Substituting Eq.~(\ref{eq:evfact}) into Eq.~(\ref{eq:fact})
\begin{eqnarray}
  \label{eq:factinit}
  F_I(x,Q^2) &=&  \sum_{jk} C_{Ij}(x,\as)\otimes 
                \Gamma_{jk}(x,\as,\as^0)
                \otimes f_k(x,Q_0^2)\,\nonumber\\
             &=& \sum_j K_{Ij}(x,\as,\as^0)
\otimes f_j(x,Q_0^2),\,
\end{eqnarray}
where the hard kernel
\be
  \label{eq:genevfact}
  K_{Ij}(x,\as,\as^0) =  
    \sum_k C_{Ik}(x,\as)\otimes\Gamma_{kj}(x,\as,\as^0),\,
\ee
may be computed in perturbation theory.

Performing many nested convolutions is numerically rather time consuming. 
However the hard kernels Eq.~(\ref{eq:genevfact}) are 
independent of the particular set of input PDFs adopted, and may thus be 
calculated once and for all at the beginning of the computation, 
interpolated, and stored. Determining the physical observables given 
by a given set of input PDFs then involves
the evaluation of only the one set of convolutions Eq.~(\ref{eq:factinit}),
which is relatively fast, these being reducible to simple sums.

\subsection{Solving the evolution equations} 
\label{sec:eveq}

\def\bgamma{\boldsymbol{\gamma}}
The QCD evolution equations are most easily solved using
Mellin moments~\cite{dflm, pegasus, bbg}, since then all the convolutions 
become simple products, and the equations can be solved in closed form.
The problem is thus reduced to the computation of the single
Mellin inversion integral. Specifically, we define
\be
\label{meldef}
\Gamma_{ij}(N,\as,\as^0)\equiv
\int_0^1 \!dx\,x^{N-1} \Gamma_{ij}(x,\as,\as^0) \ ,
\ee
where by slight abuse of notation we denote the function and its
transform with the same symbol. Equation~(\ref{evolx}) becomes  
\be
\label{evolN}
Q^2\frac{\partial}{\partial Q^2}\Gamma_{ij}(N,\as,\as^0)= 
\sum_k \gamma_{ik}(N,\alpha_s)\, \Gamma_{kj}(N,\as,\as^0) \ ,
\ee
where the anomalous dimensions $\gamma_{ij}(N,\as)$ are the 
Mellin moments of the splitting functions. Expanding perturbatively in 
powers of $\alpha_s$  
\be
\label{andim}
\gamma_{ij}(N,\alpha_s)=\as\gamma^{(0)}_{ij}(N)
+\as^2\gamma^{(1)}_{ij}(N)+\cdots,
\ee
where the dots denote higher order contributions. The anomalous dimensions 
are known at LO, NLO \cite{gNLOa,gNLOb,gNLOc,gNLOd,gNLOe,gNLOf} and 
NNLO \cite{gnnloa,gnnlob}. 

Since all the dependence on $Q^2$ of the anomalous dimension is through 
the running coupling $\as(Q^2)$, and 
\begin{equation}
  \label{runas}
  \frac{d\as}{d\ln Q^2}= \beta(\as)=
  - \as^2\beta_0
  - \as^3\beta_1+\cdots\,,
\end{equation}
we may in turn write Eq.~(\ref{evolN}) as a differential 
equation in $\as$:
\begin{equation}
  \label{eq:stdevol}
  \frac{\partial}{\partial \ln\as}
  \Gamma_{ij}(N,\as,\as^0)=
  -\sum_k R_{ik}(N,\as)\Gamma_{kj}(N,\as,\as^0).\,
\end{equation}
The matrix $R_{ij}\equiv ({\bf R})_{ij}$ has the 
perturbative expansion  
\begin{equation}
\label{rexp}
  {\bf R}(N,\as) = {\bf R}^{(0)}(N)+\as {\bf R}^{(1)}(N)+ \cdots,
\end{equation}
where in terms of the expansion Eq.~(\ref{andim}) of the anomalous 
dimension matrix $\gamma_{ij}\equiv ({\bgamma})_{ij}$
\begin{equation}
\label{rrecursion}
  {\bf R}^{(0)}(N)\equiv \frac{\bgamma^{(0)}(N)}{\beta_0},\qquad
  {\bf R}^{(k)}(N)\equiv
    \frac{\bgamma^{(k)}(N)}{\beta_0}
    - \sum_{i=1}^k \frac{\beta_i}{\beta_0}{\bf R}^{(k-i)}(N).\,   
\end{equation}
Note that Eq.~(\ref{evolN}) truncated at NLO, that is with Eq.~(\ref{andim})
and Eq.~(\ref{runas}) truncated after the first two terms, is not equivalent
to the naive truncation of Eq.~(\ref{rexp}) after two terms, but rather to
the complete series with 
\begin{equation}
\label{RNLO}
  {\bf R}^{(0)}(N)\equiv \bgamma^{(0)}(N)/\beta_0, \qquad
  {\bf R}^{(k)}(N)\equiv - b_1 {\bf R}^{(k-1)}(N),    
\end{equation}
where $b_1\equiv \beta_1/\beta_0$.

The complete matrix of anomalous dimensions $\bgamma$, and thus the matrices 
${\bf R}$ are in fact almost completely diagonal: all the flavour nonsinglet 
and valence quark distributions evolve multiplicatively, and only the singlet 
quark and gluon actually mix. Thus we only need to solve 
Eq.~(\ref{eq:stdevol}) for one by one and two by two matrices.

Consider first the simplest case of the evolution of flavour nonsinglet and 
valence quark distributions: the evolution factor then satisfies the 
simple first order equation
\begin{equation}
  \label{NSevol}
  \frac{\partial}{\partial\ln\as}
  \Gamma_{\rm NS}(N,\as,\as^0)=
  -R_{\rm NS}(N,\as)\Gamma_{\rm NS}(N,\as,\as^0).\,
\end{equation}
At LO the solution is trivial:
\begin{equation}
 \label{ns-lo}
 \Gamma_{\rm NS,LO}(N,\as,\as^0)= \left( \frac{\as}{\as^0} \right)
  ^{-R^{(0)}_{\rm NS}}, 
\end{equation}
while at NLO we need to work a little harder: using Eq.~(\ref{RNLO}) one finds
\be
\label{ns-nlo}
  \Gamma_{\rm NS,NLO}(N,\as,\as^0) = \exp \left\{-\frac{R^{(1)}_{\rm NS}}
  {b_1} \ln \left( \frac{1+b_1 \as}{1+b_1\as^0} \right) \right\} 
  \left( \frac{\as}{\as^0} \right)^{-R^{(0)}_{\rm NS}}.
\ee
This exact solution is equivalent up to subleading terms to the linearized
solution
\be 
\label{ns-nlo-lin}
  \Gamma^{\rm lin}_{\rm NS,NLO}(N,\as,\as^0)
   =\left( 1 - R^{(1)_{\rm NS}}(\as - \as^0)\right) 
  \left( \frac{\as}{\as^0} \right)^{-R^{(0)}_{\rm NS}},
\ee
which is in turn the exact solution to Eq.~(\ref{NSevol}) with $R_{\rm NS}=
R^{(0)}_{\rm NS}+\as R^{(1)}_{\rm NS}$.

Turning finally to the singlet sector, we need to solve Eq.~(\ref{eq:stdevol})
when ${\bf R}$ are two by two matrices, corresponding to coupled singlet 
quarks and gluons:
\begin{equation}
  \label{Sevol}
  \frac{\partial}{\partial \ln\as}
  {\bf \Gamma}_{\rm S}(N,\as,\as^0)=
  -{\bf R}_{\rm S}(N,\as){\bf \Gamma}_{\rm S}(N,\as,\as^0).\,
\end{equation}
At LO we can proceed by diagonalization:
\begin{equation}
\label{s-lo}
{\bf \Gamma}_{\rm S,LO}(N,\as,\as^0)\equiv  {\bf L}(N,\as,\as^0)= 
	      {\bf e}_+(N)\left(\frac{\as}{\as^0}\right)^{-\lambda_+(N)}
              +{\bf e}_-(N)\left(\frac{\as}{\as^0}\right)^{-\lambda_-(N)},
\end{equation}
where
\begin{equation}
\label{eval}
  \lambda_\pm(N)=
    \frac{1}{2\beta_0}\left[\gamma^{(0)}_{qq}(N)+\gamma^{(0)}_{gg}(N) \pm
    \sqrt{\left(\gamma^{(0)}_{qq}(N)-\gamma^{(0)}_{gg}(N)\right)^2 
                + 4\gamma^{(0)}_{qg}(N)\gamma^{(0)}_{gq}(N)}\right],
\end{equation}
are the eigenvalues of the two by two matrix 
${\bf R}_{\rm S}^{(0)}(N)$ of singlet anomalous dimensions, and 
\begin{equation}
\label{proj}
  {\bf e}_\pm(N)=\pm\frac{1}{\lambda_+(N) - \lambda_-(N)}
                 ({\bf R}_{\rm S}^{(0)}(N)-\lambda_\mp(N){\bf I})\,,
\end{equation}
are the corresponding projectors.

The full NLO solution is more complicated, and must be developed recursively
as a perturbative expansion around the LO solution ${\bf L}(N,as,\as^0)$: 
writing
\begin{equation}
\label{s-nlo}
{\bf \Gamma}_{\rm S,NLO}(N,\as,\as^0)\equiv
{\bf U}(N,\as){\bf L}(N,\as,\as^0){\bf U}(N,\as^0)^{-1},
\end{equation}
where ${\bf U}(N,\as)$ has the expansion
\begin{equation}
\label{Uexp}
{\bf U}(N,\as)= {\bf 1}+\as {\bf U}^{(1)}(N)
                      +\as^2{\bf U}^{(2)}(N)+\cdots,
\end{equation}
solves Eq.~(\ref{eq:stdevol}) provided 
\begin{equation}
  \label{eq:ukexplicit}
  {\bf U}^{(k)}=
        \frac{{\bf e}_- \widetilde{{\bf R}}^{(k)}{\bf e}_+}
                 {\lambda_+ -\lambda_- - k}
        + \frac{{\bf e}_+ \widetilde{{\bf R}}^{(k)}{\bf e}_-}
                  {\lambda_- -\lambda_+ - k}
         -\frac{1}{k}\left[{\bf e}_+\widetilde{{\bf R}}^{(k)}{\bf e}_+ 
  		        + {\bf e}_-\widetilde{{\bf R}}^{(k)}{\bf e}_-\right],
\end{equation}
where
\begin{equation}
  \label{eq:rtwiddle}
  \widetilde{{\bf R}}^{(0)} = {\bf R}_{\rm S}^{(0)},\qquad
  \widetilde{{\bf R}}^{(k)} = {\bf R}_{\rm S}^{(k)}
         +\sum_{i=1}^{k-1}{\bf R}_{\rm S}^{(k-i)}{\bf U}^{(i)}\,.
\end{equation}
By solving recursively Eqs.~\eqref{eq:ukexplicit}, \eqref{eq:rtwiddle} 
with the NLO approximation Eq.~\eqref{RNLO}, the NLO evolution factor 
Eq.~\eqref{RNLO} can be computed. Just as in the nonsinglet case, 
the exact NLO solution may be linearized to give  
\begin{equation}
\label{s-nlo-lin}
{\bf \Gamma}^{\rm lin}_{\rm S,NLO}(N,\as,\as^0)={\bf L}(N,\as,\as^0) + 
\as {\bf U}^{(1)}(N){\bf L}(N,\as,\as^0)-
\as^0 {\bf L}(N,\as,\as^0){\bf U}^{(1)}(N),
\end{equation}
which again is an exact solution to the truncated evolution equation, 
and equivalent to the full solution Eq.~(\eqref{s-nlo}) up to subleading 
terms.

In what follows we will use the exact solutions Eq.~\eqref{ns-nlo} and 
Eq.~\eqref{s-nlo} in order to be able to compare our 
results directly to those 
of $x$-space codes (see e.g.~\cite{qcdnum,Salam:2008qg}) which integrate the 
evolution equations numerically. However in the comparison to the data we 
choose the linearized solutions Eq.~\eqref{ns-nlo-lin} and 
Eq.~\eqref{s-nlo-lin}.

Generalization of the NLO solutions to NNLO and beyond is straightforward but 
tedious.

\subsection{Calculating the evolved $x$-space PDFs}
\label{sec:xevol}

The $x$--space evolution factors are obtained by taking the inverse Mellin
transforms of the solutions obtained in Eq.~(\ref{ns-nlo}) and 
Eq.~(\ref{s-nlo}). For the nonsinglets
\begin{equation}
  \label{eq:xkernelns}
  \Gamma_{\rm NS}(x,\as,\as^0)=\int_{C}
 \frac{dN}{2\pi i}x^{-N}\Gamma_{\rm NS}(N,\as,\as^0),
\end{equation}
where $C$ is taken to be the Talbot contour described in 
ref.\cite{DelDebbio:2007ee}, which goes around the singularities 
at $N=0,-1,-2,\ldots$. For the singlets we use instead
\be
  \label{eq:xkernels}
  {\bf \Gamma}_{\rm S}(x,\as,\as^0)=\int_{C+1}
 \frac{dN}{2\pi i}x^{-N}{\bf\Gamma}_{\rm S}(N,\as,\as^0)
          = x\,\int_{C}
        \frac{dN}{2\pi i}x^{-N}{\bf\Gamma}_{\rm S}(N-1,\as,\as^0),
\ee
since now the singularities are at $N=1,0,-1,\ldots$, i.e. displaced by 
one unit to the right. The contour integrals are evaluated using 
the Fixed Talbot algorithm \cite{ft}. 

However all splitting functions, except the off-diagonal entries of the
singlet matrix, diverge when $x\to 1$; this implies that the 
evolution kernels  
$\Gamma(x,\as,\as^0)$ will likewise be divergent as $x\to 1$, 
and must thus be interpreted as
distributions. Specifically, we define 
\bea
  \label{eq:gammadistns}
  \Gamma_{\rm NS}^{(+)}(x,\as,\as^0) &=& \Gamma_{\rm NS}(x,\as,\as^0)
-G_{NS}(\as,\as^0)\delta(1-x)\,, \\
  \label{eq:gammadists}
  {\bf\Gamma}_{\rm S}^{(+)}(x,\as,\as^0) 
&=& {\bf\Gamma}_{\rm S} (x,\as,\as^0)
-{\bf G}_S(\as,\as^0)\, x^{-1}\delta(1-x)\,,
\eea
where
\bea
\label{gammamomns}
  G_{\rm NS}(\as,\as^0) 
&=& \int_0^1\!dx\, \Gamma_{\rm NS}(x,\as,\as^0) 
= \Gamma_{\rm NS}(N,\as,\as^0)\vert_{N=1},
\\
\label{gammamoms}
  {\bf G}_{\rm S}(\as,\as^0) 
&=& \int_0^1\!dx\, x\,{\bf\Gamma}_{\rm S}(x,\as,\as^0) 
= {\bf\Gamma}_{\rm S}(N,\as,\as^0)\vert_{N=2},
\,
\eea
are all finite constants. The convolutions Eq.~(\ref{eq:evfact}) may then 
be evaluated as
\bea
\label{evolfullns}
f_i(x,Q^2)&=& G_{\rm NS}(\as,\as^0)f_i(x,Q_0^2) + \int_x^1
\!\frac{dy}{y}\,\Gamma_{\rm NS}^{(+)}(y,\as,\as^0)f_i\lp
\frac{x}{y},Q_0^2\rp\nonumber\\ 
&=&
\lp  G_{\rm NS}(\as,\as^0)
-\int_0^x \! dy\, \Gamma_{\rm NS}(y,\as,\as^0)\rp f_i(x,Q_0^2) 
\nonumber \\
&&\qquad + \int_x^1 \!\frac{dy}{y}\,
\Gamma_{\rm NS}(y,\as,\as^0)\lp f_i\lp \frac{x}{y},Q_0^2\rp
- y f_i(x,Q_0^2),\rp. 
\eea
for nonsinglet distributions $f_i$, and similarly
\bea
\label{evolfulls}
{\bf f}_{\rm S}(x,Q^2)&=& {\bf G}_{\rm S}(\as,\as^0){\bf f}_{\rm S}(x,Q_0^2) 
+ \int_x^1
\!\frac{dy}{y}\,{\bf \Gamma}_{\rm S}^{(+)}(y,\as,\as^0){\bf f}_{\rm S}\lp
\frac{x}{y},Q_0^2\rp\nonumber\\ 
&=&
\lp  {\bf G}_{\rm S}(\as,\as^0)
-\int_0^x \! dy\, y{\bf \Gamma}_{\rm S}(y,\as,\as^0)\rp
{\bf f}_{\rm S}(x,Q_0^2) \nonumber \\
&&\qquad + \int_x^1 \!\frac{dy}{y}\,
{\bf \Gamma}_{\rm S}(y,\as,\as^0)\lp {\bf f}_{\rm S}\lp \frac{x}{y},Q_0^2\rp
- y^2 {\bf f}_{\rm S}(x,Q_0^2)\rp, 
\eea
for singlet distributions ${\bf f}_{\rm S}$, where now all integrals 
converge and can be computed numerically, in 
practice using Gaussian integration as described in 
ref.\cite{DelDebbio:2007ee}.

\subsection{Flavour decomposition and heavy quarks}

\label{sec:heavy}

The primary quantities $f_j$ in Eq.~(\ref{eq:dglap}) 
may be thought of as the 
$2n_f$ quark and antiquark distributions $q_i$ and $\overline{q}_i$
and the gluon distribution $g$. The singlet quark distribution 
\begin{equation}
\label{singlet}
  \Sigma=\sum_{i=1}^{n_f}(q_i+\overline{q}_i)\,,
\end{equation}
and the gluon distribution mix under evolution, as in Eq.~(\ref{evolfulls}): 
a sensible basis is ${\bf f}_{\rm S} = (\Sigma,g)$. For the remaining $2n_f-1$ 
distributions we adopt a basis of charge conjugation 
eigenvectors which each evolve independently according to 
Eq.~(\ref{evolfullns}): a suitable such basis consists of the $n_f-1$ charge 
conjugation even nonsinglets
\begin{eqnarray}
  \label{eq:lincombeven}
  T_3 &=& u^+ - d^+, \nonumber\\
  T_8 &=& u^+ + d^+ - 2s^+, \nonumber\\
  T_{15} &=& u^+ + d^+ + s^+ - 3c^+, \nonumber\\
  T_{24} &=& u^+ + d^+ + s^+ + c^+ - 4b^+,\nonumber\\
  T_{35} &=& u^+ + d^+ + s^+ + c^+ + b^+ - 5t^+,
\end{eqnarray}
where $q_i^\pm=q_i\pm\overline{q}_i$, and $q_i=u,d,s,c,b,t$ are the 
various flavour distributions, which evolve with evolution factor
$\Gamma_{\rm NS}^+$, and the $n_f$ charge conjugation odd 
valence distributions 
\begin{eqnarray}
  \label{eq:lincombodd}
  V &=& u^-+d^-+s^-+c^-+b^-+t^-,\nonumber\\
  V_3 &=& u^- - d^-, \nonumber\\
  V_8 &=& u^- + d^- - 2s^-, \nonumber\\
  V_{15} &=& u^- + d^- + s^- - 3c^-, \nonumber\\
  V_{24} &=& u^- + d^- + s^- + c^- - 4b^-,\nonumber\\
  V_{35} &=& u^- + d^- + s^- + c^- + b^- - 5t^-,
\end{eqnarray}
the first of which (the singlet) evolves with evolution factor 
$\Gamma_{\rm NS}^v$, while the remainder evolve with evolution factor 
$\Gamma_{\rm NS}^-$. At LO all the quark anomalous dimensions 
are equal: 
$\gamma_{\rm NS}^{(0),+}=\gamma_{\rm NS}^{(0),-}=\gamma_{\rm NS}^{(0),v}
=\gamma_{\rm S,qq}^{(0)}$, and thus at LO
$\Gamma_{\rm NS}^{+}=\Gamma_{\rm NS}^{-}=\Gamma_{\rm NS}^{v}
=\Gamma_{\rm S}^{qq}$. However at NLO 
$\gamma_{\rm NS}^{(1),-}=\gamma_{NS}^{(1),v}$, while all the others are 
different: beyond NLO all the anomalous dimensions, and thus evolution 
factors, are different from each other. 

We regard the first three flavours $u$, $d$ and $s$ as ``light'': together 
with the gluon we thus have seven parton distributions 
$g,\Sigma,V,T_3,T_8,V_3,V_8$ which are intrinsically nonperturbative 
and are thus, at least in principle, to be determined empirically. The 
parametrization of these PDFs will be discussed in the next section. The 
remaining three flavours $c$,$b$ and $t$ are regarded as ``heavy'': this 
means that we assume that the six parton 
distributions $T_{15},T_{24},T_{35},V_{15},V_{24},V_{35}$ have a
component which may be 
computed perturbatively. Of course, it is in principle possible to 
also introduce
nonperturbative (or ``intrinsic'') contributions to these quantities.

In this paper we use the zero mass variable flavour number (ZM-VFN) scheme 
to incorporate the effects of the heavy quarks. In this, the simplest 
heavy quark scheme, the number of virtual 
flavours in the $\beta$ function and anomalous dimensions changes abruptly 
at the heavy quark thresholds: this means that while the PDFs are continuous,
their scale dependence is discontinuous. Thus for example, when computing
$\Gamma(N,\as,\as^0)$ for $Q^2>m_b^2$, we write
\be
\Gamma(N,\as,\as^0)=\Gamma(N,\as,\as^b)\Gamma(N,\as^b,\as^0),
\ee
where $\as^b\equiv\as(m_b^2)$,
and compute the two factors on the right hand side with $n_f=4$ and $n_f=5$
respectively. This scheme neglects terms above threshold which are
proportional to powers of $\frac{m^2_h}{Q^2}$, where $m_h$ is the mass
of the heavy quark, thereby losing accuracy for scales close to the
thresholds. 

The heavy quark distributions 
themselves are assumed to be zero below threshold, and then generated 
radiatively above threshold. Consider for example the charm 
distribution. For simplicity we take 
$Q_0^2=m_c^2$. For $Q^2\leq m_c^2$, $c^\pm=0$, so 
$T_{15} = \Sigma$, $V_{15} = V$, while for 
$Q^2>m_c^2$, $T_{15}$ and $V_{15}$ evolve as nonsinglet distributions:
\bea
\label{teecee}
 T_{15}(x,Q^2) &=& 
\Gamma_{\mathrm{NS}}^+(x,\as,\as^0)\otimes \Sigma(x,Q_0^2),\\ 
\label{veecee}
 V_{15}(x,Q^2) &=& 
\Gamma_{\mathrm{NS}}^-(x,\as,\as^0)\otimes V(x,Q_0^2). 
\eea
The difference between $T_{15}(x,Q^2)$ and the quark singlet $\Sigma(x,Q^2)$ 
for $m_c^2<Q^2<m_b^2$ gives the charm distribution $c^+$. Similarly $c^-$ is 
given by the difference between $V_{15}(x,Q^2)$ and $V(x,Q^2)$: however since
at NLO $\Gamma_{\rm NS}^{-}=\Gamma_{NS}^{v}$, $V_{15}=V$ 
and $c=\bar c$.

For the $b$ distribution the situation is a little more complicated since
now $m_b^2>Q_0^2$.  Assuming now that 
$b^\pm=0$ for $Q^2<m_b^2$, we have for $Q^2>m_b^2$
\bea
\label{beecee}
 T_{24}(x,Q^2) 
&=& \Gamma_{\mathrm{NS}}^+(x,\as,\as^b)\otimes \Sigma(x,m_b^2)
\nonumber\\ 
&=& \Gamma_{\mathrm{NS}}^+(x,\as,\as^b)\otimes\nonumber\\
&&\qquad\qquad\lp\Gamma_{\mathrm{S}}^{qq}(x,\as^b,\as^0)
\otimes\Sigma(x,Q_0^2)
 +\Gamma_{\mathrm{S}}^{qg}(x,\as^b,\as^0)\otimes g(x,Q_0^2)\rp.
\eea
It is thus convenient to define the evolution factors
\bea
\label{t24}
\Gamma^{24,q}_{\mathrm{S}}(N,\as,\as^0)&=&
\Gamma_{\mathrm{NS}}^+(N,\as,\as^b)\Gamma_{\mathrm{S}}^{qq}(N,\as^b,\as^0),\\
\nonumber
\Gamma^{24,g}_{\mathrm{S}}(N,\as,\as^0)&=&
\Gamma_{\mathrm{NS}}^+(N,\as,\as^b)\Gamma_{\mathrm{S}}^{qg}(N,\as^b,\as^0).
\eea
Note that these are inverted and convoluted using the {\em singlet} formulae 
Eq.~(\ref{eq:xkernels}) and Eq.~(\ref{evolfulls}) since at 
least a part of the evolution is singlet. Below threshold, i.e. 
for $Q^2< m_b^2$, $\Gamma^{24,q}_{\mathrm{S}}=\Gamma^{qq}_{\mathrm{S}}$, 
$\Gamma^{24,g}_{\mathrm{S}}=\Gamma^{qg}_{\mathrm{S}}$. 
Similarly, for evolution of the valence contribution 
$V_{24}(x,Q^2)$ it is convenient to define
\be
\label{v24}
\Gamma^{24}_{\mathrm{NS}}(N,\as,\as^0)=
\Gamma_{\mathrm{NS}}^-(N,\as,\as^b)\Gamma_{\mathrm{NS}}^{v}(N,\as^b,\as^0).
\ee
above threshold, with 
$\Gamma^{24}_{\mathrm{NS}}=\Gamma^{v}_{\mathrm{NS}}$ below.
At NLO, for all $Q^2$, $\Gamma^{24}_{\mathrm{NS}}=
\Gamma_{\mathrm{NS}}^- = \Gamma_{\mathrm{NS}}^{v}$, thus 
$V_{24}=V$ and $b=\bar{b}$. 

Precisely similar considerations apply to the top distribution: 
for $Q^2>m_t^2$ we define
\bea
\label{t35}
\Gamma_{\rm S}^{35,q}(N,\as,\as^0)&=&
\Gamma_{\mathrm{NS}}^+(N,\as,\as^t)\Gamma_{\rm S}^{qq}(N,\as^t,\as^0),\\
\nonumber
\Gamma_{\rm S}^{35,g}(N,\as,\as^0)&=&
\Gamma_{\mathrm{NS}}^+(N,\as,\as^t)\Gamma_{\rm S}^{qg}(N,\as^t,\as^0).\\
\label{v35}
\Gamma_{\rm NS}^{35}(N,\as,\as^0)&=&
\Gamma_{\mathrm{NS}}^-(N,\as,\as^t)\Gamma_{\rm NS}^{v}(N,\as^t,\as^0),
\eea
to give the evolution of $T_{35}(x,Q^2)$ and $V_{35}(x,Q^2)$. For $Q^2<m_t^2$ 
$\Gamma^{35,q}_{\mathrm{S}}=\Gamma^{qq}_{\mathrm{S}}$, 
$\Gamma^{35,g}_{\mathrm{S}}=\Gamma^{qg}_{\mathrm{S}}$, 
$\Gamma^{35}_{\mathrm{NS}}=\Gamma^{v}_{\mathrm{NS}}$, and at NLO 
for all $Q^2$, $\Gamma^{35}_{\mathrm{NS}}=
\Gamma_{\mathrm{NS}}^- = \Gamma_{\mathrm{NS}}^{v}$, 
$V_{35}=V$ and $t=\bar{t}$. Note however that all the data we 
currently use to determine the PDFs are actually below the top threshold.

\subsection{Practical implementation and benchmarks}
\label{sec:benchmarks}
 
The solution of the evolution equations
through the determination of  $x-$space evolution factors,
Eqs.~(\ref{evolfullns}) and (\ref{evolfulls}), is particularly
efficient because of the universality of the evolution factor,
i.e., its independence of the specific boundary condition which is
being evolved. This means that the evolution factors can be pre-computed and
stored, and then used during the process of parton fitting
without having to recompute them each time \cite{DelDebbio:2007ee}. 

During PDF fitting, a given PDF set
must be evolved many times up to the fixed values
of $(x,Q^2)$ at which data are available. 
It can be seen that for each $(x,Q^2)$
the numerical determination of the  
right--hand side of
Eqs.~(\ref{evolfullns},\ref{evolfulls}) 
involves the evaluation of two contributions: the first requires 
the multiplication of the PDF by a (predetermined) constant
$(G(\as,\as^0)-\int_0^x \! dy\, \Gamma(y,\as,\as^0))$
while the second requires a convolution of the (predetermined) evolution factor
$\Gamma(y,\as,\as^0)$ with the (subtracted) PDF, and thus the numerical 
evaluation of the integral over $y$. To perform this numerical integration
we use $N_{\rm quad}-$point gaussian integration 
in each of the $2^{N_{\rm iter}+1}-1$ intervals in which the 
integration range $(x,1)$
of $y$ is divided. The total number of points used
to perform the convolutions in $y$ of Eqs.~(\ref{evolfullns})
and (\ref{evolfulls}) is then given by
\be
\label{npt}
N_{\rm pt}=
N_{\rm quad}\left(2^{N_{\rm iter}+1}-1\right) \ ,
\ee 
and we determine the values of
$y$ accordingly, for each given value of $x$. We find that $N_{\rm quad}=4$ is 
precise enough for all applications, and we discuss below in detail
the choice of $N_{\rm iter}$.

\begin{table}[t!]
\begin{center}
\vskip-0.1cm
\small
\begin{tabular}{|c||c|c|c|c|c|c|}
\hline
$x$  &  $\epsilon_{\rm rel}\lp u_v\rp$ & 
 $\epsilon_{\rm rel}\lp d_v\rp$ &  $\epsilon_{\rm rel}\lp \Sigma\rp$ &
 $\epsilon_{\rm rel}\lp \bar{d}+\bar{u}\rp$ &  
 $\epsilon_{\rm rel}\lp s+\bar{s}\rp$ &
 $\epsilon_{\rm rel}\lp g\rp$ \\
\hline
\hline   
 \multicolumn{7}{|c|}{$N_{\rm iter}=6$}\\
\hline
$10^{-7}$ & $2.2~10^{-5}$ & $8.1~10^{-6}$&  $4.9~10^{-6}$
&  $1.5~10^{-5}$&  $1.2~10^{-6}$ & $2.2~10^{-5}$ \\
$10^{-6}$ & $6.3~10^{-6}$ & $3.2~10^{-6}$  & $9.8~10^{-6}$ 
& $1.1~10^{-5}$&  $5.4~10^{-6}$ & $3.0~10^{-6}$ \\
$10^{-5}$ & $1.8~10^{-5}$ & $1.4~10^{-5}$ &
 $8.3~10^{-6}$ & $3.0~10^{-6}$  &$3.6~10^{-6}$ & $1.4~10^{-6}$ \\
$10^{-4}$ & $3.1~10^{-5}$ & $1.6~10^{-5}$  &$ 3.6~10^{-5}$ &
 $4.3~10^{-5}$ & $3.3~10^{-5}$& $3.2~10^{-5}$ \\
$10^{-3}$ & $1.8~10^{-6}$ & $1.2~10^{-5}$  & 
$5.9~10^{-6}$ & $5.8~10^{-6}$ & $8.9~10^{-6}$& $ 3.6~10^{-6}$\\
$10^{-2}$ & $2.8~10^{-5}$ & $1.5~10^{-5}$ &
$ 4.7~10^{-5}$ & $4.3~10^{-5}$  &$4.6~10^{-5}$& $ 8.2~10^{-5}$\\
$0.1$ & $3.2~10^{-6}$ & $1.3~10^{-5}$  &
$3.0~10^{-6}$&  $9.4~10^{-6}$ &$ 2.1~10^{-5}$ & $5.1~10^{-7}$\\
$0.3$ & $1.9~10^{-6}$ & $2.4~10^{-5}$  
&$6.5~10^{-6}$&  $1.0~10^{-5}$& $ 3.2~10^{-6}$ & $2.6~10^{-6}$\\
$0.5$ & $1.70~10^{-5}$ & $1.3~10^{-5}$ & $1.5~10^{-5}$  &
$1.3~10^{-5}$ & $3.0~10^{-6}$ & $3.5~10^{-6}$ \\
$0.7$ & $7.0~10^{-5}$ & $8.0~10^{-6}$ & $5.9~10^{-5}$&
  $8.9~10^{-6}$&  $2.4~10^{-5}$&  $9.9~10^{-6}$ \\
$0.9$ & $1.4~10^{-5}$ & $6.2~10^{-6}$  &
$1.3~10^{-5}$& $ 7.4~10^{-4}$ & $1.8~10^{-3}$  &$5.1~10^{-5}$ \\
\hline
\hline
 \multicolumn{7}{|c|}{$N_{\rm iter}=4$}\\
\hline
$10^{-7}$ & $4.2~10^{-2}$ & $4.5~10^{-2}$  &$5.1~10^{-2}$ &
 $5.1~10^{-2}$ &$ 5.1~10^{-2}$ &$ 5.1~10^{-2}$ \\
$10^{-6}$ & $1.6~10^{-2}$ & $1.8~10^{-2}$ &
$ 2.4~10^{-2}$ &$ 2.3~10^{-2}$  &$2.4~10^{-2}$& $ 2.5~10^{-2}$ \\
$10^{-5}$ & $4.9~10^{-3}$ & $4.4~10^{-3}$  &$8.7~10^{-3}$&
 $ 8.3~10^{-3}$&$  8.7~10^{-3}$&  $9.6~10^{-3}$ \\
$10^{-4}$ & $2.3~10^{-3}$ & $2.2~10^{-3}$  &$3.9~10^{-3}$ & 
$3.7~10^{-3}$ &
$ 3.9~10^{-3}$ & $4.4~10^{-3}$ \\
$10^{-3}$ &$ 1.1~10^{-3}$ & $6.7~10^{-4}$  
&$3.5~10^{-3}$&  $3.0~10^{-3}$ & $3.4~10^{-3}$&  $4.6~10^{-3}$ \\
$10^{-2}$ & $1.5~10^{-3}$ & $8.5~10^{-4}$  &$3.4~10^{-3}$& $
 2.7~10^{-3}$ &
 $3.7~10^{-3}$ & $5.5~10^{-3}$ \\
0.1 & $3.9~10^{-6}$ & $1.3~10^{-5}$
& $ 4.3~10^{-6}$ & $1.1~10^{-5}$ & $2.4~10^{-5}$ &$ 1.0~10^{-4}$ \\
0.3 & $1.9~10^{-6}$ & $2.6~10^{-5}$ & $6.6~10^{-6}$  &
$1.610^{-5}$&  $5.9~10^{-6}$&  $7.1~10^{-7}$ \\
0.5 & $1.6~10^{-5}$ & $1.1~10^{-5}$&  $1.4~10^{-5}$ & $2.0~10^{-5}$ &
 $5.8~10^{-6}$ & $3.3~10^{-5}$ \\
0.7 & $6.8~10^{-5}$ & $1.2~10^{-5}$ & $5.7~10^{-5}$& 
$ 6.5~10^{-6}$& $ 4.6~10^{-5}$&
  $3.4~10^{-5}$ \\
0.9 & $1.4~10^{-5}$ &$ 5.1~10^{-5}$ &$ 1.6~10^{-5}$&  
$6.4~10^{-4}$ &
$ 1.7~10^{-3}$&  $1.2~10^{-4}$ \\
\hline
\end{tabular}
\end{center}
\caption{\small Comparison of the accuracy of our PDF evolution
with respect to the Les Houches benchmark tables for different
PDF combinations at NLO in the ZM-VFNS. We show results for two values
of $N_{\rm iter}$, which define the
number of points over which the gaussian integrations
are performed, as discussed in the text. 
\label{tab:lhacc}}
\vskip-0.1cm
\end{table}

The accuracy of our PDF evolution code, 
described above, has been cross-checked
against the Les Houches PDF evolution benchmark tables
\cite{lh2,heralhc}. Those tables were obtained
from a comparison of the {\tt HOPPET}
\cite{Salam:2008qg} 
and {\tt PEGASUS}
\cite{pegasus}
evolution codes, which are $x-$space and
$N-$space codes respectively. 
In order to perform a meaningful comparison, we use
the iterated solution of the $N-$space evolution
equations (see Eqs.~(\ref{ns-nlo}) and (\ref{s-nlo})), and use the same
initial PDFs and same
running coupling, following the procedure
described in detail in Ref.~\cite{lh2,heralhc}.

We show in Table~\ref{tab:lhacc} the relative difference
$\epsilon_{\rm rel}$
for various combinations of PDFs between our
PDF evolution and the benchmark tables of Refs.~\cite{lh2,heralhc} 
at NLO in the ZM-VFNS, for two different 
values of $N_{\rm iter}$, Eq.~(\ref{npt}). In the upper part 
of the table we show a very accurate evolution to prove the correctness
of our technique, with $N_{\rm iter}=6$, that is,
with approximately 500 points used to perform
the convolution integrals. As we can see, 
this choice leads to an accuracy which is enough
to reproduce the Les Houches tables with $\mathcal{O}\lp 10^{-5}\rp$
precision for all values of $x$, which is the nominal
precision of the agreement between 
{\tt HOPPET} and  {\tt PEGASUS}. 

In the lower part of Table~\ref{tab:lhacc} 
we show the accuracy results for the actual parameters which
are used in the neural network fit. We take $N_{\rm iter}=4$,
i.e. integration with 128 points, since this is enough 
to reach an accuracy of $\mathcal{O}\lp 10^{-3}-10^{-4}\rp$
in the region of $x$ relevant to the available experimental data.
Such an accuracy is enough for practical purposes, considering the
typical sizes of both experimental and theoretical
uncertainties. The use of a smaller number of points to compute the
convolutions allows a much faster
evolution, advantageous in the context of a PDF fit.

We also checked the linearized solutions Eqs.(\ref{ns-nlo-lin}) 
and (\ref{s-nlo-lin}) against {\tt PEGASUS}, and obtained a similar 
level of agreement.

\subsection{Parametrization of input PDFs}

\label{sec:pdfbas}

The non-perturbative input to the present analysis are five  
PDFs, parametrized with neural networks, at a fixed
initial evolution scale, which we choose to be
 $Q_0^2=2$ GeV$^2$. PDFs at higher values of $Q^2$
are then determined by perturbative 
evolution, as discussed in 
Sec.~\ref{sec:eveq} and Sec.~\ref{sec:heavy}.

The most unbiased approach in a PDF analysis 
would be to parametrize all seven independent light
PDFs at the initial evolution scale $Q_0^2$.
However, since the experimental data sets which are used in 
the present analysis give very little constraint on
the strange PDFs, we choose for economy to independently parametrize
only the gluon and   
the four
lightest quark flavours,  $u,\bar{u},d,\bar{d}$, and 
fix $s$ and $\bar{s}$ through two constraints.
Also,  we determine  all 
heavy quark PDFs from perturbative evolution, thereby neglecting
intrinsic heavy quark contributions.

We have the flexibility to select any basis for our PDFs: 
the neural nets have sufficient flexibility to accommodate any
reasonable choice. With the standard PDF fitting framework,
this is not necessarily the case since specific 
functional forms are chosen so that
at least some parameters have a physical interpretation:
well known examples are 
the large-$x$ parameters which are related to counting rules, and
the small-$x$ exponents typically inspired by Regge theory. Therefore,
in standard parametrization the choice of a specific basis is likely
to affect the form of the results~\cite{Thorne:2007fe}, whereas 
we will be able to verify explicitly in Sect.~\ref{sec:stabarch} the
independence of the  parametrization of our result.

The specific basis we choose at $Q_0$ is given by the
following linear combinations:
 \begin{itemize}
\item the singlet distribution, $\Sigma(x)\equiv \sum_{i=1}^{n_f}\lp
q_i(x)+\bar{q}_i(x)\rp$, 
\item the total valence, $V(x)\equiv \sum_{i=1}^{n_f}\lp
q_i(x)-\bar{q}_i(x)\rp$, 
\item the non-singlet triplet, $T_3(x) \equiv 
\lp u(x)+\bar{u}(x)\rp - \lp d(x)+\bar{d}(x)\rp$, 
\item the sea asymmetry distribution, $\Delta_S(x)\equiv \bar{d}(x)-
\bar{u}(x)= \smallfrac{1}{2}(V_3(x)-T_3(x))$,
\item the gluon, $g(x)$.
\end{itemize}

For the strange quarks we make two assumptions: 
\be
\label{eq:strangeass}
s(x)=\bar{s}(x)=\smallfrac{1}{2}C_s \lp \bar{u}(x)+\bar{d}(x) \rp \ . 
\ee
We set the constant $C_s$, the  ratio between strange and
non-strange sea,
to  the value $C_s=0.5$, which is approximately equal to the relative
size of the respective contribution to the nucleon momentum. 
Recent dimuon data~\cite{chorus-dimuon,Goncharov:2001qe}  tend   
to favor somewhat smaller
values of the momentum fraction carried by strange
quarks~\cite{Lai:2007dq}, however the choice to fix $C_s$ at 
the ratio of momentum
sum rules is {\it per se} arbitrary and it should be understood as a
rough approximation.

The assumption that all heavy quarks are generated radiatively, as 
described in Sec.~\ref{sec:heavy}, is implemented by taking
$c(x)=\bar{c}(x)=b(x)=\bar{b}(x)=t(x)=\bar{t}(x)=0$ at the initial 
scale $Q_0^2$. The vanishing of intrinsic heavy flavour contributions 
should be taken as an
approximation, justified by the fact that the   
intrinsic charm contribution is likely to be
small~\cite{Pumplin:2007wg}, 
and it is almost unconstrained by the data in our fit.
The assumption of using only five independent parton distributions is
thus a source of theoretical uncertainty. As for all theoretical
uncertainties, the only way of accurately assessing its impact is to
study how results change when a more accurate theory is used. 

On top of the constraints from experimental data,
PDFs have to satisfy a set of sum rules which
follow from conservation laws.
The sum rules implemented in our analysis will be
the momentum sum rule,
\be
\label{eq:momentumsr}
\int_0^1 dx~ x\lc \Sigma(x) + g(x) \rc=1 \ ,
\ee
and the valence sum rules 
\be
\label{eq:valencesr}
\int_0^1 dx~(u(x)-\bar{u}(x)) =2 \ , \quad 
\int_0^1 dx~(d(x)-\bar{d}(x)) =1 \ .
\ee
Note that once the sum rules are satisfied at the initial
evolution scale $Q_0^2$, they will be satisfied
for any other values of $Q^2$. The implementation of the sum 
rules in our approach will be described in Sec.~\ref{sec:net-param}.

\subsection{Hard cross-sections and physical observables}
\label{subsec:hard}

To determine the input PDFs we must not only be able to evolve them to 
a particular scale, but we must then compute physical observables to compare 
to experimental data. This involves convolution of the evolved PDFs with 
hard coefficient functions Eq.~(\ref{eq:fact}). As explained in 
Sec.~\ref{ltev}, this may be done most efficiently by pre-computing the 
hard kernels Eq.~(\ref{eq:genevfact}), which can then be convoluted with 
the initial PDFs as in Eq.~(\ref{eq:factinit}). In Mellin space 
\be
\label{eq:genevfactMel}
K_{Ij}(N,\as,\as^0) = \sum_k C_{Ik}(N,\as)\Gamma_{kj}(N,\as,\as^0),
\ee
so the most efficient procedure is to compute $K_{Ij}(N,\as,\as^0)$, and invert
the Mellin transform using formulae corresponding to Eqs.~(\ref{eq:xkernelns},
\ref{eq:xkernels}), i.e. 
\be
  \label{xks}
  K_{Ij}(x,\as,\as^0)=\begin{cases}
\int_{C}
 \frac{dN}{2\pi i}x^{-N}K_{Ij}(N,\as,\as^0),\quad &{\rm if} \quad j=T,V,\\
  x\,\int_{C}
        \frac{dN}{2\pi i}x^{-N}K_{Ij}(N-1,\as,\as^0)\quad 
&{\rm if} \quad j=\Sigma,g.
\end{cases}
\ee

The convolutions Eq.~(\ref{eq:factinit}) can then be performed using 
formulae analogous to Eqs.~(\ref{evolfullns},\ref{evolfulls}): 
writing $F_I(x,Q^2)=\sum_j F_{Ij}(x,Q^2)$, for the 
nonsinglet contributions (i.e. $j=T,V$) 
\bea
\label{kernconns}
F_{Ij}(x,Q^2)
&=&
\lp  \kappa_{Ij}(\as,\as^0)
-\int_0^x \! dy\, K_{Ij}(y,\as,\as^0)\rp f_{j}(x,Q_0^2) 
\nonumber\\
&&\qquad + \int_x^1 \!\frac{dy}{y}\,
K_{Ij}(y,\as,\as^0)\lp f_{j}\lp \frac{x}{y},Q_0^2\rp
- y f_{j}(x,Q_0^2),\rp. 
\eea
while for the singlets (i.e. $j=\Sigma,g$)
\bea
\label{kerncons}
F_{Ij}(x,Q^2) 
&=&
\lp  \kappa_{Ij}(\as,\as^0)
-\int_0^x \! dy\, yK_{Ij}(y,\as,\as^0)\rp
f_{Ij}(x,Q_0^2) \nonumber \\
&&\qquad + \int_x^1 \!\frac{dy}{y}\,
K_{Ij}(y,\as,\as^0)\lp f_{j}\lp \frac{x}{y},Q_0^2\rp
- y^2 f_j(x,Q_0^2)\rp, 
\eea
where
\be
\label{Kmoms}
  \kappa_{Ij}(\as,\as^0) 
=
\begin{cases} 
\int_0^1\!dx\, K_{Ij}(x,\as,\as^0) 
= K_{Ij}(N,\as,\as^0)\vert_{N=1},\quad &{\rm if}\quad j=T,V,
\\
\int_0^1\!dx\, x\,K_{Ij}(x,\as,\as^0) 
= K_{Ij}(N,\as,\as^0)\vert_{N=2},\quad &{\rm if}\quad j=\Sigma,g,
\end{cases}
\ee
are all finite constants. The convolutions in 
Eqs.~(\ref{kernconns},\ref{kerncons}) are evaluated in precisely the same 
way as those in Eqs.~(\ref{evolfullns},\ref{evolfulls}), i.e. as described in 
Sec.~\ref{sec:benchmarks}, with all the kernels pre-computed.
It remains to give expressions for the 
Mellin space kernels Eq.~(\ref{eq:genevfactMel}).
There are very many of these, roughly the number of observables times 
the number of PDFs. Complete expressions for all the 
observables and PDFs used in our current analysis may be found in 
Appendix~\ref{sec:kernels}.

Structure functions  computed using the kernels of
Appendix~\ref{sec:kernels} can be compared to experimental data
directly, or after having been combined into  reduced cross-sections as
discussed in Sect.~\ref{sec:obscuts}, with the only addition of
target--mass corrections, to be discussed below. Besides direct
experimental information, a further constraint on the input PDFs comes from
the requirement of positivity. Indeed, even though,
as  well known,
PDFs are not positive-definite beyond LO,
cross sections must remain positive, and this constrains the set of
admissible PDFs~\cite{Altarelli:1998gn}. The implementation of
positivity constraints is nontrivial, because in principle one should
require positivity of all observables, regardless of the fact that
they are measurable in a realistic experiment. In practice, we will
only impose a positivity constraint which has an immediate implication
on the admissible gluon distribution. Namely, we  impose  (in a way to be described in
Sec.~\ref{sec:nn_ga_minim} below) 
positivity of the longitudinal
structure function $F_L(x,Q^2)$ for $Q^2 \ge Q_0^2$ and $x \ge
10^{-5}$. This has the effect of vetoing gluon distributions which
become too negative at small $x$, though a negative gluon remains allowed.
Imposing such a constraint for even smaller values of $x$ is delicate
since $F_L$ has a perturbative instability in this region, which could
only be cured through small-$x$ resummation (see
Ref.~\cite{Altarelli:2008aj} and  
references therein).

\subsection{Target mass corrections}\label{sec:tmcht}
\label{sec:tmc}

We compute all physical observables using leading twist perturbation
theory, and higher twist corrections are kept under control by our
choice of a relatively high kinematic cut, as discussed in
Sect.~\ref{sec:obscuts}. However, we do include Target Mass
Corrections (TMCs)
up to twist four, since these are of purely kinematic origin and can
be determined exactly~\cite{tmc}. The implementation of TMCs  in the
present analysis is different to that in~\cite{DelDebbio:2007ee}: here we
rearrange the TMC so that it is explicitly factorised into the 
hard kernel, and can thus be pre-computed along with the perturbative 
evolution and coefficient functions.

To see how this works, consider first the structure function $F_2(x,Q^2)$.
From Eq.~(4.19) of Ref.~\cite{tmc}, $\widetilde{F}_2$ at twist four is 
given in terms of the leading twist $F_2$ by
\begin{equation}
  \label{eq:tmcformula}
  \widetilde{F}_2(\xi,Q^2)=
          \frac{x^2}{\tau^{3/2}}\frac{F_2(\xi,Q^2)}{\xi^2}
	  + 6\frac{M_N^2}{Q^2}\frac{x^3}{\tau^{2}}I_2(\xi,Q^2)
\end{equation}
where 
\begin{equation}
  \label{tauxi}
    \tau = 1\,+\,\frac{4M_N^2x^2}{Q^2},\qquad
    \xi= \,\frac{2x}{1+\sqrt{\tau}},
\end{equation}
where $M_N$ is the mass of the target, and
\begin{equation}
  \label{eq:i2}
    I_2(\xi,Q^2)=\int_{\xi}^1\!\frac{dz}{z^2}\,F_2(z,Q^2).
\end{equation}
Taking Mellin transforms with respect to $\xi$:
\begin{equation}
  \label{eq:fslt}
  F_2(\xi,Q^2)=
  \sum_j\int_C\! \frac{dN}{2\pi i}\,\xi^{-N}C_{2,j}(N,\as)f_j(N,Q^2),
\end{equation}
while
\be
  \label{eq:i2N}
  \begin{split}
    I_2(N,Q^2)&=
    \int_0^1  d\xi\,\xi^{N-1}
    \int_{\xi}^1\! \frac{dz}{z^2}\,F_2(z,Q^2)\nonumber\\
    &= \bigg[ \frac{\xi^N}{N}\,
    \int_{\xi}^1\,dz\,\frac{F_2(z,Q^2)}{z^2} \bigg]_0^1
    + \frac{1}{N}\int_0^1 \! d\xi\,\xi^{N-2}\,
              F_2(\xi,Q^2)\nonumber\\
    &= \frac{1}{N} F_2(N-1,Q^2),
\end{split}
\ee
so
\be  
\label{eq:i2xi}
I_2(\xi,Q^2) 
    = \int_{C+1}\! \frac{dN}{2\pi i}\,\frac{\xi^{-N}}{N}\,F_2(N-1,Q^2)
    = \frac{1}{\xi}\,\int_C\! \frac{dN}{2\pi i}\,\frac{\xi^{-N}}{N+1}\,
       F_2(N,Q^2).  
\ee
Now, by substituting Eqs.~(\ref{eq:fslt},\ref{eq:i2xi}) 
into Eq.~(\ref{eq:tmcformula}) we obtain
\be
  \label{eq:tmcformulaNEW}
  \widetilde{F}_2(\xi,Q^2)=
  \,\int_C\frac{dN}{2\pi i}\,\xi^{-N}\,
  \left(\frac{x^2}{\tau^{3/2}\xi^2} +
  \frac{6M_N^2}{Q^2}\frac{x^3}{\xi\tau^{2}}\frac{1}{(N+1)}\right)
  \sum_j C_{2,j}(N,\as)f_j(N,Q^2).
\ee
We can reinterpret the factor in front of $C_{2.j}(N,\as)$
as the new target mass corrected coefficient function:
\begin{equation}
  \label{eq:newgamma}
  \widetilde{C}_{2,j}(N,\as,\tau)=
  \frac{(1+\tau^{1/2})^2}{4\tau^{3/2}}
  \left(1+\frac{3\left(1-\tau^{-1/2}\right)}{N+1}\right)C_{2,j}(N,\as).
\end{equation}
The target mass corrected hard kernel is then simply
\begin{equation}
  \label{eq:ktmc}
  \widetilde{K}_{F_2,j}(\xi,\as,\as^0)=
  \sum_k\int_C\frac{dN}{2\pi i}\,\xi^{-N}
\widetilde{C}_{2,k}(N,\as,\tau)\Gamma_{kj}(N,\as,\as^0).
\end{equation}

The same procedure can be applied to find the target mass 
corrections to the $F_3$ and $F_L$ structure functions. For $F_3$, from 
Ref.\cite{tmc} we have
\begin{equation}
  \label{eq:tmc3}
  \widetilde{F}_3(\xi,Q^2)=  \frac{x}{\tau}\frac{F_3(\xi,Q^2)}{\xi}
	  + \frac{4M_N^2}{Q^2}\frac{x^2}{\tau^{3/2}}
            \int_{\xi}^1\! \frac{dz}{z}\,F_3(z,Q^2),
\end{equation}
whence we deduce the target mass corrected coefficient function
\be
  \label{eq:newgamma3}
  \widetilde{C}_{3,j}(N,\as,\tau)=
\frac{1+\tau^{1/2}}{2\tau}
  \left(1\,+\,2\frac{1-\tau^{-1/2}}{N}\right)C_{3,j}(N,\as),
\ee
and thus $\widetilde{K}_{3,j}(\xi,\as,\as^0)$ using an equation analogous to 
Eq.~(\ref{eq:ktmc}). Finally, from Ref.~\cite{tmc}
\begin{equation}
  \label{eq:tmcL}
  \widetilde{F}_L(x,Q^2)= F_L(x,Q^2)+
          \frac{x^2(1-\tau)}{\tau^{3/2}}\frac{F_2(\xi,Q^2)}{\xi^2}
	  + 2\frac{M_N^2}{Q^2}\frac{x^3(3-\tau)}{\tau^{2}}I_2(\xi,Q^2),
\end{equation}
whence
\begin{eqnarray}
  \label{eq:newgamma2}
  \widetilde{C}_L(N,\as)&=&
  C_L(N,\as)+\nonumber\\ &&
  \frac{(1+\tau^{1/2})^2(1-\tau)}{4\tau^{3/2}}
  \left(1-\frac{(3-\tau)(1+\tau^{1/2})}{4\tau^2}\frac{1}{N+1}\right)C_2(N,\as).
\end{eqnarray}

Note that in the limit $M_N^2/Q^2\rightarrow 0$, $\tau \to 1$, 
$\xi\to x$, $\widetilde{C}_{I,j}(N,\as,\tau)\to C_{I,j}(N,\as)$, and 
$\widetilde{K}_{I,j}(\xi,\as,\as^0)\to K_{I,j}(x,\as,\as^0)$ for each of
$I=2,3,L$.

%----------------------------------------------------------------

% --------------------------------------------
%
%\section{Minimization strategy}
%
%-------------------------------------------------
%----------------------------------------------------

\section{Neural networks and fitting strategy}
\label{sec:minim}

In this Section we discuss the parametrization used to represent
the parton densities at the initial scale, the training (i.e. fitting) 
strategy used in our analysis, and the method used to determine the
best fit.

Our approach to the parametrization of 
PDFs is rather different from that which is most commonly adopted. 
Instead of choosing an optimized basis of functions with a relatively
small number of physically motivated parameters,  our PDFs use 
an unbiased basis of functions (provided by neural networks), 
parametrized by a  very large and
redundant set of parameters. 
As a consequence, the determination of the best fit form of the
functions which give the PDF
is not trivial since it is not just given by the absolute minimum of
some figure of merit. Indeed, a redundant parametrization  
may accommodate
not only the smooth shape of the ``true'' underlying PDFs, but also the 
random fluctuations of the experimental data about it. In fact, it is
the possibility of further decreasing the figure of merit which
guarantees that the best fit is not driven by the form of the
parametrization. The best fit is then given by an optimal 
training, beyond which the figure of merit improves because one is 
fitting the statistical 
noise in the data.

 This raises the question of how this best
fit is determined. We do this through the so-called
cross-validation method~\cite{Bishop:1995}, based on the
random separation of the data into  training and validation sets.
Namely, the  PDFs are trained on a fraction of the data
and  validated on the rest of the data.  A stopping criterion
for the whole process emerges when the quality of the fit to
validation data deteriorates while the quality of the fit to
training data keeps improving: this corresponds to the onset of a regime
where neural networks  start to fit random fluctuations rather than the
underlying physics.  

As explained below, fitting the neural networks to the data is
performed by minimization of a suitably defined figure of merit. This is
a complex task for two reasons: we need to find a
minimum in a  very large parameter space, and the figure of merit 
is a nonlocal functional of the set of functions which are being
determined in the minimization.  Carefully tuned genetic 
algorithms turn out to provide
an efficient solution to this  minimization problem.

In summary, the main ingredients of our fitting procedure are:
\begin{enumerate}
\item Neural network parametrization, in order to have a flexible,
redundant parametrization of PDFs.
\item Genetic Algorithm minimization, which allows an efficient
minimization on a large parameter space.
\item Determination of the best fit by cross-validation, in order to
  determine the smooth physical law which underlies statistical fluctuations.
\end{enumerate}
We shall now discuss each of these aspects in turn. The results of the
application of the method to the construction of the NNPDF1.0 parton
set and in particular tests of its stability will then be discussed
in Section~\ref{sec:results}, in particular Section~\ref{sec:stabarch}.

\subsection{Neural network parametrization}
\label{sec:net-param}

Each of the independent PDFs in the evolution basis introduced in
Section~\ref{sec:pdfbas} ($\Sigma,V,T_3,\Delta_S,g$) is parametrized
using a multi-layer feed-forward neural network~\cite{Bishop:1995}
supplemented with a polynomial preprocessing; this procedure is a
straightforward generalization of the framework used in
Ref.~\cite{DelDebbio:2007ee} to the case of several parton
distributions. As explained in Refs.~\cite{f2ns,f2p,DelDebbio:2007ee},
neural networks provide a very flexible and unbiased parametrization
of the PDFs, the only theoretical assumption being smoothness.

The neural networks we use are chosen to have all the same
architecture, namely 2-5-3-1. This corresponds  to 37 free parameters for
each PDF, i.e. a  total of 185 free parameters, to be compared
to less than a total of 30 free parameters for parton fits based on
standard functional
parametrizations~\cite{Pumplin:2002vw,Martin:2002aw,Alekhin:2006zm}.
This choice of architecture is motivated by our previous
studies~\cite{f2ns,f2p}, where it was found that it is adequate for
a fit of the full structure function $F_2(x,Q^2)$, in which both the $x$
and the $Q^2$ dependence are fitted. It is thus surely very redundant
for the fit of a single PDF as a function of $x$ at a fixed initial
scale. Because the aim is to have a redundant parametrization, we do
not find it necessary to use a smaller architecture even for parton
distributions which are poorly known and will thus carry little
information,  such as the light sea asymmetry $\Delta_S$.
The use of a redundant architecture reduces a priori the possibility
of a
functional bias. Lack of bias will be checked a posteriori  
in  Section~\ref{sec:stabarch}, by verifying the independence of
results on the choice of architecture.

The neural network parametrization is then supplemented with a preprocessing
polynomial.  Large enough neural networks can reproduce any functional
form given sufficient training time. However, the training  can be made
more efficient by adding a preprocessing step, i.e. by multiplying the
output of the neural networks by a fixed function. The neural network
then only fits the deviation from this function, which
improves the speed of the minimization procedure if the preprocessing
function is suitably chosen.

We thus write the input PDF basis in terms of neural networks as
follows
\begin{eqnarray}
  \label{eq:PDFbasisnets}
  \Sigma(x,Q_0^2)
  &=&{(1-x)^{m_{\Sigma}}}{x^{-n_{\Sigma}}}{\rm NN}_{\Sigma}(x) \ ,
  \nonumber\\
  V(x,Q_0^2)&=&A_{V}{(1-x)^{m_{V}}}{x^{-n_{V}}}
  {\rm NN}_{V}(x)  \ , \nonumber\\
  T_3(x,Q_0^2)&=& {(1-x)^{m_{T_3}}}{x^{-n_{T_3}}}
  \label{eq:pdfdef}
  {\rm NN}_{T_3}(x)  \ , \\
  \Delta_S(x,Q_0^2)&=&A_{\Delta_S} {(1-x)^{m_{\Delta_S}}}{x^{-n_{\Delta_S}}}
  {\rm NN}_{\Delta_S}(x)  \ ,\nonumber\\
  g(x,Q_0^2)&=&A_g{(1-x)^{m_{g}}}{x^{-n_{g}}}{\rm NN}_g(x) \ .
\nonumber
\end{eqnarray}
The values of the preprocessing exponents $m$ and $n$ for each
PDFs are summarized in Table~\ref{tab:prepexps}. They are chosen by
comparison to the result of available
fits\cite{Pumplin:2002vw,Martin:2002aw,Alekhin:2006zm} based on
functional forms. Results should be independent of them, provided the
training is sufficiently long, when they take reasonable
values. Example 
of unreasonable values would be those which lead
to the divergence of sum rules if the function ${\rm NN}_i(x)$ is constant: the
neural network should then compensate for the divergence, which
eventually would happen, but would lead to very inefficient
training. This leads to the constraints $n<2$ for the singlet and
gluon and $n<1$ for the valence and triplet.
  Independence of our global fit on these choices will be
discussed in
Section~\ref{sec:stabarch}. We have further verified on individual
replicas that results are stable upon removal of the
preprocessing function, provided only the length of training is
greatly increased.

% Randomization of the preprocessing?

\begin{table}
  \begin{center}
    \begin{tabular}{|c|c|c|}
      \hline PDF & $m$ & $n$ \\
      \hline
      $\Sigma(x,Q_0^2)$  & 3 & 
      1.2 \\
      \hline
      $g(x,Q_0^2)$  & 4 & 
      1.2 \\
      \hline
      $T_3(x,Q_0^2)$  & 3 & 
      0.3 \\
      \hline
      $V_T(x,Q_0^2)$  & 3 & 
      0.3 \\
      \hline
      $\Delta_S(x,Q_0^2)$  & 3 & 
      0 \\
      \hline
    \end{tabular}
    \caption{\small \label{tab:prepexps} The preprocessing exponents
      used in the present analysis, defined in
      Eq.~(\ref{eq:pdfdef}).}
  \end{center}
\end{table}

For three of the basis PDF parametrizations
in Eq.~(\ref{eq:pdfdef}), namely $g,V$ and $\Delta_S$, we have
factored out an overall normalization constant. The value of this constant
is determined by requiring that the valence and momentum sum rules 
Eq.~(\ref{eq:valencesr}) and Eq.~(\ref{eq:momentumsr}) be satisfied.
The valence sum rules fix the value of the total valence and sea asymmetry
normalizations to be
\begin{eqnarray}
  A_{V}&=&\frac{3}{ \int_0^1 dx~ \lc (1-x)^{m_{V}} {\rm NN}_{V}(x)
    /x^{n_{V}} \rc }  \ ,\nonumber\\ 
  A_{\Delta_S}&=&\frac{1-\int_0^1 dx~ \lc (1-x)^{m_{T_3}}  {\rm NN}_{T_3}(x)
    /x^{n_{T_3}} \rc
  }{
    2\int_0^1 dx~ \lc (1-x)^{m_{\Delta_S}}  {\rm NN}_{\Delta_S}(x)
    /x^{n_{\Delta_S}} \rc } \ , 
  \label{eq:sumrules1}
\end{eqnarray}
while the momentum sum rule constrains the normalization of the gluon 
density
\begin{equation}
  \label{eq:sumrules2}
  A_{g} = \frac{
    1-\int_0^1 dx~x  \lc (1-x)^{m_{\Sigma}}{\rm NN}_{\Sigma}(x)
    /x^{n_{\Sigma}} \rc 
  }{\int_0^1 dx~x 
    \lc (1-x)^{m_g}{\rm NN}_{g}(x) /x^{n_g}\rc }  \ .
\end{equation}
The integrals are computed numerically each time the parameters of 
the PDF set are modified. We demand an accuracy of $\mathcal{O}(10^{-3})$ 
for these integrals, enough for practical purposes, so this is the
accuracy to which the sum rules will be satisfied.

\subsection{Genetic algorithm minimization}
\label{sec:nn_ga_minim}

As extensively discussed in Ref.~\cite{DelDebbio:2007ee}, the fitting of the
neural networks on the individual replicas is performed by minimizing 
the error function
\begin{equation}
  \label{eq:errfun}
  E^{(k)}[\omega]=\frac{1}{N_{\mathrm{dat}}}\sum_{i,j=1}^{N_{\rm dat}}
                 \left(F_i^{(\mathrm{art})(k)}-F_i^{(\mathrm{net})(k)}\right)
                 \left(\left({\overline{\mathrm{cov}}}^{(k)}\right)^{-1}\right)_{ij}
                 \left(F_j^{(\mathrm{art})(k)}-F_j^{(\mathrm{net})(k)}\right) \ ,
\end{equation}
where the value $F_i^{(\rm net)}$ of the observable corresponding to
the $i-$th data point is computed from the PDFs as
discussed in detail in Section~\ref{evolution}.  
The covariance matrix used for the
minimization is defined in Eq.~(\ref{eq:covmatnn}).

Due to the non--local nature of the error function~(\ref{eq:errfun})
and the complex structure of the parameter space genetic algorithms
turn out to be the most efficient method  for its minimization.  
The procedure we adopted follows closely that
of Ref.~\cite{DelDebbio:2007ee}, to which we refer for a  general
discussion, while here we concentrate on improvements introduced in
the present work. The first of these is  that we
 allow , for each PDF 
$j=1,\ldots,N_{\rm pdf}$, different values of the  mutation 
rates $\eta_{i,j}$, $i=1,\ldots,N_{\rm mut}^{j}$.
This is motivated by the fact that each PDF functionality is
different, and
thus best approached using a specific learning rate.

Furthermore, all mutation rates are adjusted dynamically during the
fitting procedure as a function of the number of iterations $N_{\rm
ite}$
\begin{equation}
  \eta_{i,j} = \eta_{i,j}^{(0)}/N_{\rm ite}^{r_{\eta}} \ . 
\end{equation}
As the algorithm gets closer to the minimum, large mutations become more
likely to increase the value of the error function: they would then
be rejected thus making the algorithm highly inefficient. This is
prevented by
the reduction of the mutation rate as the minimization proceeds.

The initial values of the mutation rates for each PDF are collected in
Table~\ref{tab:ga_params}, together with the other parameters which
control the genetic algorithm.  It has been found that the choice of two
mutations per PDF,  $N_{\rm mut}^{j}=N_{\rm mut}=2$, is  optimal.
To see how this works, consider for instance
the neural network  ${\rm NN}_\Sigma$ for the singlet PDF. 
This is trained
with a genetic algorithm with two mutations which are initially set to
$\eta_{1,\Sigma}=10$ and $\eta_{2,\Sigma}=1$. Both mutation rates then
decrease as $1/{N_{ite}}^{1/3}$. The
learning rates for the remaining PDFs are given in
Table~\ref{tab:ga_params}.

At each iteration of the genetic algorithm we generate $N_{\rm cop}$
copies of the PDF parameters, and perform
$N_{\rm mut}\times N_{\rm pdf}$ mutations on each of the copies. The
copy which yields the lowest value of the figure of merit
Eq.~(\ref{eq:errfun}) is then chosen as a starting point for the following
iteration. We have found no advantage in using probabilistic
methods for the selection of the best PDF parameters. This is because
we are not looking for the absolute minimum of the error
function: rather, the training procedure must be stopped at some point to
avoid overlearning as we shall explain later. The selection of the
copy with the lowest error is then best suited for our strategy.

\begin{table}
  \centering
  \begin{tabular}{|c|c|c|c|c|c|c|c|c|c|}
    \hline 
    $\eta^{(0)}_{i,\mathrm{\Sigma}}$ & $\eta^{(0)}_{i,\mathrm{g}}$ 
    & $\eta^{(0)}_{i,T_3}$  & 
    $\eta^{(0)}_{i,\mathrm{V_T}}$  & $\eta^{(0)}_{i,\mathrm{\Delta_S}}$
    & $N_{\rm ite}^{\rm max}$ & $r_{\eta}$ & $N_{\rm cop}$
    & $E_{\rm sets}$ &$N_{\rm update}$\\
    \hline
    $[10,1]$ & $[10,1]$ & $[1,0.1]$ & $[1,0.1]$ & $[1,0.1]$ & 5000
    & 1/3 & 120 & 3 &10 \\
    \hline
  \end{tabular}
  \caption{Parameters controlling the genetic algorithm minimization.
    Since we work with $N_\mathrm{mut}=2$ there are two entries in
    each column for the values of $\eta^{(0)}$.}
  \label{tab:ga_params}
\end{table}

Since we start from
a random configuration, and since the neural networks allow for 
great flexibility, it may turn out that some or
all the integrals which appear in Eqs.~(\ref{eq:sumrules1}-\ref{eq:sumrules2}) are divergent, especially at earlier 
stages of the fitting. Similarly, some configurations may lead 
to negative values of $F_L$, as 
discussed in Sec.~\ref{subsec:hard}. To suppress these unphysical
configurations we added a large penalty to the error
function Eq.~(\ref{eq:errfun}), which means  that they are never 
selected.

In order to deal more efficiently with the needs of fitting data from
a wide variety of different experiments and different data sets within
an experiment we adopt a weighted fitting technique, following our
earlier study in Ref.~\cite{DelDebbio:2007ee}.  The aim of the
technique is to let the minimization procedure converge rapidly
towards a configuration for which the final $\chi^2$ is even among all
the experimental sets. Weighted fitting consists of adjusting the
weights of the data sets in the determination of the error function
during the minimization procedure according to their individual figure
of merit: data sets that yield a large contribution to the error
function get a larger weight in the total figure of merit. In order to
avoid any source of bias, however, weighted fitting is only used at
intermediate stages and it is switched off when approaching the
minimum. 

The way weighted fitting is implemented is by minimizing the error function
\begin{equation}
  \label{eq:weight_errfun}
  E_{\rm wt}^{(k)}=\frac{1}{N_{\mathrm{dat}}}
  \sum_{j=1}^{N_{\mathrm{sets}}}p_j^{(k)} N_{\mathrm{dat},j}E_j^{(k)}\,,
\end{equation}
where $N_{\mathrm{dat},j}$ is the number of data points of the $j-$th set and 
$E_j^{(k)}$ the error function defined in Eq.~(\ref{eq:errfun}) but restricted 
to the points of the $j-$th dataset (see
Table~\ref{tab:exps-sets}). This is therefore a weighted version of
the original error function $E^{(k)}$  Eq.~(\ref{eq:errfun}).
The weights $p_j^{(k)}$ are determined as 
\begin{equation}
  \label{eq:weights}
  p_j^{(k)}=\left(\frac{E_j^{(k)}}{E_{\max}^{(k)}}\right)^2 \ ,
\end{equation}
with $E_{\max}^{(k)}$ being the highest among the $E_j^{(k)}$ at the given GA 
generation. Their values are updated every $N_{\rm update}$
generations, with default $N_{\rm update}=10$

An important feature of the implementation of weighted training is
that weights are given  to individual data sets, as identified in
Table~\ref{tab:exps-sets}, and not just to experiments. This is
motivated by the fact that typically each data set covers a distinct,
restricted kinematic region. Hence, the weighting takes care of the
fact that the data in different  kinematic regions carry different
amounts of  information and thus require unequal amounts of training.

This procedure 
poses the problem that  different sets coming from the same
experiment are correlated with each other, as discussed in
Sect.~\ref{sec:data}, and  these correlations are neglected in the  
evaluation of Eq.~(\ref{eq:weight_errfun}).
To deal with this problem, the weighted training is divided 
in two
stages. In the first stage the
weighted error function Eq.~(\ref{eq:weight_errfun}) is minimized.  
When the total figure of merit is below
a threshold $E_{\rm wt}^{(k)} \le E_{\mathrm{sets}}$, weighted
training is switched off by setting $p^{(k)}_j=1$, and the
unweighted  figure of merit Eq.~(\ref{eq:errfun}) which retains all
correlations
is then minimized until stopping
 (convergence).  
The procedure ensures that first,  a uniform quality of the fit
for all data sets  is achieved, and then the fit is refined using the correct
figure of merit which includes all the information on correlated
systematics.

A final improvement of the minimization procedure makes use of the
stopping criterion which will be described in detail in
Sect.~\ref{sec-dynstop}. Indeed, it might happen that the stopping
criterion Eqs.~(\ref{eq:dec-train}-\ref{eq:dec-valid}) is met
 for one or more  individual experiments. If the criterion were
 generally satisfied, the fit would have reached convergence and
 further training would lead to overlearning, i.e. the fitting of
 fluctuations. Thus, if the criterion is met by a single experiment,
in order to avoid overlearning of that  experiment,
its weight in Eq.~(\ref{eq:weight_errfun}) is
temporarily set to zero, so the experiment is effectively removed from the
training set. The behaviour of the training and validation error
functions for the experiment are then monitored and, and 
 if
it exits the overlearning regime its weight is restored to a nonzero
value.

\subsection{Determination of the optimal fit}
\label{sec-dynstop}

We now come to the formulation of the stopping criterion, which
is designed to stop the fit at the point
where it reproduces the information contained in the data but not its
statistical fluctuations, that is, before the training of the neural networks 
enters the overlearning regime.  The criterion is based on the
cross-validation method, widely used
in the context of neural network training~\cite{Bishop:1995}. Its
application to our case has
been described in detail in Ref.~\cite{DelDebbio:2007ee}.

\begin{table}
  \centering
  \begin{tabular}{|c|c|c|c|c|c|}
    \hline 
    $N_{\mathrm{smear}} $ & $\Delta_{\mathrm{smear}}$ & $\delta_{\rm tr}$
    & $\delta_{\rm val}$ &  $E_{\mathrm{thres}}$ & $N_{\rm gen}^{\rm max}$\\
    \hline
    $45$ & $13$ & $10^{-4}$&  $10^{-4}$ & 6 & 5000\\
    \hline
  \end{tabular}
  \caption{Parameters controlling the best-fit stopping criterion.}
  \label{tab:dynstop}
\end{table}
First, the  data set is partitioned into  training and
validation subsets with fraction $f_{\rm tr}^{(j)}$ and $f_{\rm
val}^{(j)}=1-f_{\rm tr}^{(j)}$ of the data points respectively. The
values of the fractions can in 
general be different for each experiment.  The
points in each set are chosen randomly out of the total dataset. In
our fit, this random partitioning of the data is different for each
replica, thereby ensuring that on average all the information in the
original data set is retained.
Then, the figure of merit
Eq.~(\ref{eq:errfun}) or Eq.~(\ref{eq:weight_errfun}) is minimized
for the points
in the training set, while the corresponding figure of merit for the
points 
in the validation acts as a control: it is computed, but not used for
minimization.  The minimization is stopped when  the figure of merit
keeps improving for  the training set, but it deteriorates for the
validation set.  This
behaviour signals the fact that we are fitting the statistical
fluctuations of the points in the training set rather the underlying
physics which is supposed to describe both the training and the validation
data.

The implementation of this method here follows closely 
Ref.~\cite{DelDebbio:2007ee}.  We take the same value of the training
fraction for all data sets, $f_{\rm tr}^{j}=f_{\rm
tr}=\frac{1}{2}$. In Section~\ref{sec:stabdata} we shall also consider a lower
value $f_{\rm tr}=\frac{1}{4}$ 
and show that it leads to essentially unchanged
results.
The partitioning  of data points into training and validation sets is
done on each data set independently. This ensures that
all data sets (and thus essentially all kinematic regions) are
represented in the training and validation sets for each replica.

\begin{figure}[t!]
  \centering
  \epsfig{width=1.0\textwidth,figure=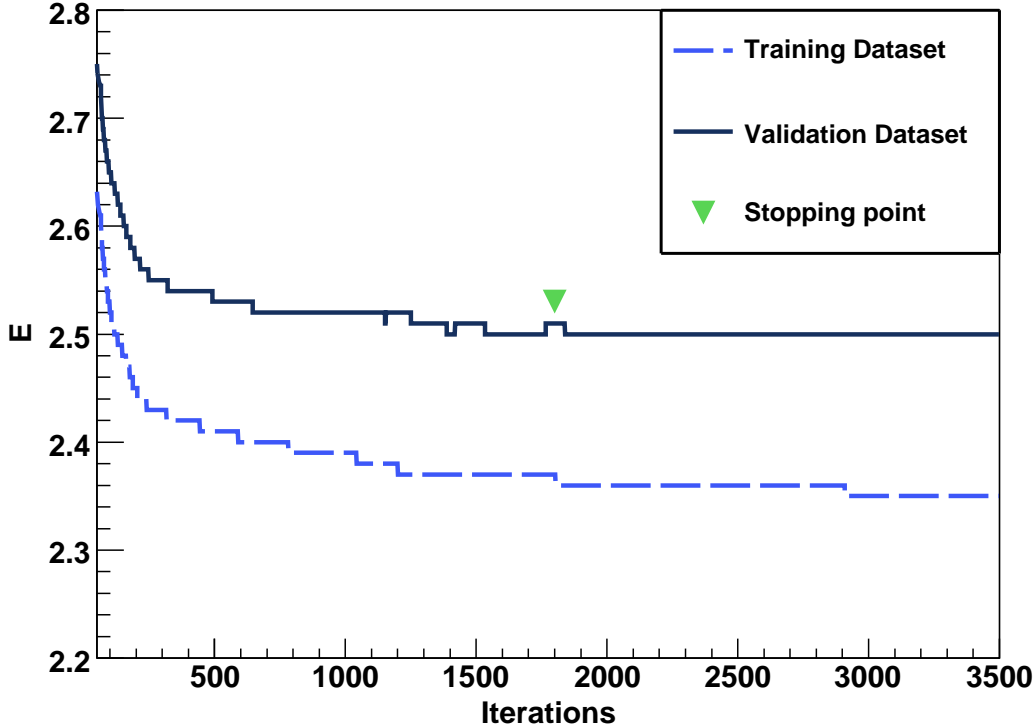}
  \caption{Training and validation error functions as a function of the number 
           of iterations for one of the replicas in the reference fit.}
  \label{fig:stopping}
\end{figure}
Whereas usually when using a genetic algorithm the figure of merit cannot
increase during the minimization, with the
weighted training algorithm discussed in
Sect.~\ref{sec:nn_ga_minim}  the value of the figure of merit during
minimization oscillates due to  updating of the
weights. The wavelength of this oscillation is set by the value of
$N_{\rm update}$, i.e. the frequency
with which weights are updated.
In order to avoid spurious stopping induced by these oscillations, we
apply the stopping criterion to moving averages computed 
over a given number of
iterations, namely
\begin{equation}
  \label{eq:smearing}
  \langle E_{\mathrm{tr,val}}(i)\rangle\equiv
  \frac{1}{N_{\mathrm{sm}}}
\sum_{l=i-N_{\mathrm{sm}}+1}^iE_{\mathrm{tr,val}}(l)\,.
\end{equation}
We take as a  default value for the averaging (``smearing'')
$N_\mathrm{sm}=45$. The value is chosen to be unequal to a multiple of
the wavelength and larger than two full periods, in order to minimize
spurious fluctuations.

The stopping criteria are then
satisfied 
if the averaged training error function is decreasing
\begin{equation}
  \label{eq:dec-train}
  \frac{\langle E_{\mathrm{tr}}(i)\rangle}
  {\langle E_{\mathrm{tr}}(i-\Delta_{\mathrm{smear}})\rangle} 
< 1-\delta_{\rm tr}\, ,
\end{equation}
while the averaged validation error function increases
\begin{equation}
  \label{eq:dec-valid}
  \frac{\langle E_{\mathrm{val}}(i)\rangle}
       {\langle E_{\mathrm{val}}(i-\Delta_{\mathrm{smear}})\rangle} >
 1+\delta_{\rm val}\,.
\end{equation}
The parameters $\delta_{\rm tr}$, $\delta_{\rm val}$ set the accuracy
to which the increase and decrease is required in order to be
significant. Their value has been determined as
$\delta_{\rm tr}=\delta_{\rm val}=10^{-4}$ by verifying that with much
larger values (more than an order of magnitude) 
a significant fraction of fits never stops, while
with much smaller values a sizable fraction of fits stops due to
fluctuations. Results are unchanged upon moderate variations of 
the values of  $\delta_{\rm
  tr},\>\delta_{\rm val}$.

A graphical example of how the stopping criterion works in practice 
is given in Fig.~\ref{fig:stopping}, where the moving averaged
training and validation error functions Eq.~(\ref{eq:smearing}) 
are plotted as a function of the number of generation for one particular
replica of our reference fit, whose training has been artificially
prolonged beyond stopping point. Overlearning is apparent as
a rather small though visible effect: beyond the stopping point the
training figure of merit keeps decreasing steadily while the
validation flattens out and actually rises by a small amount. The
smallness of the rise is a consequence of the fact that the data set
is  very large and  mostly quite
consistent with itself.

\begin{figure}[t!]
  \centering
  \epsfig{width=0.8\textwidth,figure=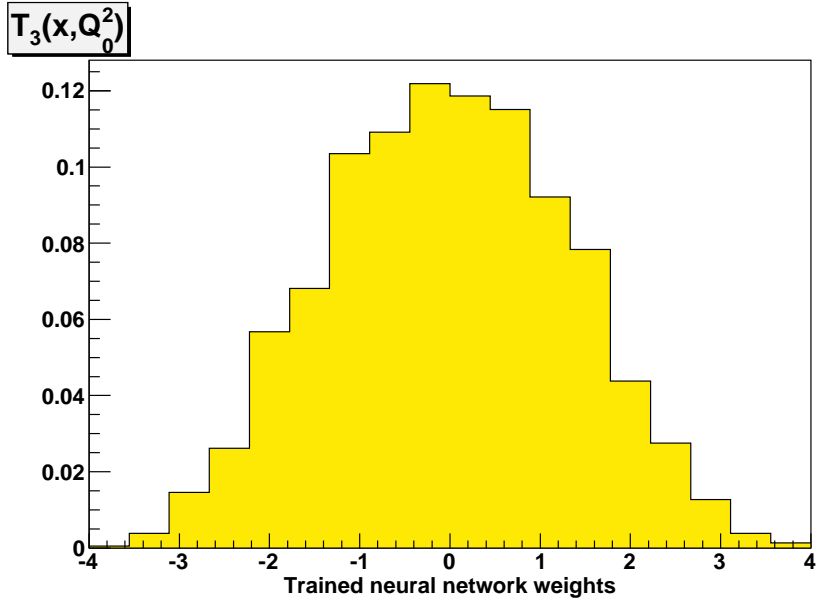}
  \caption{Distribution of neural network weights 
for $N_{\rm rep}=100$ replicas of  the PDF $T_3(x,Q^2)$. The
plot shows the percentage of weights which take the value given in abscissa.}
  \label{fig:nnweights}
\end{figure}
In order to avoid unacceptably long fits,  when
a very large number of 
iterations $N_{\rm gen}^{\rm max}$ is reached the training is stopped
anyway, even if the stopping
conditions Eqs.~(\ref{eq:dec-train},\ref{eq:dec-valid}) are not 
satisfied. This of course leads to loss of accuracy of the
corresponding fits, and it is acceptable provided it only happens for
a small fraction of replicas. We will verify that this is indeed the
case in 
Section~\ref{sec:reffit}.
The full set of stopping  parameters is
summarized in Table~\ref{tab:dynstop}.
We have verified that our final fit is stable against variations of
these parameters as well as those of Table~\ref{tab:ga_params}.

The set of neural nets at stopping provides our best fit, but it is
otherwise impossible to endow  the best fit values of their parameters
with a physical interpretation. In fact, 
since the nets are redundant, the values of most of these parameters
is unconstrained or zero. As an example, 
the distribution of neural network weights at stopping for 100
replicas of the triplet neural network $T_3(x,Q_0^2)$ is displayed
in 
Fig.~\ref{fig:nnweights}.  The well-balanced distribution of weights
around zero in Fig.~\ref{fig:nnweights} shows that the individual
neurons in the neural network operate in their natural range of
sensitivity.

\clearpage

% --------------------------------------------
%
%\section{Results}
%
%-------------------------------------------------
%%--------------------------------------------------------

\section{Results}
\label{sec:results}

We have used the methodology discussed in the previous sections to
produce a set of parton distributions, which we refer to as the
NNPDF1.0 parton set. As discussed in Section~\ref{sec:pdfbas}, this
parton set is based on five independent parton distributions,
corresponding to the two light flavours and the gluon; the strange
distribution is assumed to be proportional to the light sea according
to Eq.~(\ref{eq:strangeass}), while heavy flavours are generated
dynamically using a ZM-VFN scheme, as discussed in
Section~\ref{sec:heavy}. Evolution is performed at NLO as discussed
in Sections~\ref{sec:eveq}-\ref{sec:xevol}, with
$\alpha_s(M_Z^2)=0.119$. The heavy 
quark thresholds are at $m_c=Q_0=\sqrt{2}$ GeV, 
$m_b=4.3$ GeV and $m_t=175$ GeV.

We now present the NNPDF1.0 PDFs, describe some general statistical 
features of the fits,
compare our PDFs to other available parton sets, investigate their statistical
uncertainties by discussing their stability with respect to changes in some of 
the underlying assumptions, and discuss 
the theoretical uncertainties related
to the perturbative order and value of $\alpha_s$. We shall then
present some results obtained using the NNPDF1.0 set for DIS
observables as well as for 
a few benchmark LHC observables, and compare them to
those found using other available sets.
The methodology used to obtain estimates for central values and errors
from our parton set is summarized and compared to that used with
other sets in Appendix~\ref{sec:app-pdferr}.

\subsection{The NNPDF1.0 parton set: statistical features}
\label{sec:reffit}

Our full parton set consists of an ensemble of
$N_{\rep}=1000$ sets of five PDFs. The general statistical features of the
global fit are summarized in  Tables~\ref{tab:est-fin}-\ref{tab:est-fin-tot}. Here $\langle {\rm
  TL}\rangle$ denotes the average number of  iterations of the genetic
algorithm at stopping,
as defined in Sect.~\ref{sec:nn_ga_minim}-\ref{sec-dynstop}. The other 
statistical estimators are defined in Appendix~B
of Ref.~\cite{DelDebbio:2007ee} 

%------------------------------------------------------------
\begin{figure}[b!]
\centering
\epsfig{width=0.49\textwidth,figure=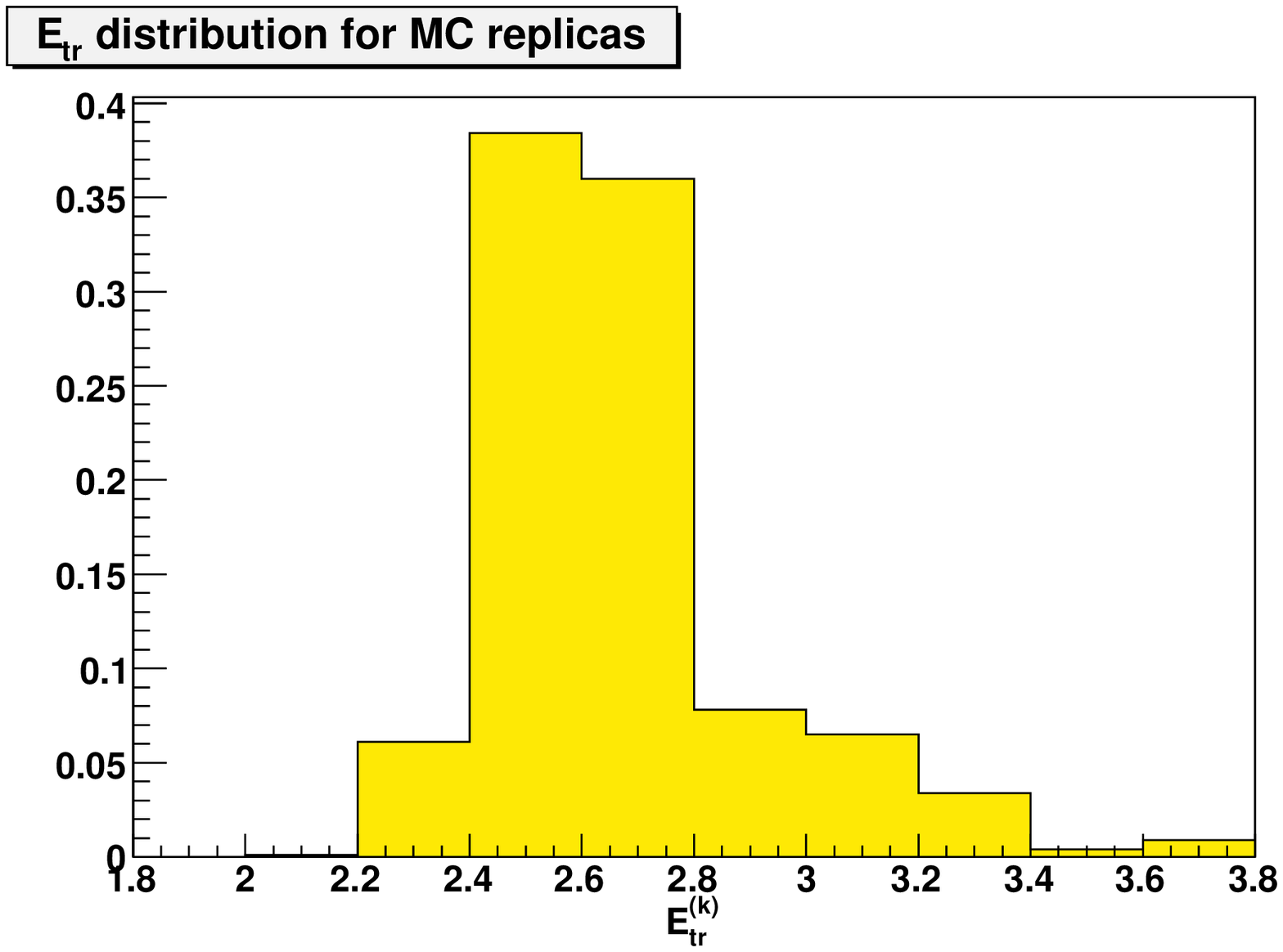} 
\epsfig{width=0.49\textwidth,figure=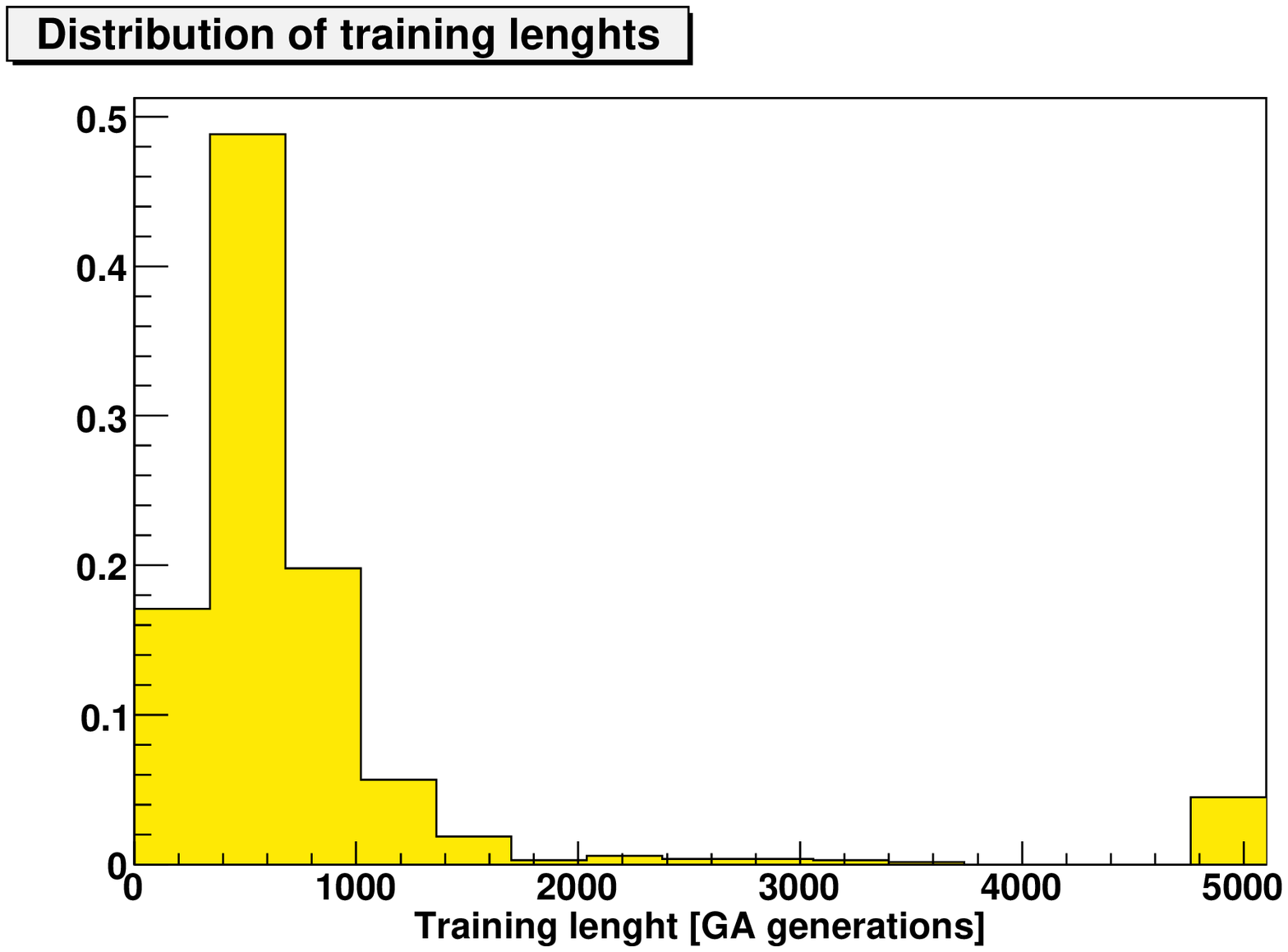} 
\caption{\small Left: distribution of $E^{(k)}_{\rm tr}$ and
of ${\rm TL}^{(k)}$ among the set of trained replicas. The percentage
value in given in the $y$ axis.}
\label{fig:ertot}
\end{figure}
%------------------------------------------------------------------

\begin{figure}[t!]
\centering
\epsfig{width=0.49\textwidth,figure=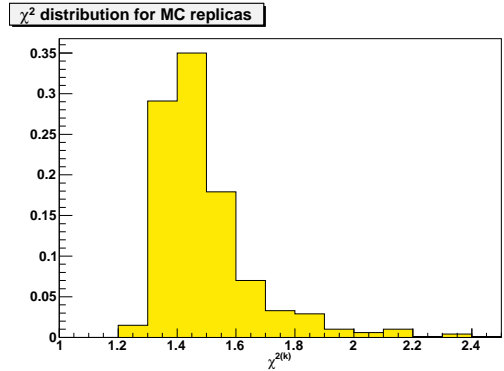} 
\caption{\small Distribution of $\chi^2$ to the fit to actual data of
  each replica. The percentage
value in given in the $y$ axis.}
\label{fig:chittot}
\end{figure}

The distribution of values of the error function Eq.~(\ref{eq:errfun})
at stopping is displayed in
Fig.~\ref{fig:ertot} along with the
distribution of training lengths. The former appears to be
approximately gaussian and the latter approximately poissonian, with a
long tail which causes a slight accumulation of points at $N_{\rm
  gen}^{\rm max}=5000$ iterations (see Table~\ref{tab:dynstop}). This
may cause a loss in accuracy in outlier fits, which however make up fewer
than 5\% of the total sample. For completeness, we also show in
Fig.~\ref{fig:chittot} the
distribution of values of the $\chi^2$ of a global fit to the data
obtained by using each individual replica PDF set, instead of the
best-fit PDF set, whose shape is very similar to that of the
distribution of error functions.

The features of the fit can be summarized as follows:
\begin{itemize}
\item The $\chi^2$ of the central fit (obtained averaging over all PDFs
  in the sample) is about half of the average value $\langle E\rangle$
  of the error function Eq.~(\ref{eq:errfun}) which measures
  the quality of the fit of each set to its corresponding replica. This
  is as expected for a fit which correctly reproduces a  ``true''
  behaviour underlying a set of experimental
  data~\cite{f2ns,f2p,DelDebbio:2007ee}, given that replicas 
  fluctuate about the experimental measurements, which in turn
  fluctuate about ``true'' values. 
It thus supports the use of this
  central fit as our best fit (see
  Appendix~\ref{sec:app-pdferr}). A similar conclusion is obtained by
  comparing the distribution of values of the error function of
Fig.~\ref{fig:ertot} with the distribution of values of the $\chi^2$
shown in 
Fig.~\ref{fig:chittot}.
\item The quality of the central fit as measured by its $\chi^2=1.34$
  is close to its expected value. With several thousands of data points, the
  statistical fluctuations of $\chi^2$ are of order of a few percent;
  moreover, the inaccuracy of the NLO theory used here is likely to
  contribute 5-10\% percent to the value of $\chi^2$. Therefore,
  this value of $\chi^2$ might indicate some minor data inconsistency.
 We shall come back to the issue of the self-consistency of the data in
 Sect.~\ref{sec:stabdata}.

\item The quality of the fit to individual experiments is fairly
  uniform. The $\chi^2$ only grows somewhat larger for the NMC
  experiment, which is known to have consistency
  problems~\cite{f2p,Pumplin:2002vw}.
\item The uncertainty of 
 the fit, as measured by the standard deviation 
$\langle\sigma\rangle$ is rather smaller than 
that of the experimental data: $0.014$ versus $0.055$. 
Correspondingly, the correlation $\langle\rho\rangle$
 between pairs of data points is larger by almost the same factor in the
 fit than in the experimental data. This can be either a consequence 
of the fact that the fit is
 biased by an exceedingly restrictive parametrization, or that it is
 reproducing an underlying physical law which the data consistently
 follow~\cite{f2ns,f2p,Pumplin:2002vw}. We will show in
 Sect.~\ref{sec:stabarch} that the former possibility is strongly
 disfavoured.  Therefore, we conclude that the latter is the case:
 this reinforces the conclusion that our central fit provides an
 optimal best fit to the data.
\end{itemize}

\begin{table}[t!]
\begin{center}
\begin{tabular}{|c|c|}
\hline 
$\chi^{2}_{\tot}$ &      1.34 \\
\hline
\hline
$\la E \ra $   &       2.71      \\
$\la E_{\rm tr} \ra $   &       2.68      \\
$\la E_{\rm val} \ra $   &       2.72      \\
$\la{\rm TL} \ra $   &     824      \\
\hline
 $\la \sigma^{(\exp)}
\ra_{\dat}$ &  $5.6~10^{-2}$\\
 $\la \sigma^{(\net)}
\ra_{\dat}$&  $1.4~10^{-2}$ \\
\hline
 $\la \rho^{(\exp)}
\ra_{\dat}$ &  0.15\\
 $\la \rho^{(\net)}
\ra_{\dat}$&  0.40\\
\hline
 $\la {\rm cov}^{(\exp)}
\ra_{\dat}$ &  $1.0~10^{-3}$\\
 $\la  {\rm cov}^{(\net)}
\ra_{\dat}$&  $1.6~10^{-4}$\\
\hline
\end{tabular}

\end{center}
\caption{\small Statistical estimators for the
final PDF set 
 with $N_{\rep}=1000$ for the total data set. \label{tab:est-fin-tot}}
\end{table}

\begin{table}[t!]
\begin{center}
{
\tiny
\begin{tabular}{|c|c||c|c|c|c|c|c|c|}
\hline 
Experiment    & $\chi^{2}_{\tot}$  & $\la E\ra $   & $\la \sigma^{(\exp)}\ra_{\dat}$ & $\la \sigma^{(\net)}\ra_{\dat}$ & $\la \rho^{(\exp)}\ra_{\dat}$ & $\la \rho^{(\net)}\ra_{\dat}$ & $\la \rm cov^{(\exp)}\ra_{\dat}$ & $\la \rm cov^{(\net)}\ra_{\dat}$\\
\hline
SLAC     &    1.27&   3.23& $1.0~10^{-2}$& $6.0~10^{-3}$& 0.31& 0.69& 
$3.1~10^{-5}$& $2.6~10^{-5}$\\
\hline
BCDMS     &   1.59&   3.10& $7.0~10^{-3}$& $3.9~10^{-3}$& 0.47& 0.55&
$ 2.9~10^{-5}$& $9.3~10^{-6}$\\
\hline
NMC      &    1.70&   3.01& $1.7~10^{-2}$& $7.2~10^{-3}$& 0.16& 0.67&
 $4.4~10^{-4}$& $3.4~10^{-5}$\\
\hline 
NMC-pd   &    1.53&   2.98& $1.5~10^{-2}$& $1.2~10^{-2}$& $3.3~10^{-2}$& 0.40& 
$6.5~10^{-6}$& $5.0~10^{-5}$\\
\hline
ZEUS      &   1.11&   2.53& $6.1~10^{-2}$& $1.0~10^{-2}$& 
$7.9~10^{-2}$& 0.28& $1.5~10^{-4}$& $2.6~10^{-5}$\\
\hline
H1        &   1.03&   2.41& $4.7~10^{-2}$& $1.0~10^{-2}$& $2.7~10^{-2}$& 0.27& 
$4.9~10^{-2}$& $2.5~10^{-5}$\\
\hline
CHORUS    &   1.40&   2.77& 0.11& $2.7~10^{-2}$& $9.4~10^{-2}$&0.43 &
 $2.2~10^{-3}$& $3.3~10^{-4}$\\
\hline
FLH108   &    1.62&   2.61& 0.17& $1.5~10^{-2}$& 0.65& 0.77& 
$2.0~10^{-2}$ & $1.8~10^{-4}$\\
\hline
\end{tabular}
}

\end{center}
\caption{\small Statistical estimators for the
final PDF set 
 with $N_{\rep}=1000$ for individual experiments. \label{tab:est-fin}}
\end{table}

%\end{minipage}
%\end{center}

\subsection{The NNPDF1.0 parton set: results}
\label{sec:resfit}

%------------------------------------------------------------
\begin{figure}[t!]
\centering
\epsfig{width=0.49\textwidth,figure=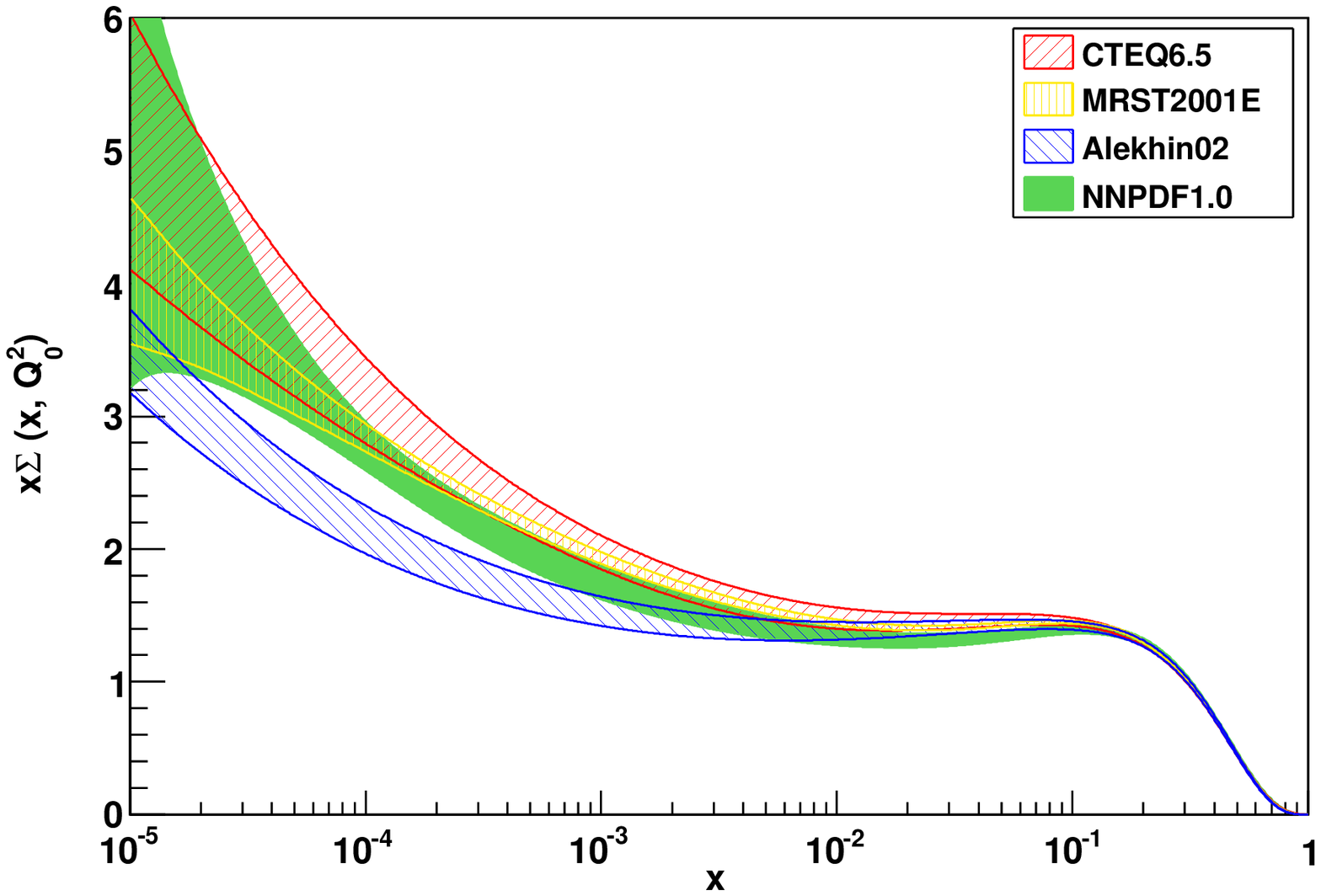}
\epsfig{width=0.49\textwidth,figure=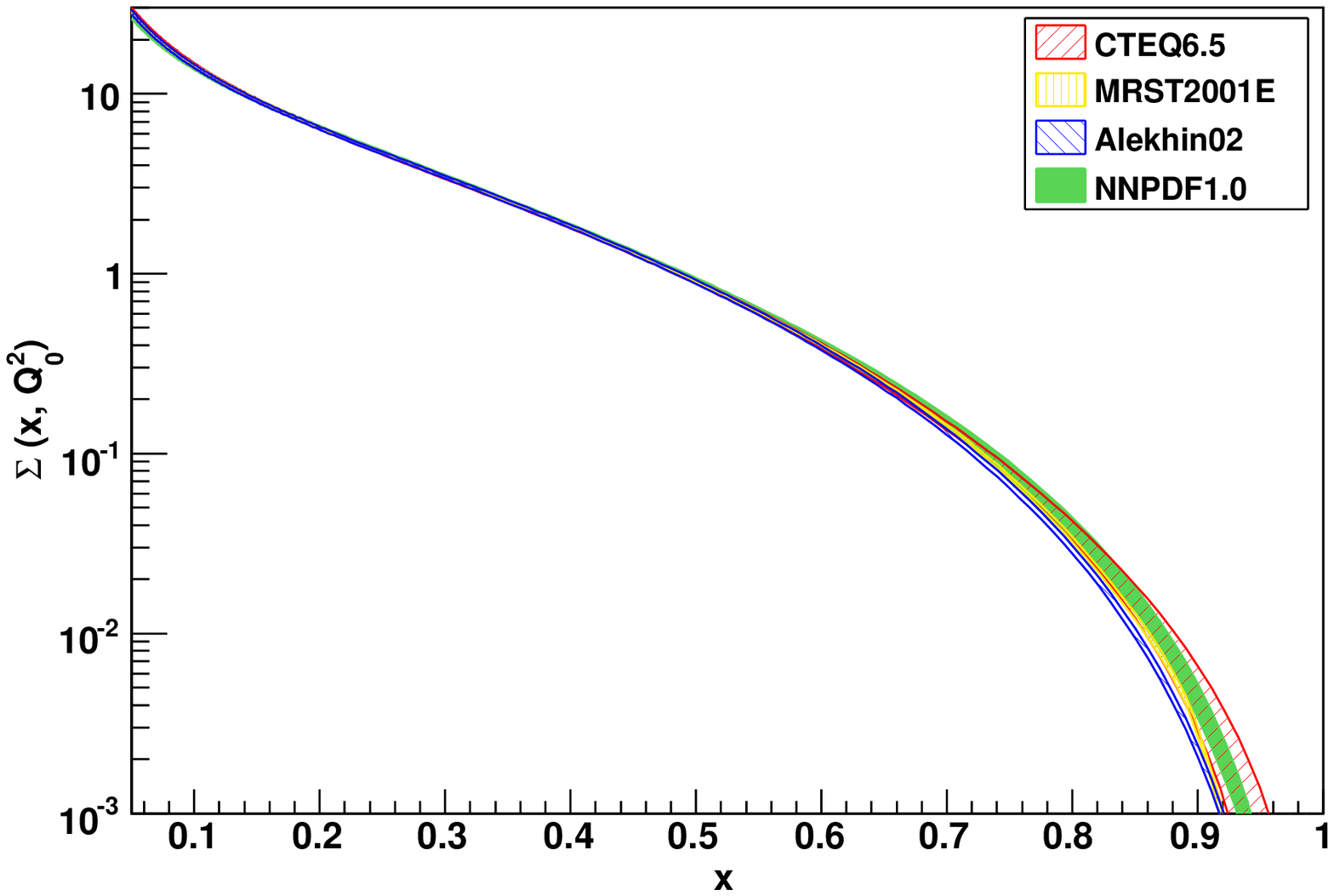}
\epsfig{width=0.49\textwidth,figure=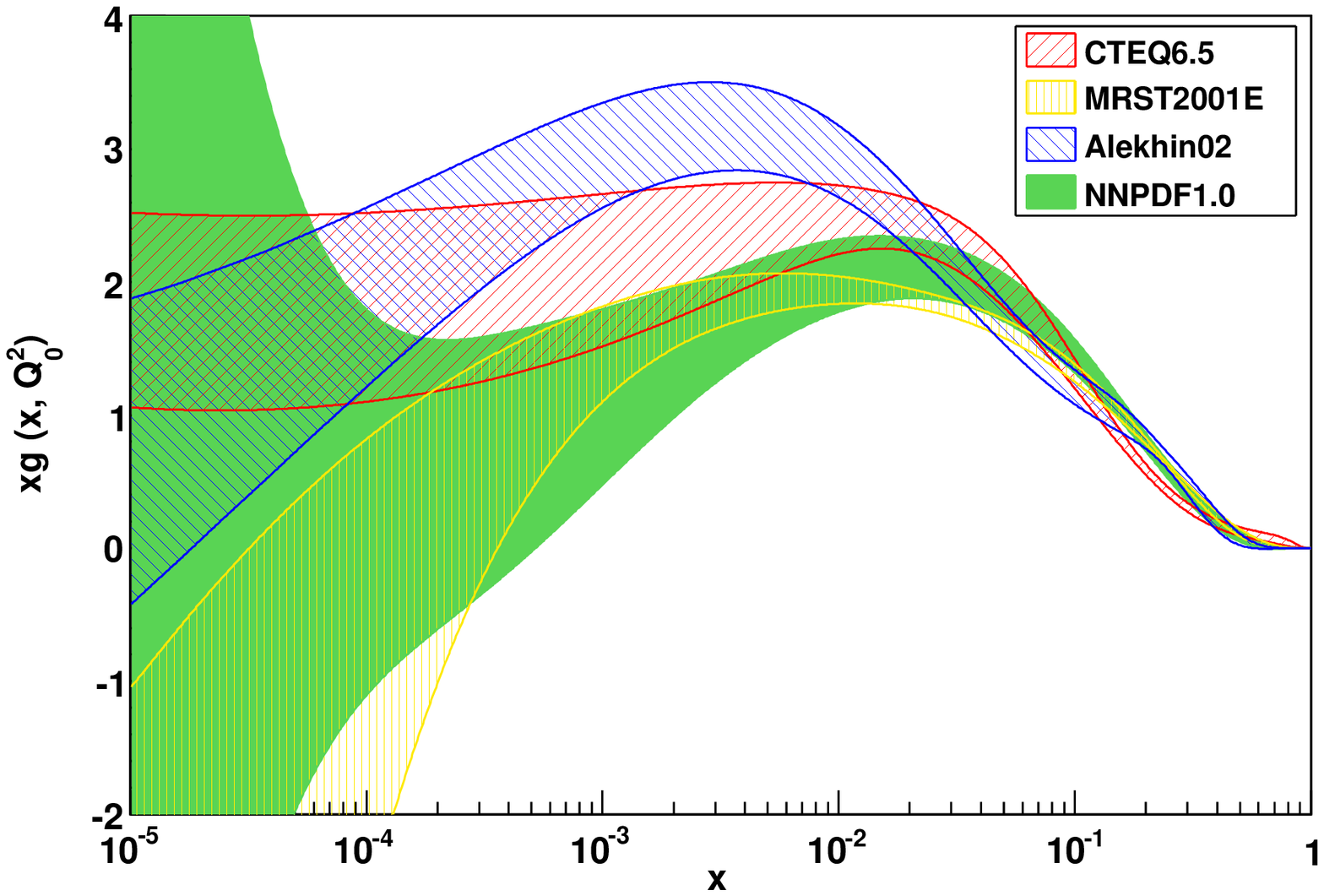}
\epsfig{width=0.49\textwidth,figure=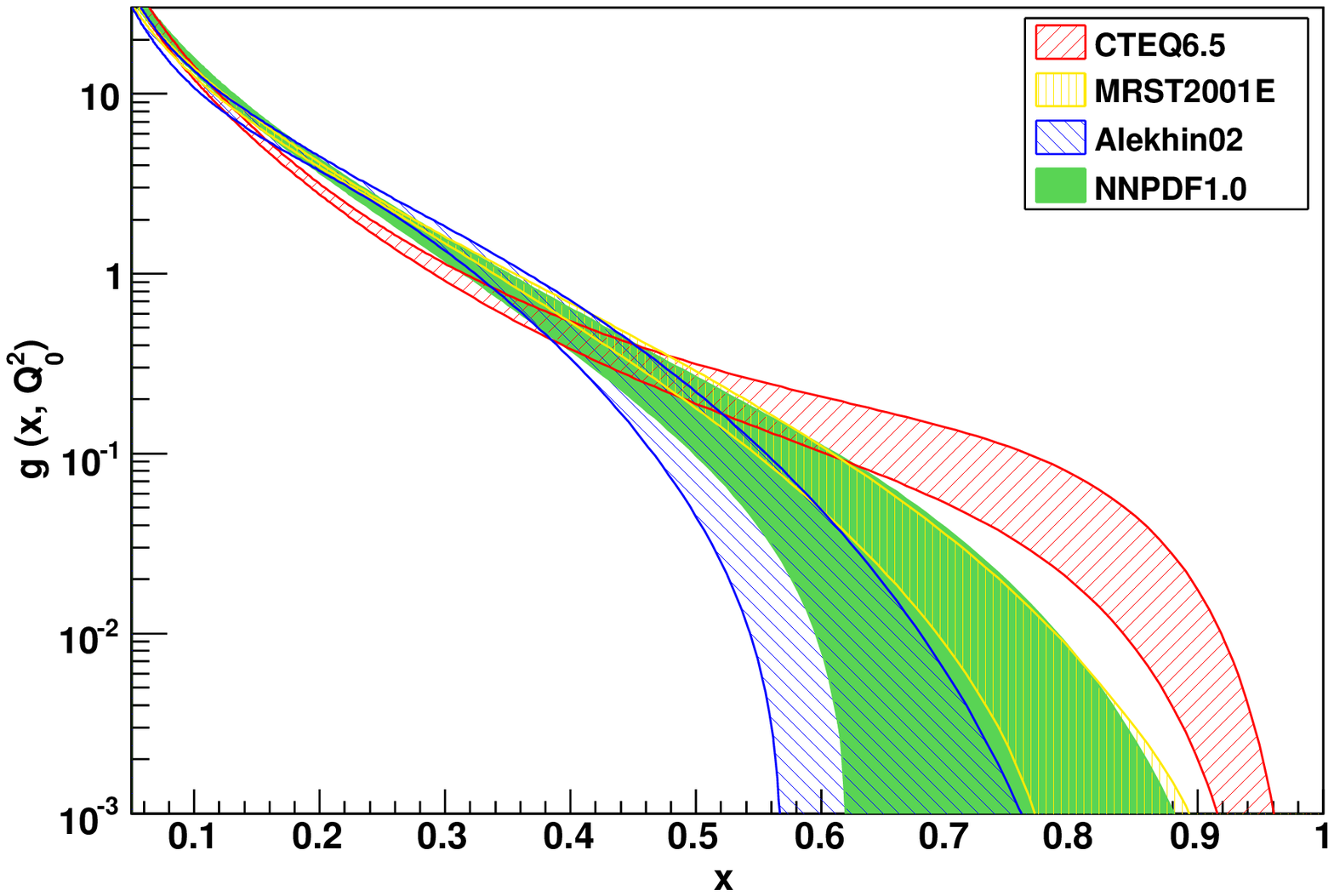}
\caption{\small The singlet and gluon PDF at the starting  scale
$Q_0^2 = 2$ GeV$^2$, plotted versus $x$  on a log (left) or 
 linear
scale (right).}
\label{fig:final-pdfs}
\end{figure}
%------------------------------------------------------------------

%------------------------------------------------------------
\begin{figure}[t!]
\centering
\epsfig{width=0.49\textwidth,figure=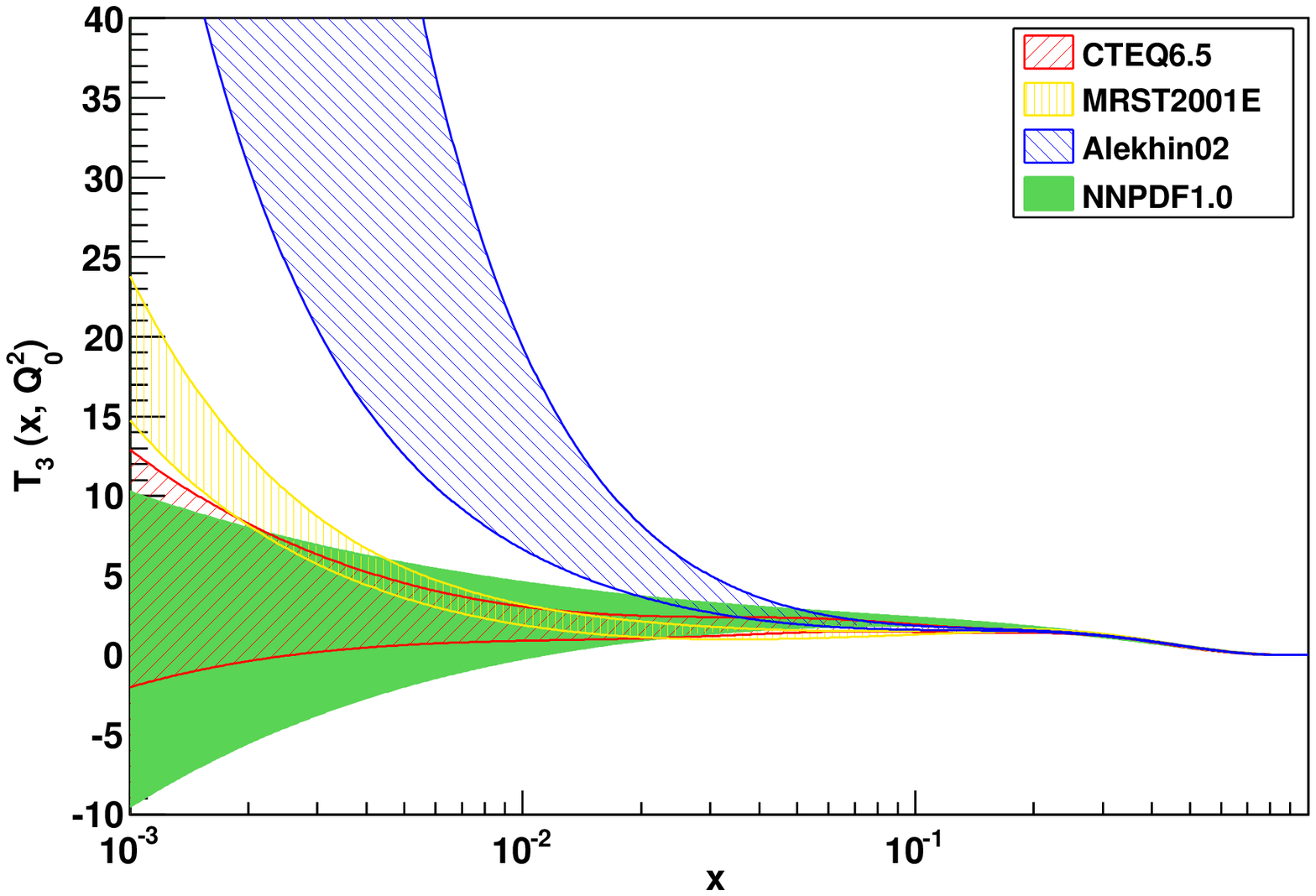}
\epsfig{width=0.49\textwidth,figure=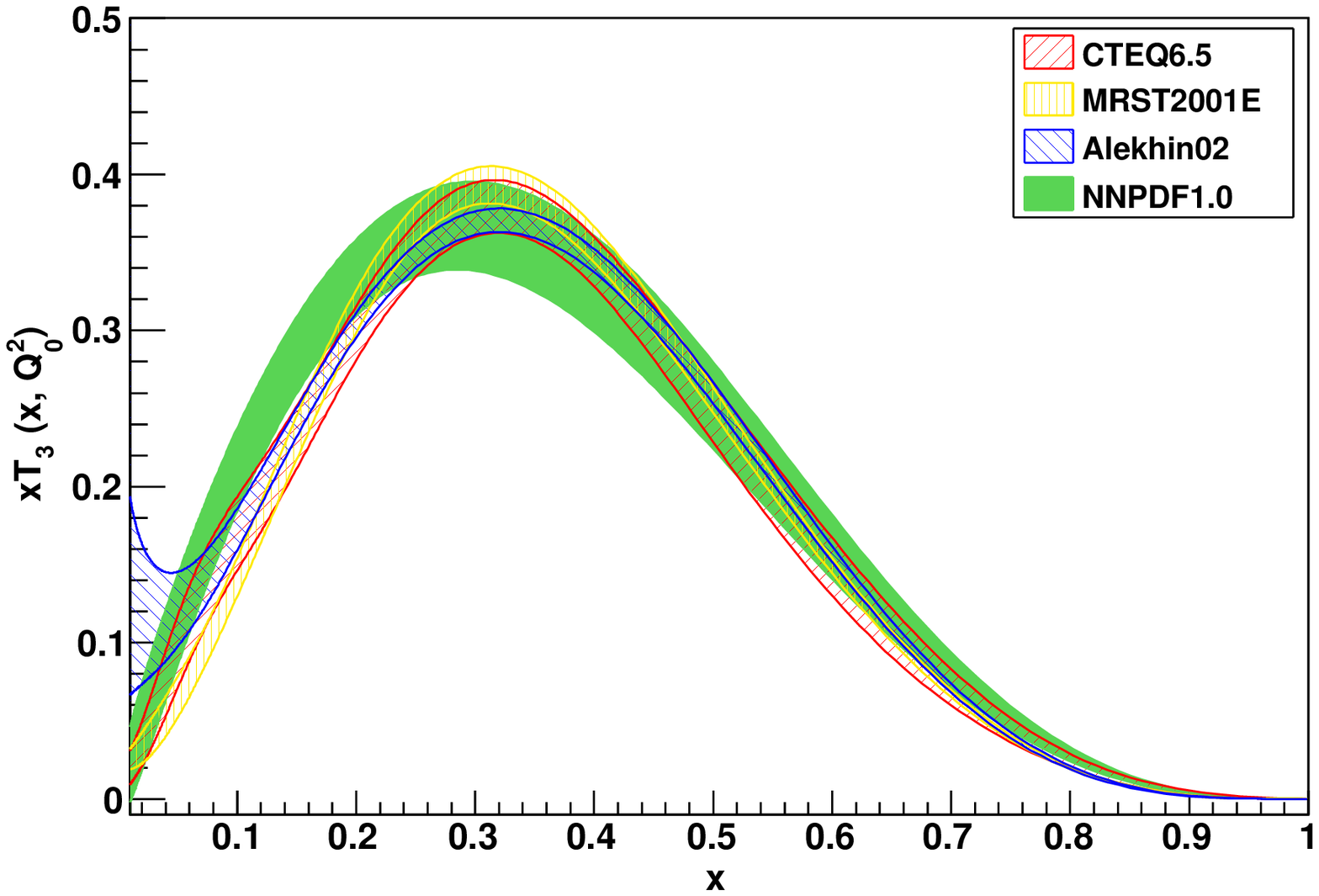}
\epsfig{width=0.49\textwidth,figure=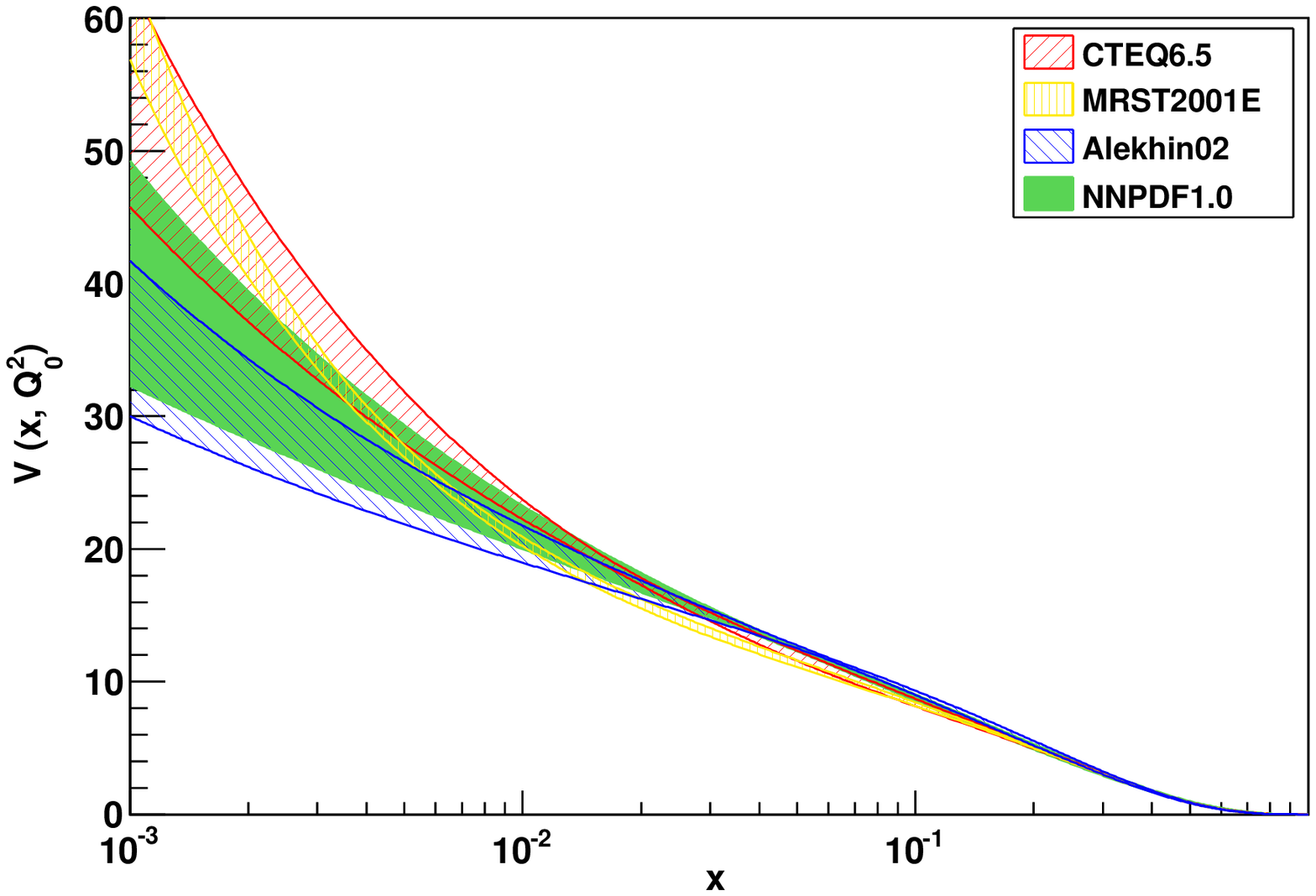}
\epsfig{width=0.49\textwidth,figure=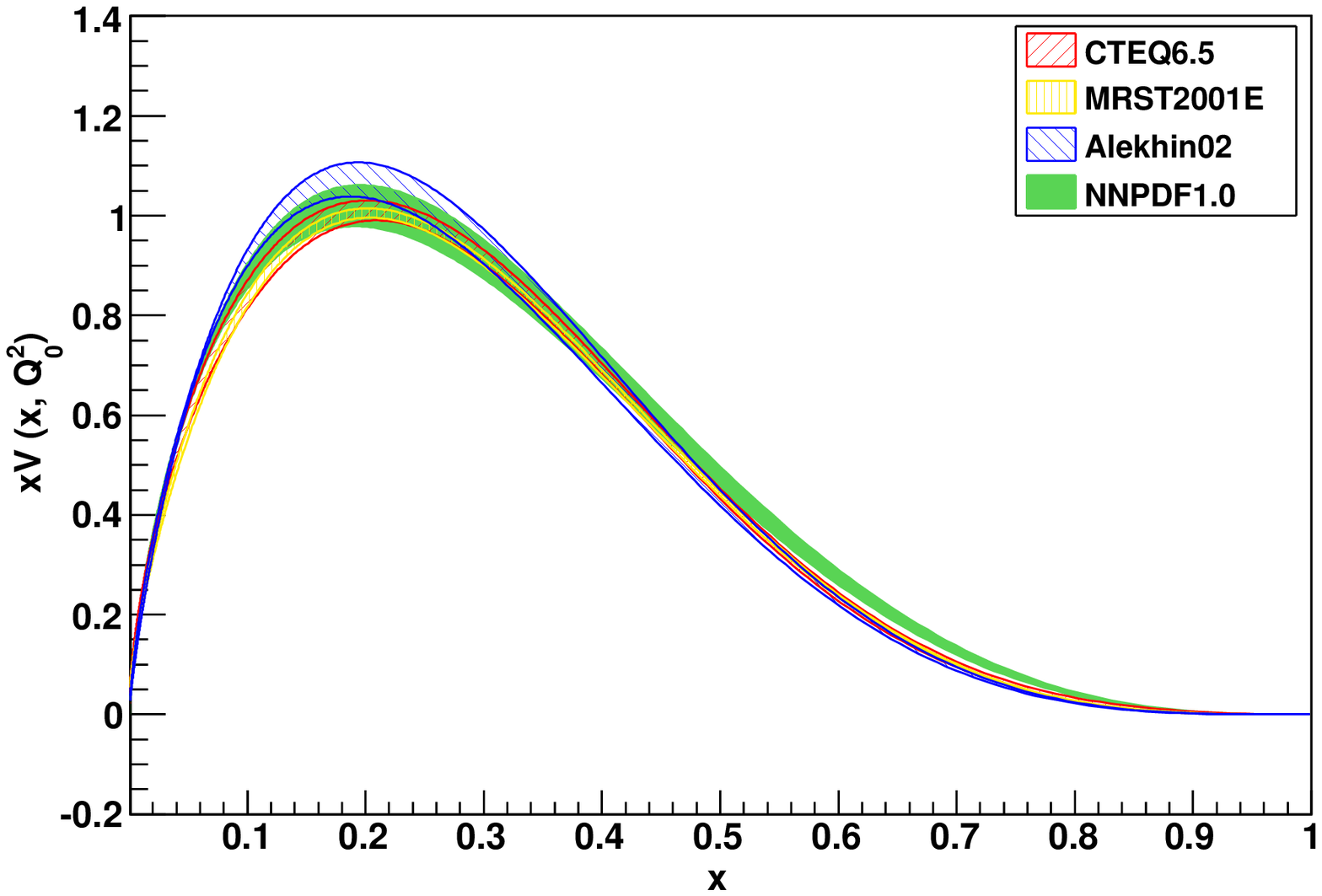}
\epsfig{width=0.49\textwidth,figure=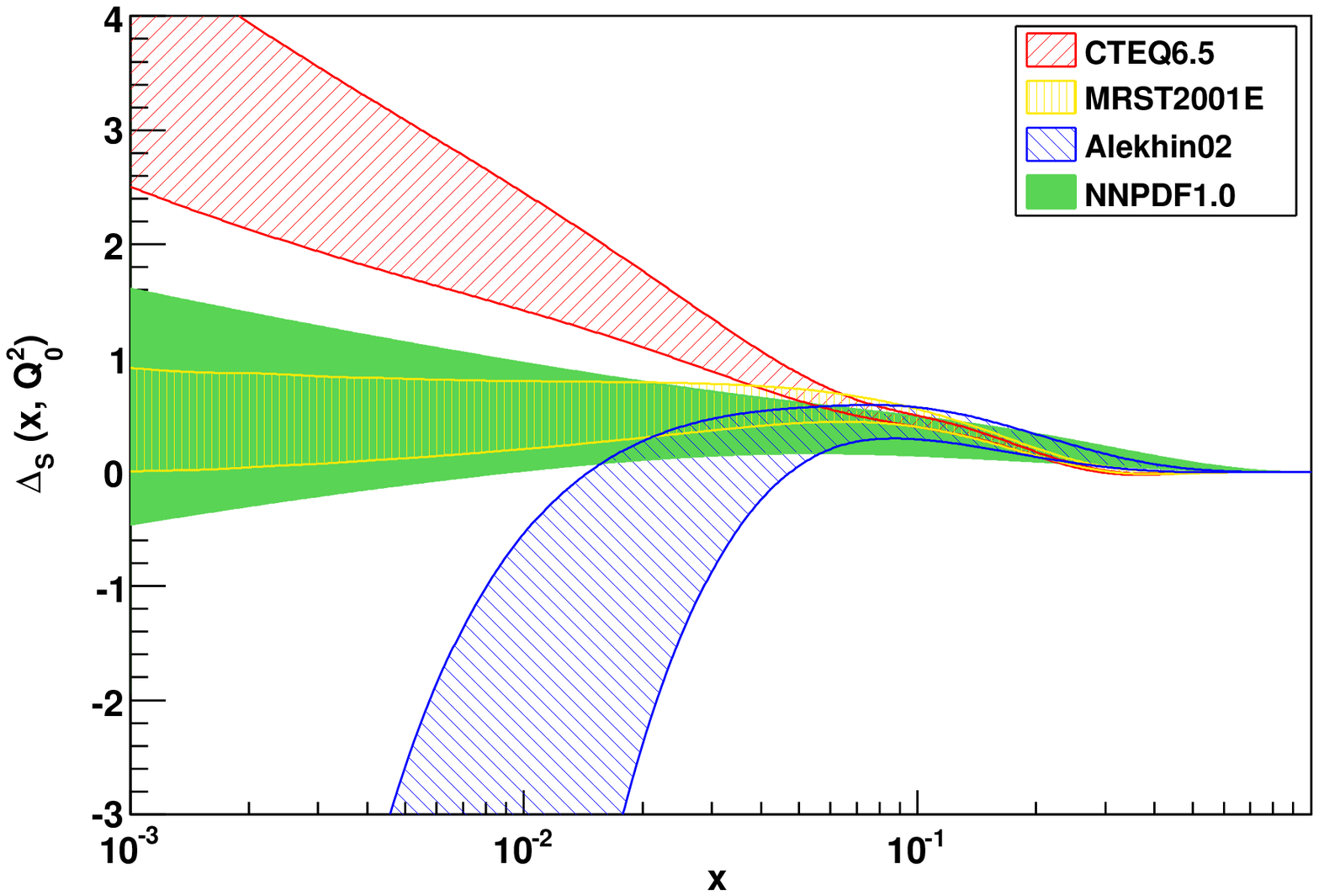}
\epsfig{width=0.49\textwidth,figure=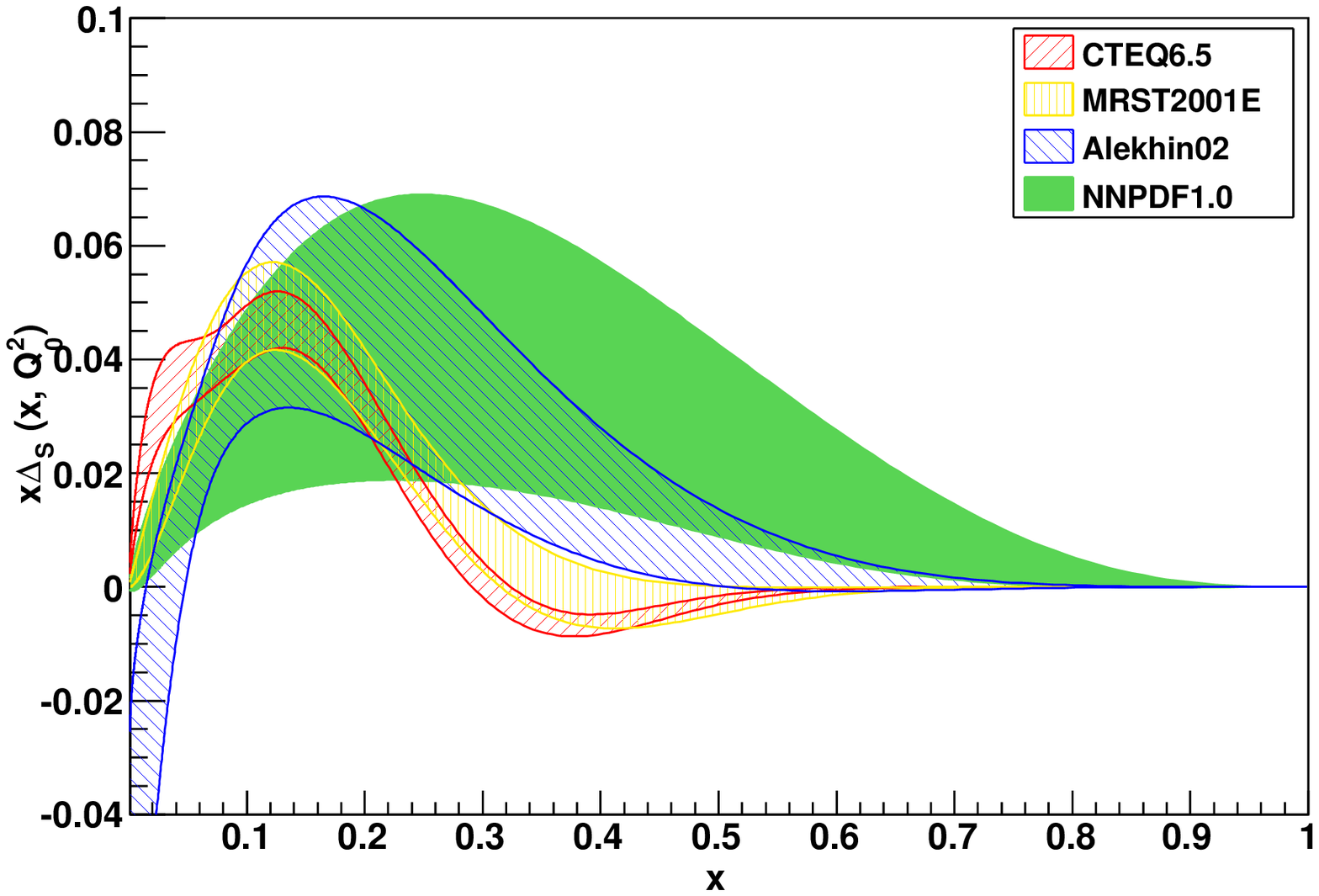}
\caption{\small The valence and nonsinglet
 PDFs at the starting scale
$Q_0^2 = 2$ GeV$^2$, plotted versus $x$  on a log (left) or 
 linear
scale (right).}
\label{fig:final-pdfs2}
\end{figure}
%------------------------------------------------------------------

%-------------------------------------------------------- 
\begin{figure}[t!]
\centering
\epsfig{width=0.48\textwidth,figure=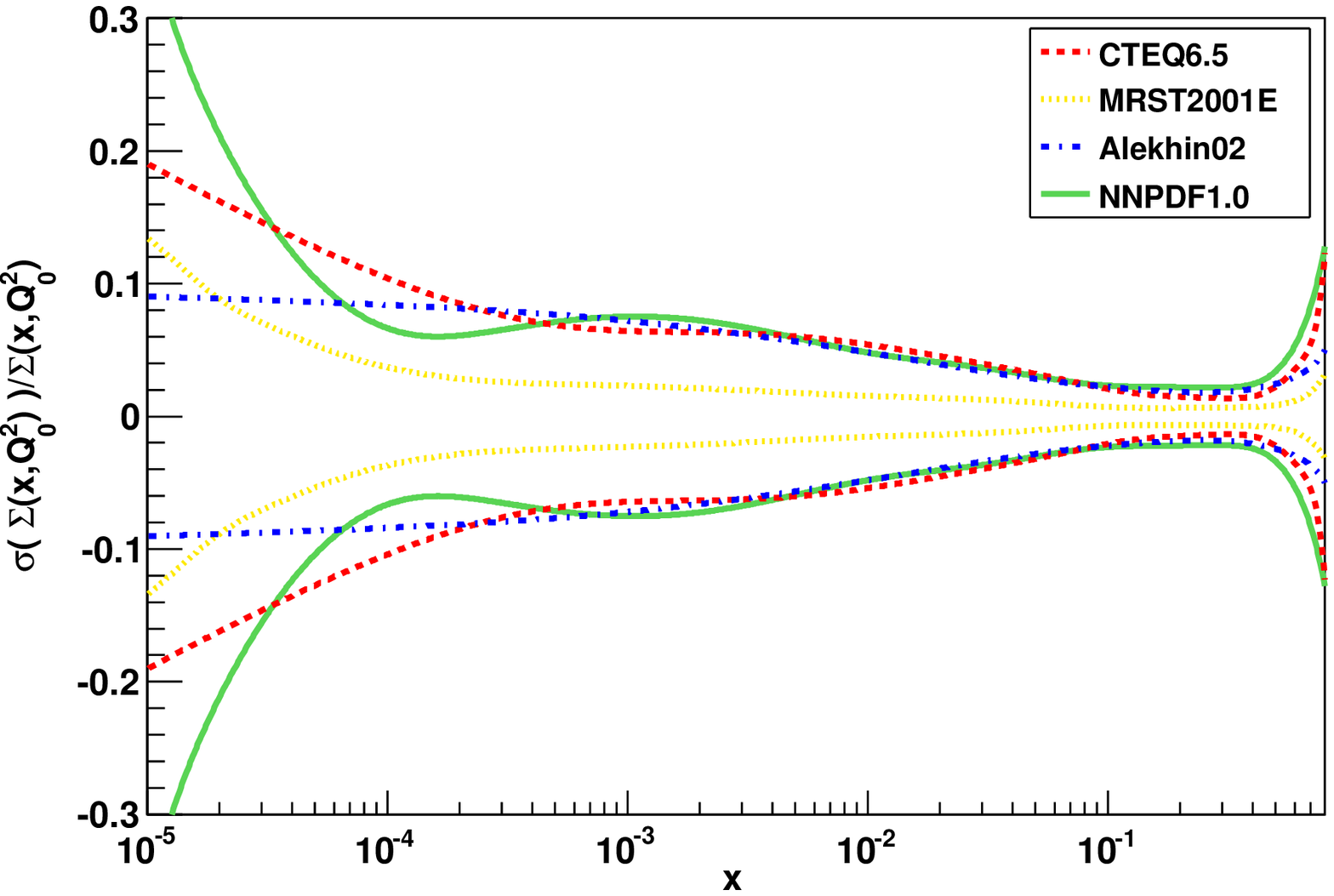} 
\epsfig{width=0.48\textwidth,figure=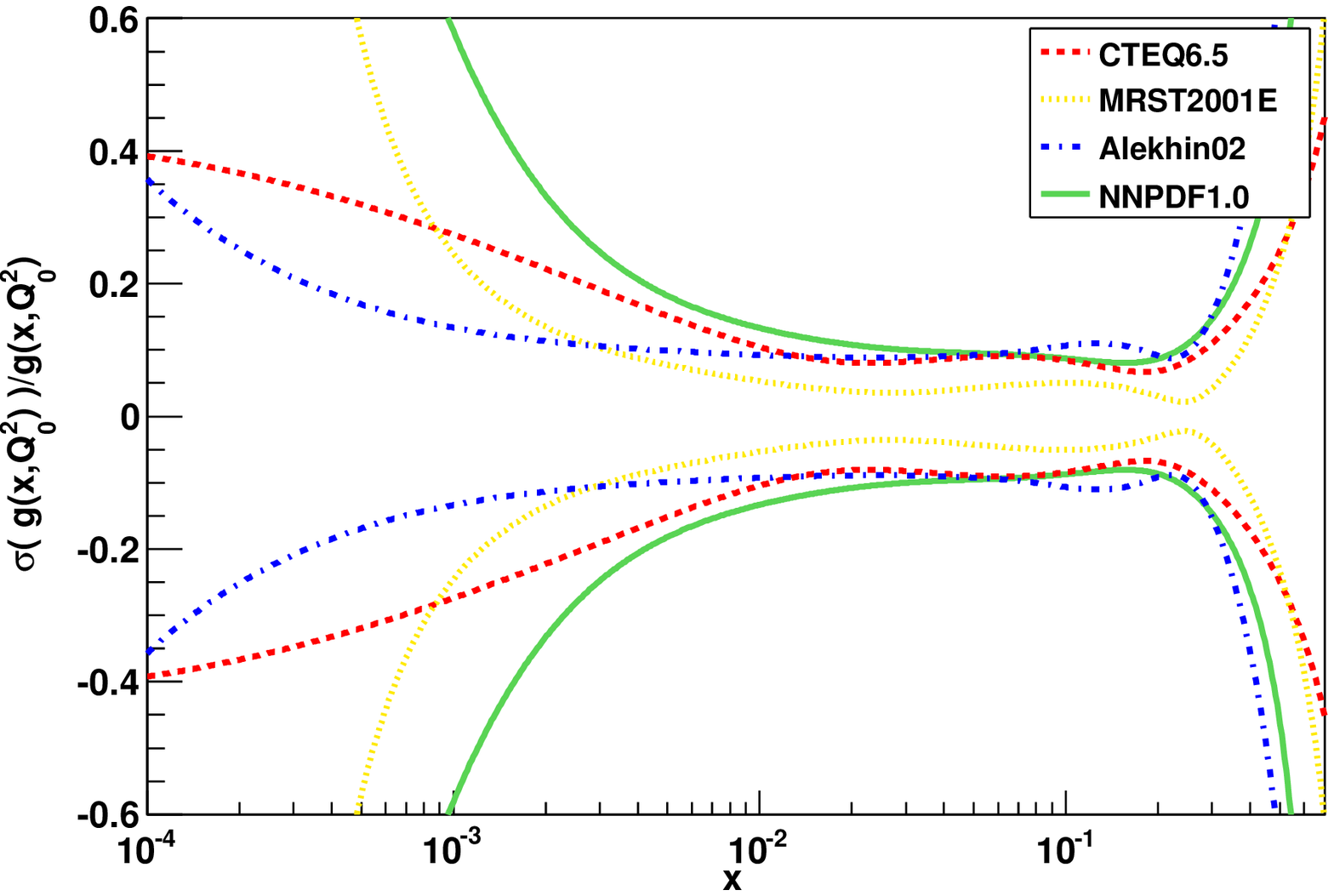} 
\epsfig{width=0.48\textwidth,figure=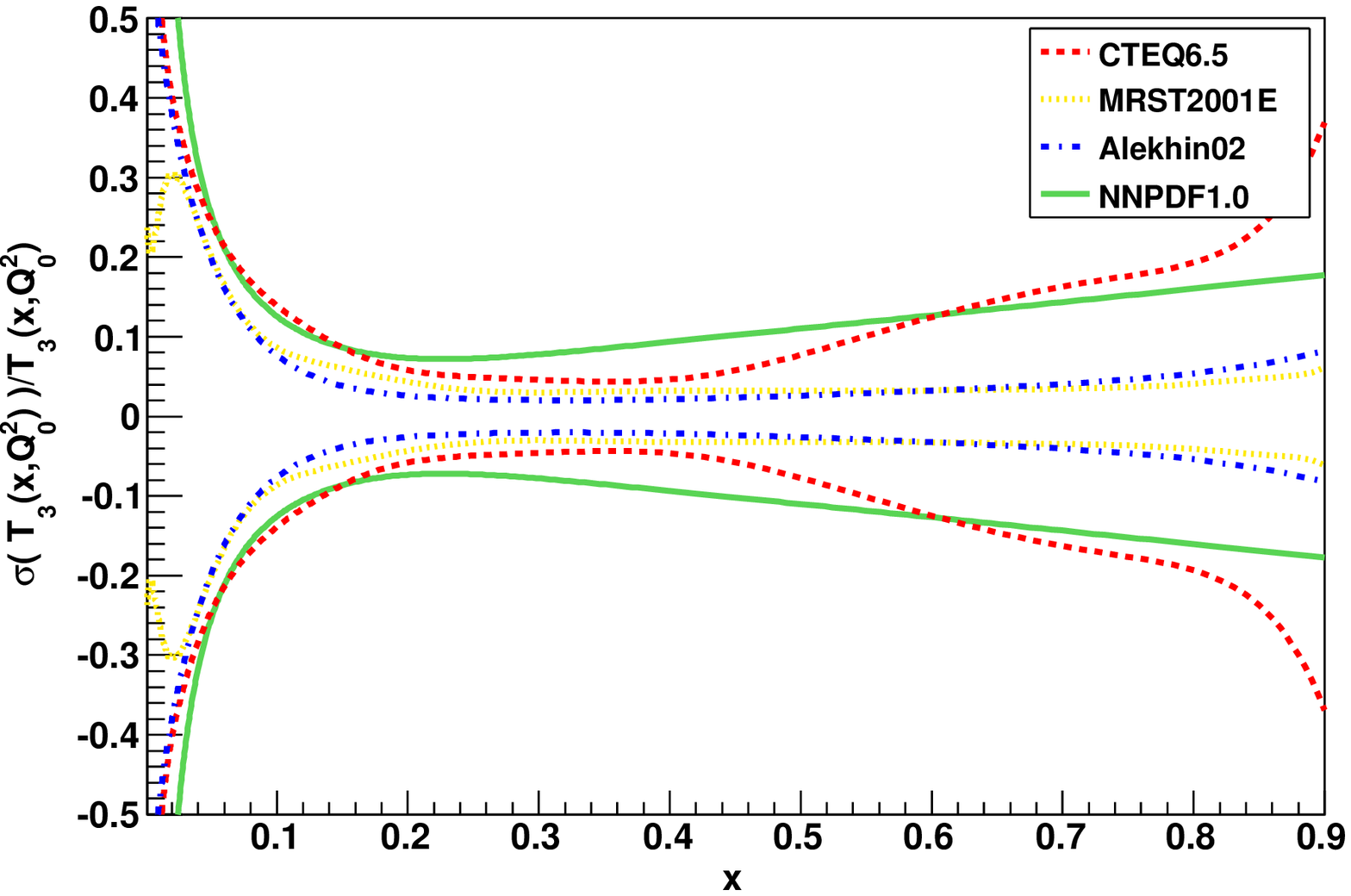}
\epsfig{width=0.48\textwidth,figure=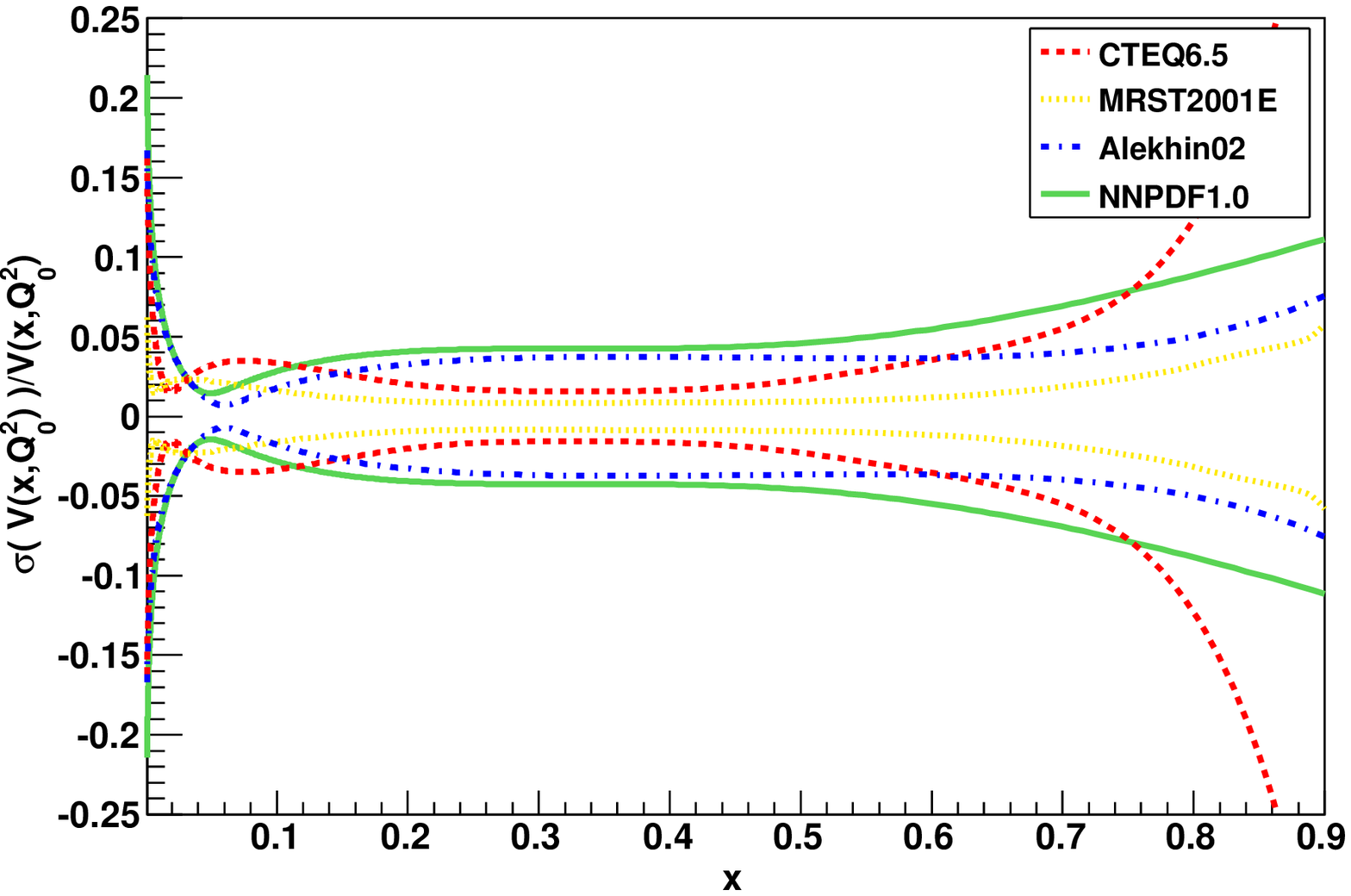} 
\epsfig{width=0.48\textwidth,figure=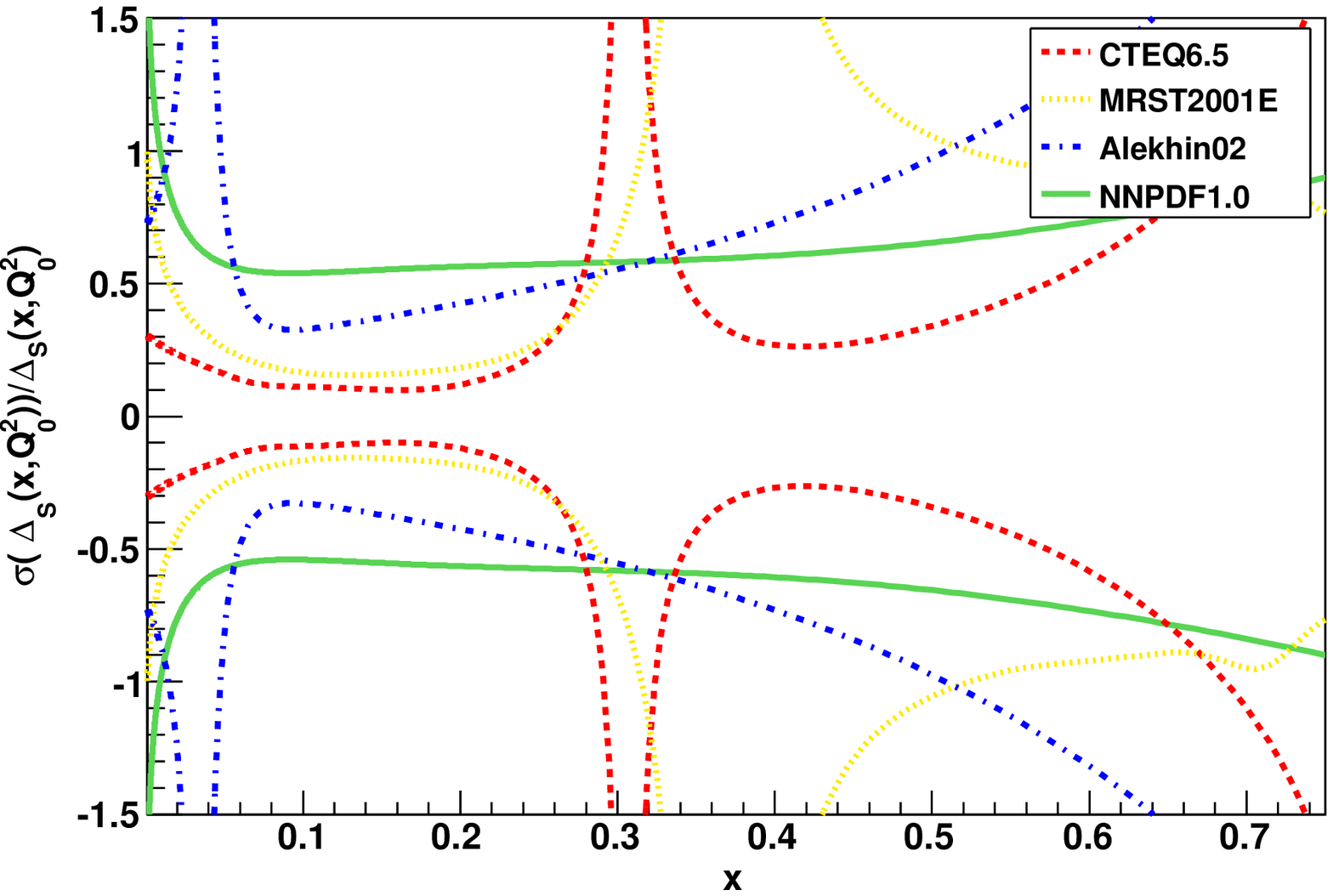}  
  \caption{Relative uncertainty on
 PDFs. All bands correspond to one $\sigma$
 (see
 Appendix~\ref{sec:app-pdferr}).}
  \label{fig:relerr}
\end{figure}
%-----------------------------------

%-------------------------------------------------------- 
\begin{figure}[t!]
\centering
\epsfig{width=0.49\textwidth,figure=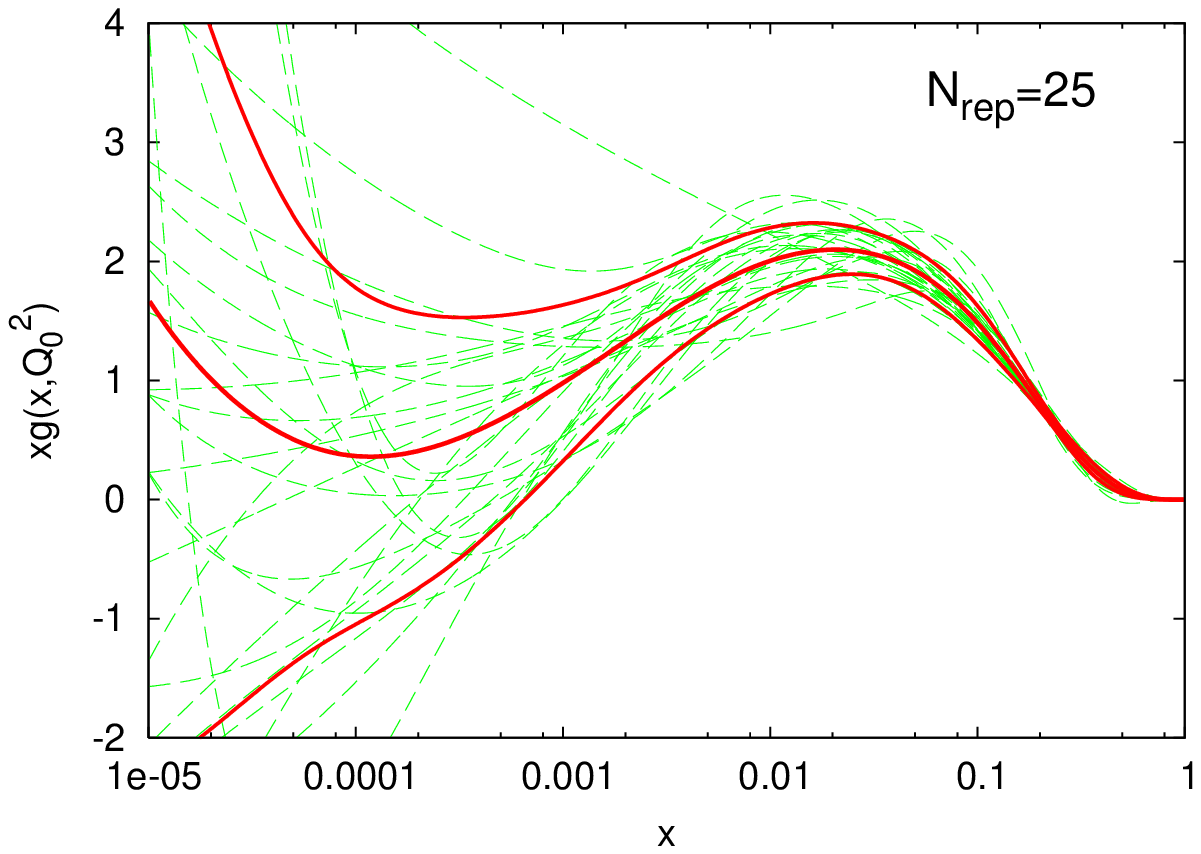} 
\epsfig{width=0.49\textwidth,figure=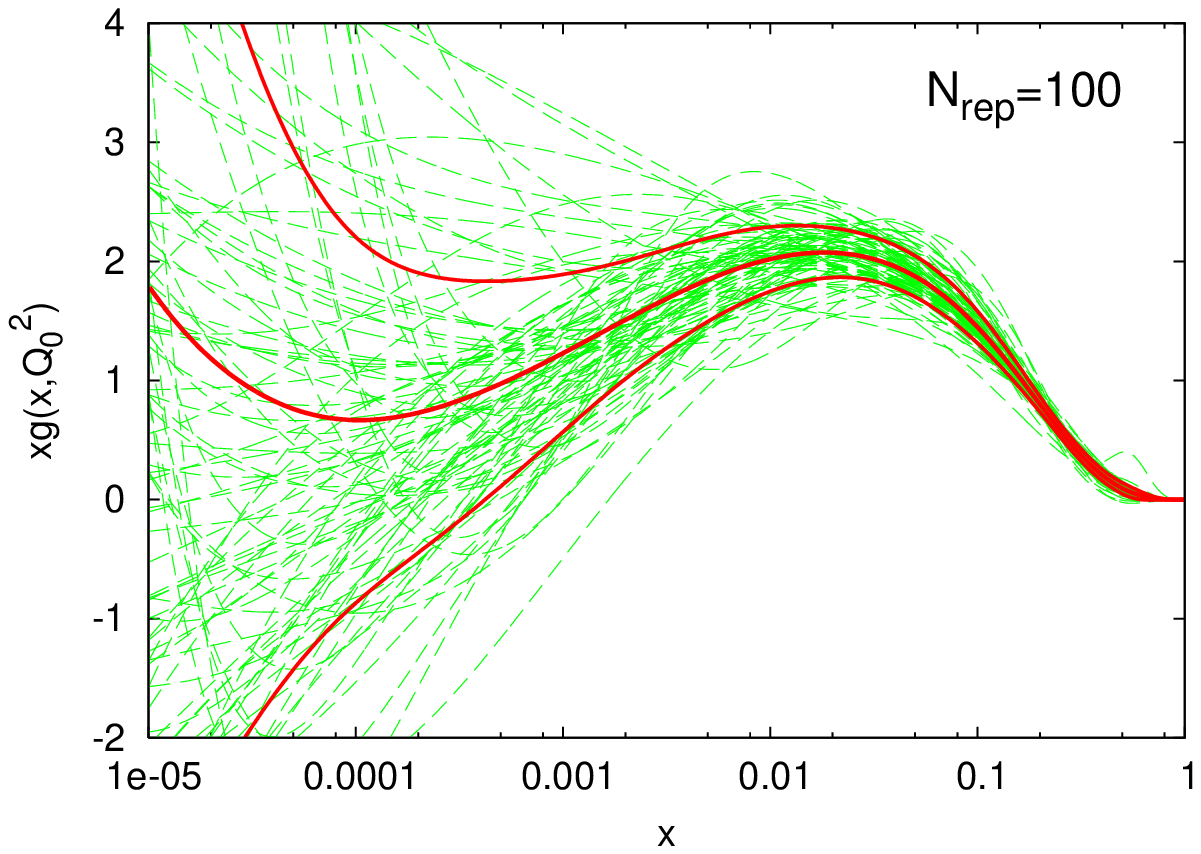} 
  \caption{Sets of 25 replicas (left) and 100 replicas (right) of the gluon
    distribution. The solid red (dark) lines show the average and
    one-sigma intervals computed from the given sets.}
  \label{fig:replicas}
\end{figure}
%-----------------------------------

The  NNPDF1.0 set of parton distributions at the  starting scale
$Q_0^2 = 2$ GeV$^2$ is displayed in Fig.~\ref{fig:final-pdfs} (singlet
sector) and Fig.~\ref{fig:final-pdfs2} (valence
sector), as a function of $x$ 
both on a linear and a logarithmic scale. The PDFs shown are those
which we adopt as a basis set, given in 
Eq.~(\ref{eq:pdfdef}), from which other PDFs can be obtained
by linear combination. Relative uncertainties on these PDFs are shown
in Fig.~\ref{fig:relerr}. Results are compared to those of the other
parton sets CTEQ6.5~\cite{Tung:2006tb}, MRST2001E~\cite{Martin:2002aw}
and Alekhin02~\cite{Alekhin}, which are the most recent NLO sets from
the respective groups available
through the LHAPDF interface~\cite{LHAPDFurl,Bourilkov:2006cj}.
All uncertainties in these plots (including those of other parton
sets)
correspond to nominal one-$\sigma$ error bands.
In Fig.~\ref{fig:replicas} we also show the shape of individual
replicas and the error band computed for two sets of 25 and 100
replicas of the gluon PDF.

The values of the momentum and valence sum rules computed from
NNPDF1.0 using Eqs.~(\ref{eq:momentumsr}-\ref{eq:valencesr})
are correct to within one part per mille, as expected from the 
accuracy of the numerical integration noted in Sec.~\ref{sec:net-param}. 

The general features of the PDF set can be summarized as follows:
\begin{itemize}
\item Even though individual replicas may fluctuate significantly
  (see Fig.~\ref{fig:replicas})
thereby showing the flexibility of the neural parametrization, this
appears to be due to statistical fluctuations of the underlying
replica data: average quantities, such as the central value and error
band shown in the figure,  are  smooth, all the more so as
stability is reached as the
number of replicas increases. It is clear from the plot that with
$N_{\rm rep}=100$ replicas the results for central values and error
bands have stabilized already, as we shall discuss  in more detail in
Sect.~\ref{sec:stabarch} below.
\item The central values of all PDFs are in reasonable agreement with those
  from other parton sets, especially in the region where data are
  available.  
\item Even though the uncertainty band on the gluon allows for a
  negative PDF for $x \lsim 10^{-4}$, positivity of $F_L$ holds for
$x \ge 10^{-5}$ and $Q^2\ge Q_0^2$.
\item Uncertainties on PDFs in the region where data are available
  tend to be generally a little larger than those of the 
CTEQ6.5 and Alekhin02 sets,
  and rather larger than those of the MRST2001E sets.
Note that the one-$\sigma$ uncertainty band that we find without having to
introduce a tolerance is thus comparable to or larger than the one-$\sigma$
uncertainty bands obtained by the CTEQ and MRST-MSTW groups with 
their respective
tolerance criteria, which amount to an upward rescaling of all experimental
uncertainties by a factor between four and six.
\item Uncertainties on PDFs in the region where no data are available
  tend to be larger than with any other set:
  this applies to the singlet and gluon  at very small $x$, to the
  valence and triplet at small $x$, and to the gluon at large $x$.
\item Some aspects of the NNPDF1.0 PDFs are unconstrained by data
  which are used in the CTEQ and MRST global fits but not in the
  data set of Tab.~\ref{tab:exps-sets}. This includes the $\bar u-\bar d$
  asymmetry, which in global fits 
  is mostly constrained by Drell-Yan data, and the
  large $x$ gluon, which in global fits is constrained by the
  large-$E_T$ jet data. Hence, the wider uncertainty  bands found for
  these quantities are likely to be at least partly due to the smaller data set used in
  the present fit.
\end{itemize}

\subsection{Parton-parton correlations}
\label{sec:correlations}

\begin{figure}[t!]
\centering
\epsfig{width=0.44\textwidth,figure=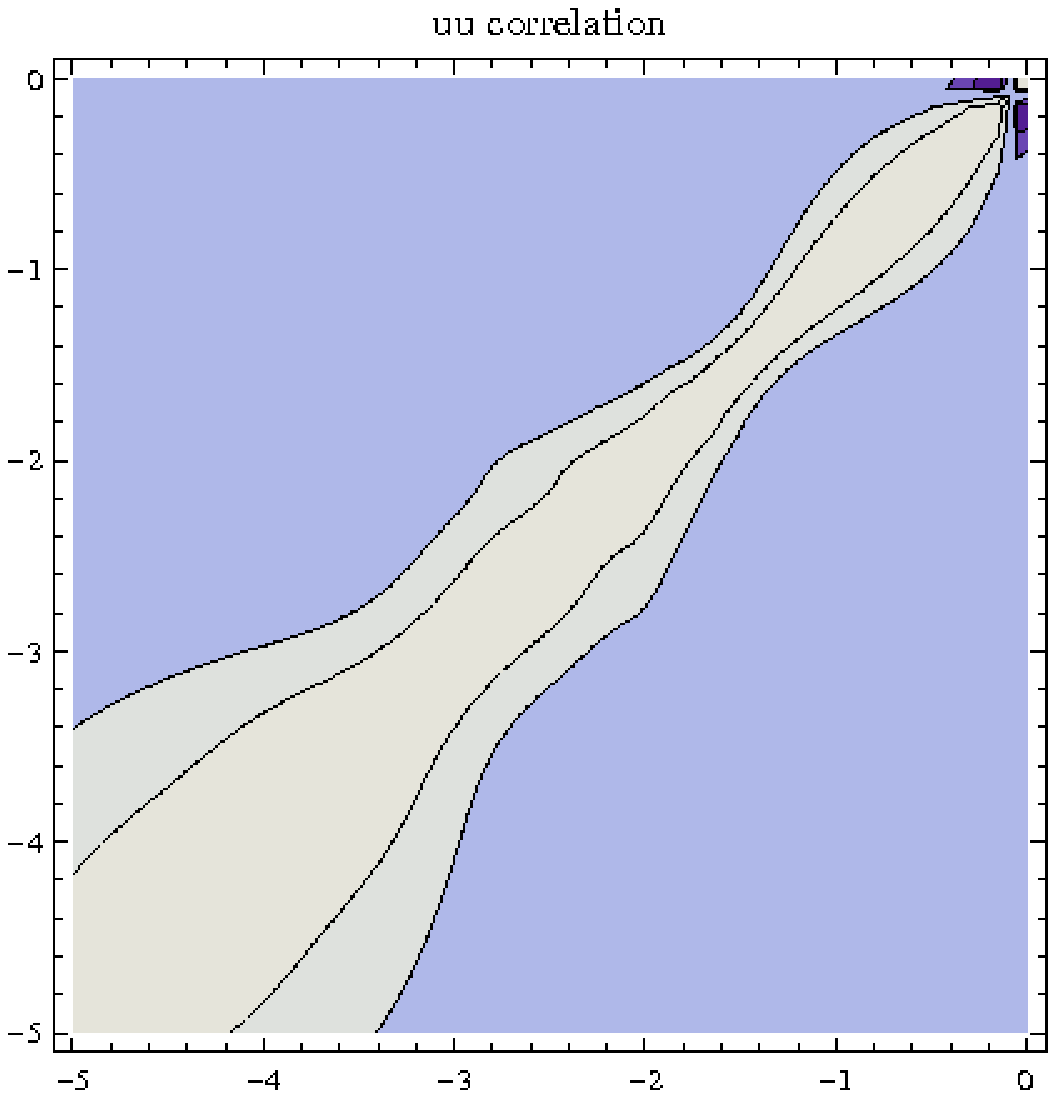} 
\epsfig{width=0.44\textwidth,figure=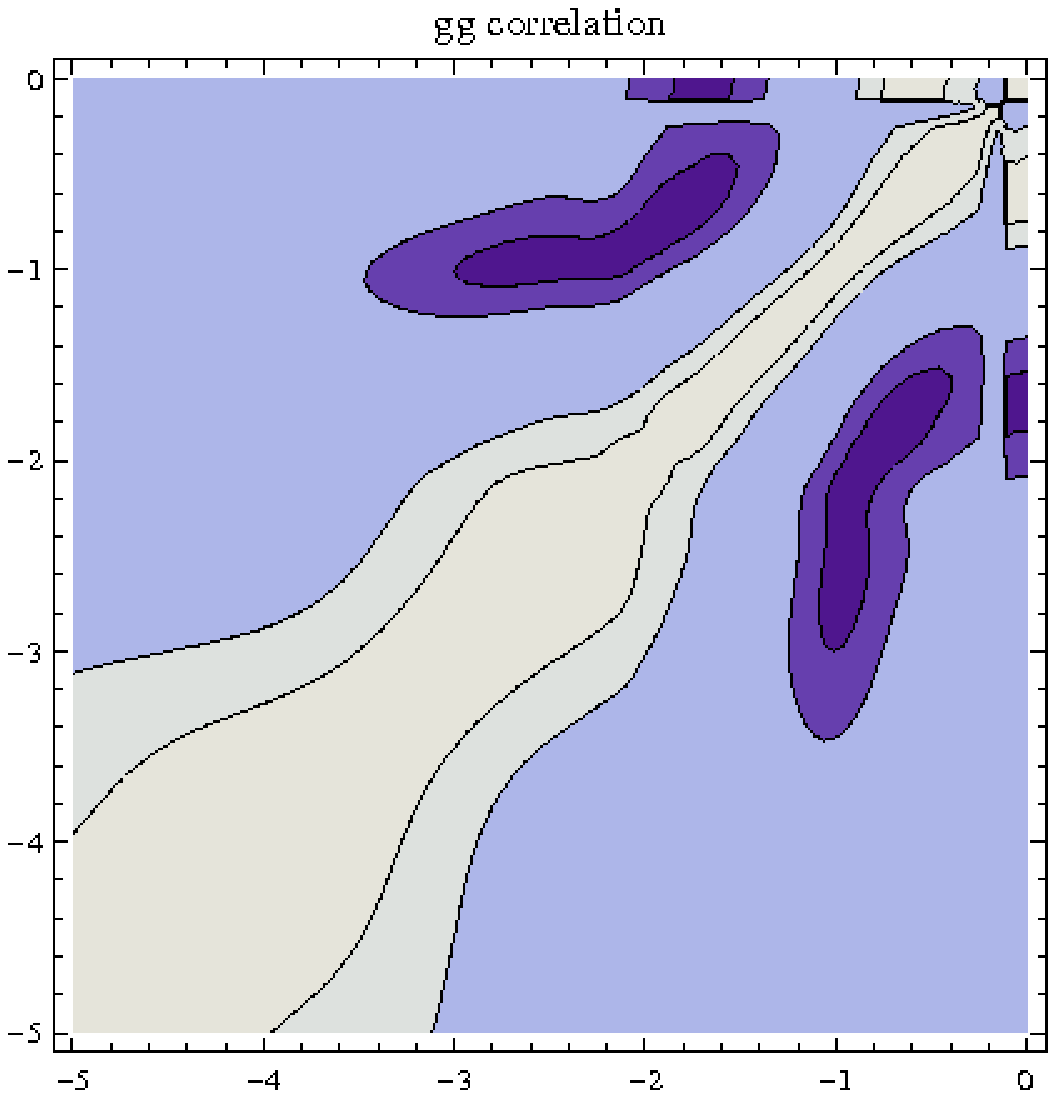}
\epsfig{width=0.10\textwidth,figure=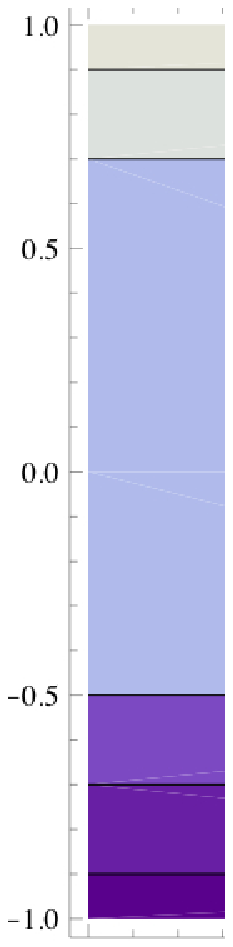} 
\epsfig{width=0.44\textwidth,figure=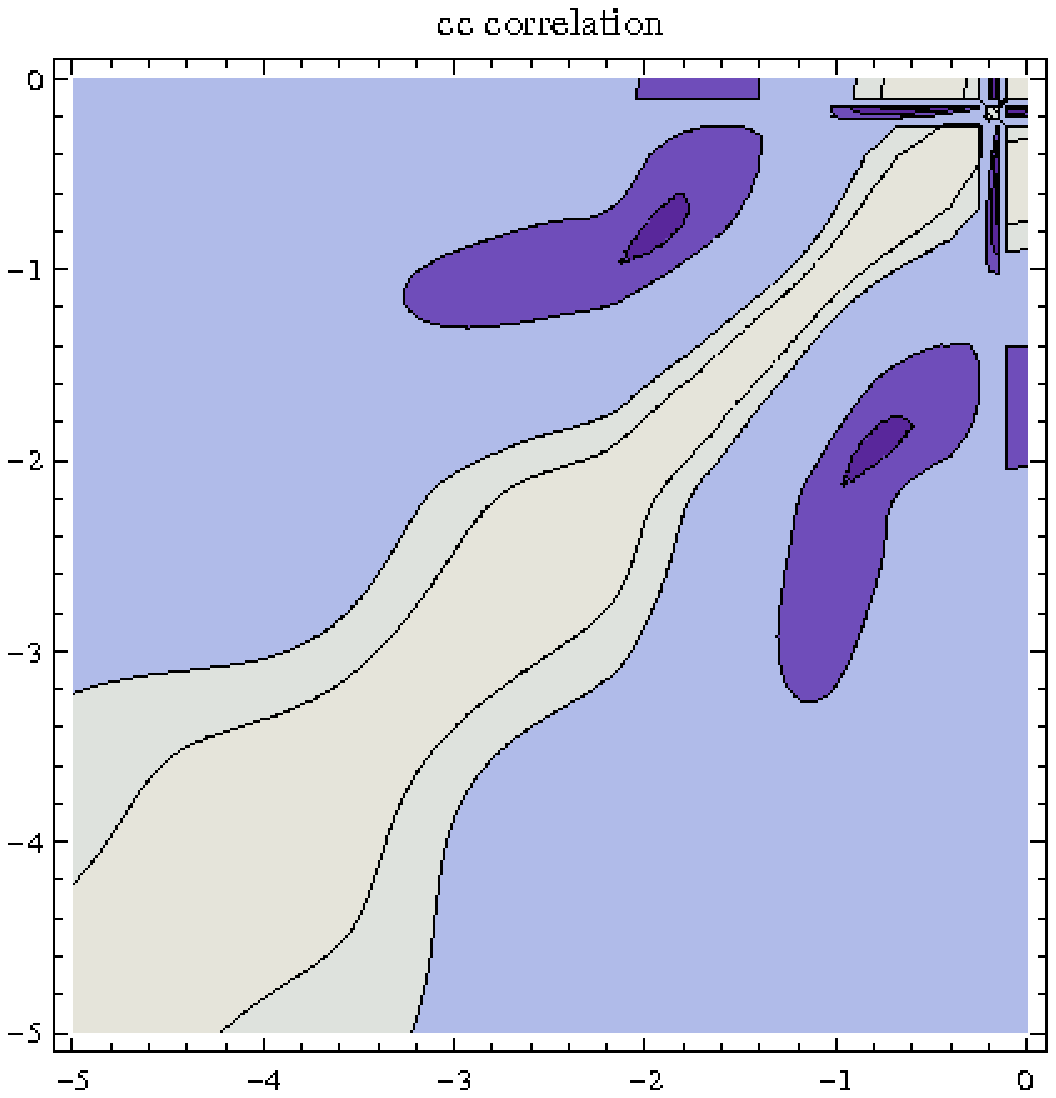} 
\epsfig{width=0.44\textwidth,figure=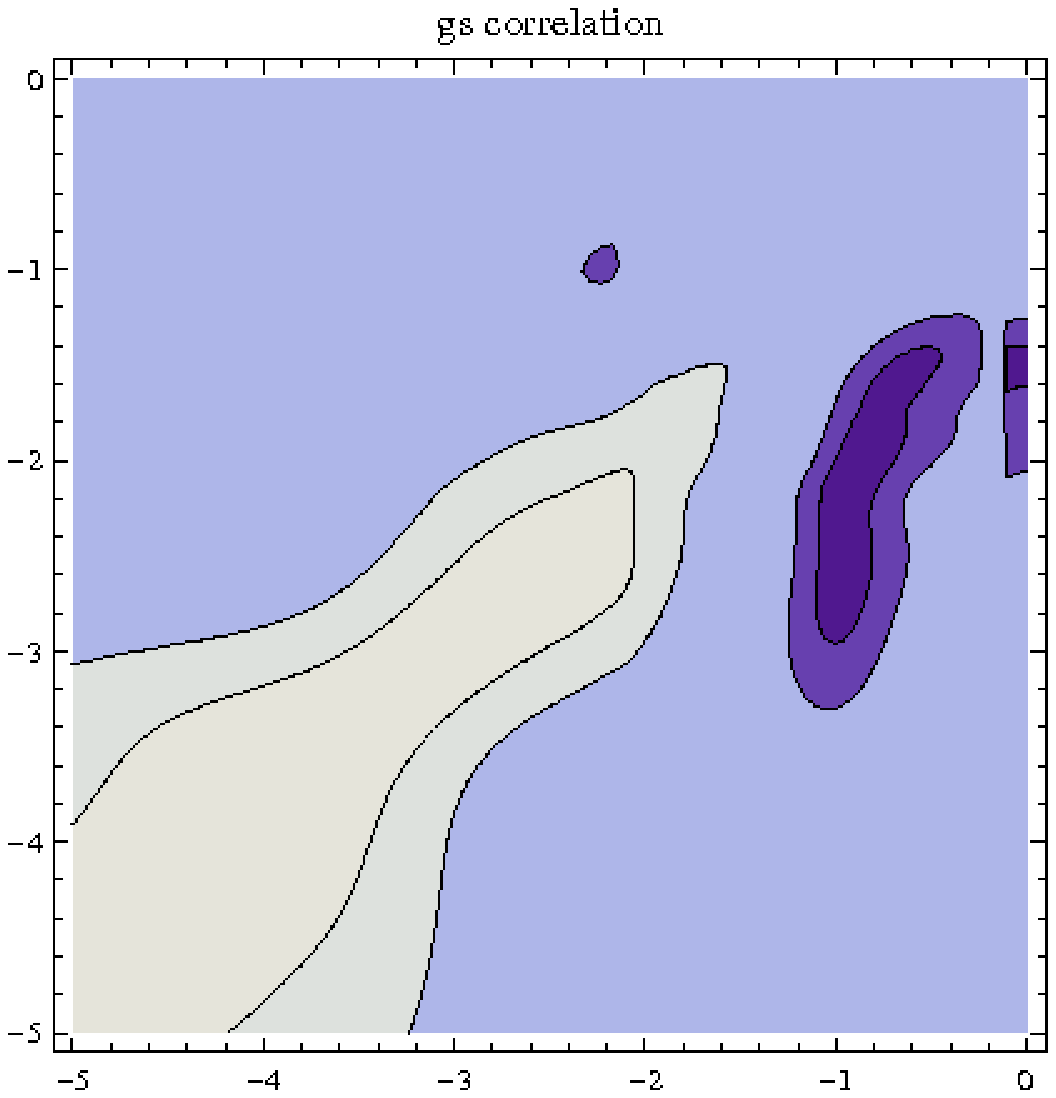}
\epsfig{width=0.10\textwidth,figure=plots/corr_legend_final.eps} 
  \caption{Correlations between
PDFs,  computed using Eq.~(\ref{eq:corrPDF}). 
The correlations shown are
$u-u$ (upper left), $g-g$ (upper right), $c-c$
(lower left)  and $g-s$
(lower right, $g$ in abscissa). The labels on the axes are the logarithm 
to base ten of $x_1$ and $x_2$. 
All the correlations are computed at  $Q=85$ GeV.  
  \label{fig:pdfcorr}}
\end{figure}
Besides the uncertainty bands on the PDFs, it is interesting to also determine
the correlations between different PDFs in a given set: the relevance
of these studies for the assessment of the uncertainties on LHC has
been recently emphasized in Ref.~\cite{Nadolsky:2008zw}. The
determination of the correlation between any pair of quantities depending
on PDFs  using NNPDF1.0 is
straightforward, as the correlation can be simply calculated using the
covariance as its estimator. For example, the correlation between the
two PDF values $q(x_1,Q_1^2)$ and $\widetilde{q}(x_2,Q_2^2)$ is
determined as
\be
\label{eq:corrPDF}
\rho \lc q(x_1,Q_1^2) \widetilde{q}(x_2,Q_2^2)\rc
=\frac{\la 
 q(x_1,Q_1^2)  \widetilde{q}(x_2,Q_2^2)
\ra_{\rm rep}- \la  q(x_1,Q_1^2) \ra_{\rep}
\la \widetilde{q}(x_2,Q_2^2)\ra_{\rm rep} 
}{\sigma_{q}(x_1,Q_1^2)\sigma_{\widetilde{q}}(x_2,Q_2^2)} ,
\ee
with all averages performed on the ensemble of replicas.
Results for some selected 
correlations are displayed in Fig.~\ref{fig:pdfcorr}. For ease of
comparison, we have chosen the same PDFs and scales that were studied
in Ref.~\cite{Nadolsky:2008zw}. The pattern of correlations is similar
to that found and discussed in that reference.

\subsection{Statistical uncertainties: parametrization independence}
\label{sec:stabarch}

As discussed in Section~\ref{sec-nnpdfapp}, an advantage of the NNPDF
approach is that various features of the PDF set can be assessed using
standard statistical tools. In this Section, we assess the stability
of the fit and the reliability of the error estimate provided with it.

The main estimator used in this analysis is the distance
$d[q]$. This is defined as the difference in values between two different
predictions for some quantity $q$ obtained from two different ensemble
of PDFs, measured in units of the sum in quadrature of their
uncertainties: thus if two determinations of $q$ have
$d=1$ this means that their difference is equal to the sum in
quadrature of their respective uncertainties.
(see appendix B of Ref.~\cite{DelDebbio:2007ee} for a more
extensive discussion). We compute this distance 
for each PDF in turn and for the uncertainty on it, at
the starting scale $Q_0^2=2$~GeV$^2$.
Specifically, we quote values for the distance in various ranges of $x$,
by computing it at ten
equally spaced points in $x$ in the given range and
averaging the results. In order to guarantee accuracy of the
results, the computation is repeated a large number of times over
different sets of replicas (of order
of a hundred). In practice, since we only have a limited total ensemble of
PDFs, the repetitions are performed by taking different subsets
of the full ensemble.

At first, we check that the results of the fit are stable when using
different subsets of 100 replicas out of the full ensemble of 1000
replicas. Results are shown in Table~\ref{tab:stabtab-selfstab} for
all PDFs and corresponding uncertainties, in two kinematical $x$
regions which are schematically identified as a region where the PDF
is mostly controlled by data, and a region where it is mostly extrapolated.
It is clear that the results of the fits are indeed behaving as they
ought to. It is important to notice that distances listed in the 
table refer to an average over 100 replicas, whose standard deviation  is
by a factor $\sqrt{N_{\rm rep}}$ smaller than the standard deviation
of each
of the replicas. Hence, the fact that $d\sim 1$ for two sets of 100
PDFs means that the central values of these PDFs differ by
a tenth of the sum in quadrature of their standard
deviations. It follows that Table~\ref{tab:stabtab-selfstab} also
verifies that indeed the fluctuation of results obtained as an average
over $N_{\rm rep}$ replicas scales as $1/\sqrt{N_{\rm rep}}$ as
it ought to.
\begin{table}[t!]
\begin{center}
\small
\begin{tabular}{|c|c|c|}
\hline 
 &  Data  & Extrapolation \\
\hline 
$\Sigma(x,Q_0^2)$ & $5~10^{-4} \le x \le 0.1$  & $10^{-5} \le x \le 10^{-4}$  \\
\hline
$\la d[q]\ra$  & 0.99 &  0.87\\
$\la d[\sigma]\ra$  & 0.96& 0.95 \\
\hline 
\hline 
$g(x,Q_0^2)$ & $5~10^{-4} \le x \le 0.1$  & $10^{-5} \le x \le 10^{-4}$  \\
\hline
$\la d[q]\ra$  & 0.98 &  0.82\\
$\la d[\sigma]\ra$  & 1.00&  0.96\\
\hline 
\hline 
$T_3(x,Q_0^2)$ & $0.05 \le x \le 0.75$  & $10^{-3} \le x \le 10^{-2}$  \\
\hline
$\la d[q]\ra$  & 0.88 & 0.76 \\
$\la d[\sigma]\ra$  & 0.94 & 0.82 \\
\hline 
\hline 
$V(x,Q_0^2)$ & $0.1 \le x \le 0.6$  & $3~10^{-3} \le x \le 3~10^{-2}$  \\
\hline
$\la d[q]\ra$  &  0.95&  0.82\\
$\la d[\sigma]\ra$  &  0.94&  0.79\\
\hline 
\hline 
$\Delta_S(x,Q_0^2)$ & $0.1 \le x \le 0.6$  & $3~10^{-3} \le x \le 3~10^{-2}$  \\
\hline
$\la d[q]\ra$  & 0.75 & 0.81\\
$\la d[\sigma]\ra$  & 0.88 &  0.82 \\
\hline 
\end{tabular}

\end{center}
\caption{\small Distance between results obtained from two different
  sets of 100 PDFs out of the full ensemble of 1000 PDFs.
\label{tab:stabtab-selfstab}}
\end{table}

Next, we study the dependence of  results on the architecture of
PDFs. Specifically, we reduce the architecture
from 2-5-3-1 to 2-4-3-1, thereby decreasing the number of parameters
of each PDF from 37 to  31.
Results are given in Table~\ref{tab:stabtab-arch}. We find remarkable
stability: fluctuations are at most at the two-$\sigma$ level
in poorly controlled quantities, such as the value of the
light quark sea asymmetry
in the extrapolation region, or the uncertainty on the isotriplet
combination in the extrapolation region. This shows that results are
indeed independent of the number of parameters.
\begin{table}[t!]
\begin{center}
\small
\begin{tabular}{|c|c|c|}
\hline 
 &  Data  & Extrapolation \\
\hline 
$\Sigma(x,Q_0^2)$ & $5~10^{-4} \le x \le 0.1$  & $10^{-5} \le x \le 10^{-4}$  \\
\hline
$\la d[q]\ra$  & 0.98 & 1.25 \\
$\la d[\sigma]\ra$  & 1.14&  1.34\\
\hline 
\hline 
$g(x,Q_0^2)$ & $5~10^{-4} \le x \le 0.1$  & $10^{-5} \le x \le 10^{-4}$  \\
\hline
$\la d[q]\ra$  & 1.52 &  1.15\\
$\la d[\sigma]\ra$  & 1.16&  1.07\\
\hline 
\hline 
$T_3(x,Q_0^2)$ & $0.05 \le x \le 0.75$  & $10^{-3} \le x \le 10^{-2}$  \\
\hline
$\la d[q]\ra$  & 1.00 & 1.11 \\
$\la d[\sigma]\ra$  &  1.76&  2.27\\
\hline 
\hline 
$V(x,Q_0^2)$ & $0.1 \le x \le 0.6$  & $3~10^{-3} \le x \le 3~10^{-2}$  \\
\hline
$\la d[q]\ra$  &  1.30& 0.90 \\
$\la d[\sigma]\ra$  & 1.10 &  0.98\\
\hline 
\hline 
$\Delta_S(x,Q_0^2)$ & $0.1 \le x \le 0.6$  & $3~10^{-3} \le x \le 3~10^{-2}$  \\
\hline
$\la d[q]\ra$  &  1.04&  1.91\\
$\la d[\sigma]\ra$  &  1.44&  1.80\\
\hline 
\end{tabular}

\end{center}
\caption{\small Distance between results obtained from a
  sets of 100 PDFs with neural network architecture 2-5-3-1 and a
  sets of 100 PDFs with neural network architecture 2-4-3-1.
\label{tab:stabtab-arch}}
\end{table}

A more subtle issue related to the parametrization is the choice of
preprocessing functions, introduced in the relation
Eq.~(\ref{eq:pdfdef}) between PDFs and their neural network
parametrizations. These functions, as discussed in
Sect.~\ref{sec:net-param}, are introduced in order to speed up the
training but should not affect final results. We have thus checked the
stability of result upon variation of the preprocessing exponents away from 
 their default values, listed in Table~\ref{tab:prepexps}.

\begin{table}[!]
\begin{center}
\begin{tabular}{|c|c|c||c|c|c|}
\hline 
Valence sector  &   &  &  Singlet sector &  &  \\
\hline 
\hline
 & $\chi^2 $& $\la \rm TL \ra$  & &  $\chi^2 $& $\la \rm TL \ra$ \\
\hline
$n_{T_3}=n_{V}=0.1$ & 1.38 & 771 & $n_{\Sigma}=n_{g}=0.8$ & 1.39
& 1002 \\
$n_{T_3}=n_{V}=0.5$ & 1.34 & 1629 & $n_{\Sigma}=n_{g}=1.6$ & 1.52 
&  2287 \\
$m_{T_3}=m_{V}=2$ & 1.55 & 1186 & $m_{\Sigma}=m_{g}-1=2$ & 1.37 & 647 \\
$m_{T_3}=m_{V}=4$ & 1.28 & 1311 & $m_{\Sigma}=m_{g}-1=4$ & 1.41 & 1306 \\
\hline
\end{tabular}

\end{center}
\caption{\small Values of the total $\chi^2$
and of the average training length $\la \rm TL \ra$
when  preprocessing exponents are varied away from
their default values 
Tab.~\ref{tab:prepexps}. The value of the exponents which are varied
is given for each row.
\label{tab:stabtab-prep1}}
\end{table}
First, in Table.~\ref{tab:stabtab-prep1} 
we display the dependence on preprocessing exponents of
 the values of the total $\chi^2$
and of the average training length $\la \rm TL \ra$. The table shows
that in some cases the quality of the fit deteriorates and the average
length of training increases significantly
 when the
preprocessing exponents are varied. This implies that in these cases a
satisfactory fit has not been obtained: indeed, in such cases there is
a noticeable increase in the
fraction of replicas which reach the maximum training length of $N_{\rm
  gen}^{\rm max}=5000$ without achieving convergence. Hence, an
increase of $N_{\rm
  gen}^{\rm max}$, or possibly a more efficient or differently
tuned minimization algorithm would be required to obtain a
satisfactory fit. Thus, in these cases we expect reduced  stability of
results: this applies to the case of increase of the small $x$
preprocessing exponent $n$ for the singlet, and to a lesser extent to
the decrease of the large $x$ preprocessing exponent $m$ for the
nonsinglet and valence.

\begin{table}[t!]
\begin{center}
%\tiny
\footnotesize
\begin{tabular}{|c|c|c|c|c|c|c|c|c|}
\hline
 ~Data~region~~  &&& &&&& & \\
\hline
\hline 
 &  $n_{v}=0.1$ & $n_{v}=0.5$ & $m_{v}=2$ & $m_{v}=4$ &
$n_{s}=0.8$ &  $n_{s}=1.6 $ & $m_{s}=2$ & $m_{s}=4$  \\
\hline 
$\Sigma(x,Q_0^2)$ &   &  &  &&&& &\\
\hline
$\la d[q]\ra$  & 1.34 &  1.25 & 1.37 & 2.14& 1.72& 1.38 & 1.45 & 1.64\\
$\la d[\sigma]\ra$  & 1.45 & 1.44 & 1.25  & 1.44& 2.03& 2.66& 0.95 & 1.35\\
\hline 
\hline 
$g(x,Q_0^2)$ &   &   & &&&& & \\
\hline
$\la d[q]\ra$  & 1.31 &  1.30 & 2.69 & 1.15& 3.06& 2.08& 1.20 & 1.74 \\
$\la d[\sigma]\ra$  & 1.34& 1.60 & 1.56 & 1.37& 3.21& 2.44& 0.98 & 1.72 \\
\hline 
\hline 
$T_3(x,Q_0^2)$ &   &  &  &&&& &\\
\hline
$\la d[q]\ra$  & 1.97 & 2.48 & 8.35  & 9.74& 1.31& 3.23& 1.03 & 1.41 \\
$\la d[\sigma]\ra$  &  1.10& 1.47 & 1.98 & 1.53& 1.10& 2.66& 1.76 & 1.99 \\
\hline 
\hline 
$V(x,Q_0^2)$ &   &   &  &&&& &\\
\hline
$\la d[q]\ra$  & 11.03 & 1.55 & 3.61 & 5.60& 0.94& 2.12& 1.25 & 3.54 \\
$\la d[\sigma]\ra$  & 3.57 &  4.74& 4.04& 3.09& 1.03& 1.10 &  0.66 & 1.98\\
\hline 
\hline 
$\Delta_S(x,Q_0^2)$ & & &   &&&& &\\
\hline
$\la d[q]\ra$  & 2.00 & 2.29 & 7.51 & 2.36& 1.14& 1.70& 0.76  & 0.92 \\
$\la d[\sigma]\ra$  & 1.25 & 5.20 & 1.17 & 3.50& 1.00& 1.98& 0.97 & 2.05 \\
\hline 
\end{tabular}

% --------------------------------------------------

\footnotesize
\begin{tabular}{|c|c|c|c|c|c|c|c|c|}
\hline
 Extrapolation  &&& &&&& &\\
\hline
\hline 
 &  $n_{v}=0.1$ & $n_{v}=0.5$ & $m_{v}=2$ & $m_{v}=4$ &
$n_{s}=0.8$ &  $n_{s}=1.6 $ & $m_{s}=2$ & $m_{s}=4$  \\
\hline 
$\Sigma(x,Q_0^2)$ &   &  &  &&&& &\\
\hline
$\la d[q]\ra$  & 1.06 & 1.69  & 1.49 & 1.84& 7.72& 4.67& 0.87 & 3.15\\
$\la d[\sigma]\ra$  & 1.12& 1.84 &  2.11 & 1.52& 2.47& 3.66& 0.82 & 2.34\\
\hline 
\hline 
$g(x,Q_0^2)$ &   &   & &&&& & \\
\hline
$\la d[q]\ra$  & 1.41 & 2.32 & 2.33 & 1.34& 1.62& 4.73& 1.04 & 3.49\\
$\la d[\sigma]\ra$  & 1.41& 1.86 & 1.95 & 1.30& 2.15& 2.72& 0.81 & 2.38\\
\hline 
\hline 
$T_3(x,Q_0^2)$ &   &  & &&&& & \\
\hline
$\la d[q]\ra$  & 1.71 & 2.70 & 7.40 & 1.60& 1.36 & 2.37 & 0.78 & 0.91 \\
$\la d[\sigma]\ra$  & 4.83 & 4.54 & 2.89 & 5.09& 1.00 & 1.65& 0.92 & 1.26\\
\hline 
\hline 
$V(x,Q_0^2)$ &   &   & &&&& & \\
\hline
$\la d[q]\ra$  & 14.85 & 3.23 & 3.75 & 2.55&0.86& 2.52& 1.26 &  1.34\\
$\la d[\sigma]\ra$  & 2.65 & 5.08 & 3.94 & 2.78& 1.20& 0.87 & 0.62 & 2.25\\
\hline 
\hline 
$\Delta_S(x,Q_0^2)$ & & &  &&&& & \\
\hline
$\la d[q]\ra$  & 1.25 & 2.50 & 7.75  & 2.48& 1.09& 1.47& 1.09 & 0.83\\
$\la d[\sigma]\ra$  & 1.80 & 2.85 & 1.50 & 2.28& 0.90& 2.01& 0.90 & 1.64\\
\hline 
\end{tabular}

\end{center}
\caption{\small Distance  between results found with a set of 100
PDFs with the default preprocessing exponents
Table~\ref{tab:stabtab-prep1}, 
and 100 PDFs obtained with a
different value of some of the preprocessing exponents. The data
and extrapolation
regions for each PDF are the same as in 
Table~\ref{tab:stabtab-selfstab}. The preprocessing exponents which
are varied are listed in the table, with
 $n_v\equiv n_{T_3}=n_{V}$,
$m_v\equiv m_{T_3}=m_{V}$, $n_s\equiv n_{\Sigma}=n_{g}$ 
and
$m_s\equiv m_{\Sigma}=m_{g}-1$.
\label{tab:stabtab-prep2}}
\end{table}
In Table~\ref{tab:stabtab-prep2} we show the distance from the default
of results obtained
as the preprocessing exponents are varied. In almost all cases,
remarkable stability is found, with variation of results well within
the 90\% confidence level. Even in the case of increase of the singlet
$n$ exponent mentioned above reasonable stability is found.
A notable
exception is the behaviour of the triplet and the valence upon
variation of the large or small $x$ exponents. In these cases, the
distance is of order ten without significant deterioration of the fit
quality: this
 (with 100 replicas) means that
central values differ by about 1.4~$\sigma$ in units of the respective
standard deviations. We must conclude that in these cases we do not
find complete independence of results on the preprocessing function,
and therefore that 
uncertainties on our quark distributions in the valence region are
likely to be underestimated by a factor between
one and two. A more faithful estimate of these uncertainties would
require for example random variation of the preprocessing exponents
within the range where a good fit quality obtains. However,  
it should be noted that uncertainties on valence PDFs are going to be greatly
reduced when Drell-Yan data are included in our data set.

\subsection{Statistical uncertainties: dependence on the data set}
\label{sec:stabdata}

\begin{table}[t!]
\begin{center}
\small
 \begin{tabular}{|c|c|c|c|c|c|}
  \hline
  Set & $N_{\rm dat}$ & $x_{\rm min}$ &  $x_{\rm max}$ 
 &  $Q^2_{\rm min}$ &  $Q^2_{\rm max}$ \\ 
 \hline BCDMSp   &  322 &  $7~10^{-2}$ & 0.75 &    10.3 &   230\\
 \hline
   NMC      &  95 &  0.028 & 0.48 &    9 &    6 \\
\hline
   NMC-pd   &  73 & 0.035 & 0.67 &    11.4 &    99  \\
 \hline Z97NC    &  206 & $1.6~10^{-4}$ & 0.65 &   10 &  $2~10^4$ \\
 \hline  H197low$Q^2$ &   77 & $3.2~10^{-4}$ & 0.2 &   12 &   150  \\
 \hline
\end{tabular}

\end{center}
\caption{\small Reduced data set used for benchmarking.
\label{tab:exps-sets-bench}}
\end{table}
Having verified  the reliability and independence of our results
on the parametrization, we perform a series of checks to see how
they behave when data are excluded from the fitting set. 
The aim of these tests is to make sure that our results are not
fine-tuned to our choice of a specific data set, and thus that 
the same approach can
lead to satisfactory results with data sets of different sizes and 
with different features.

%------------------------------------------------------------
\begin{figure}[t!]
\centering
\epsfig{width=0.49\textwidth,figure=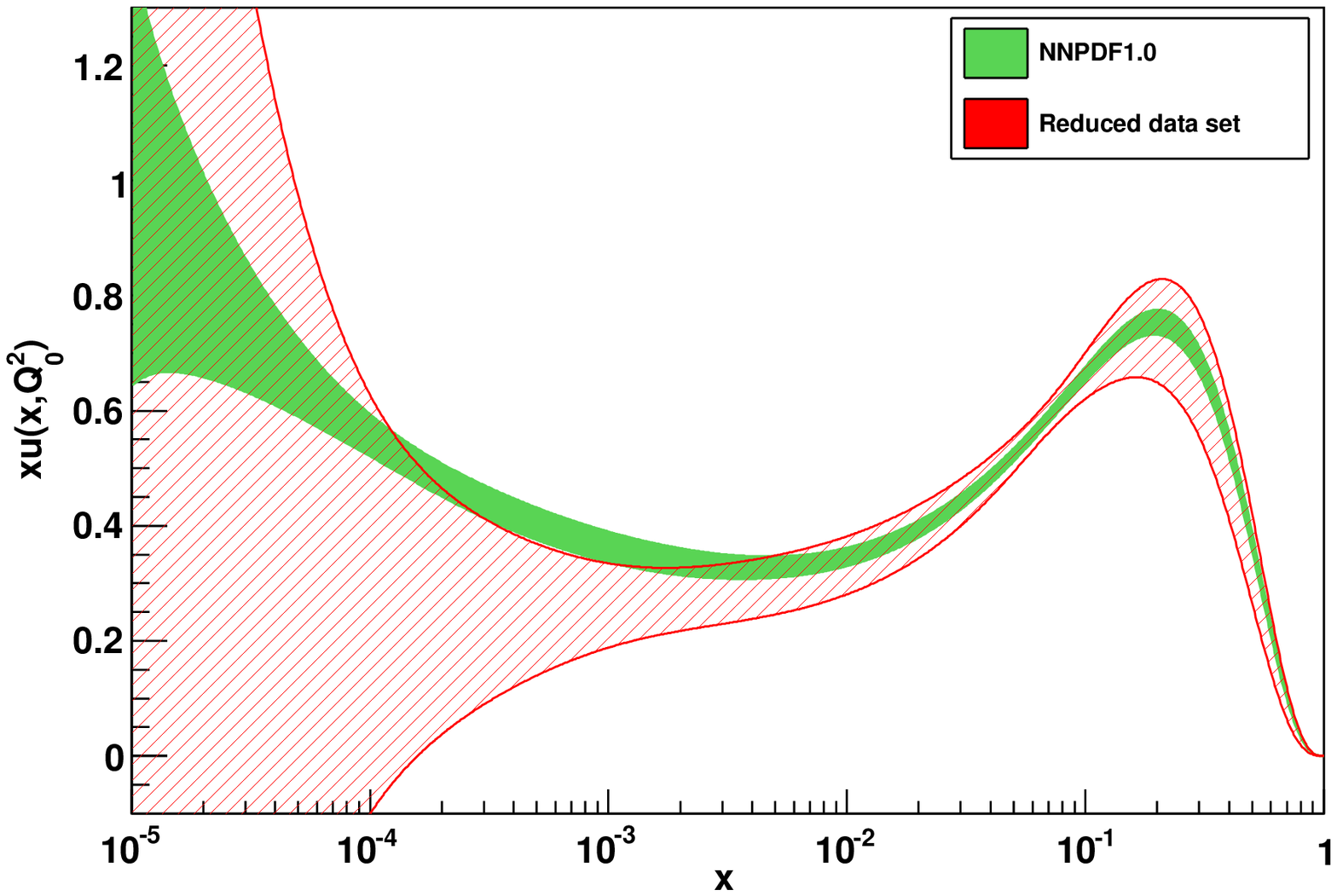}
\epsfig{width=0.49\textwidth,figure=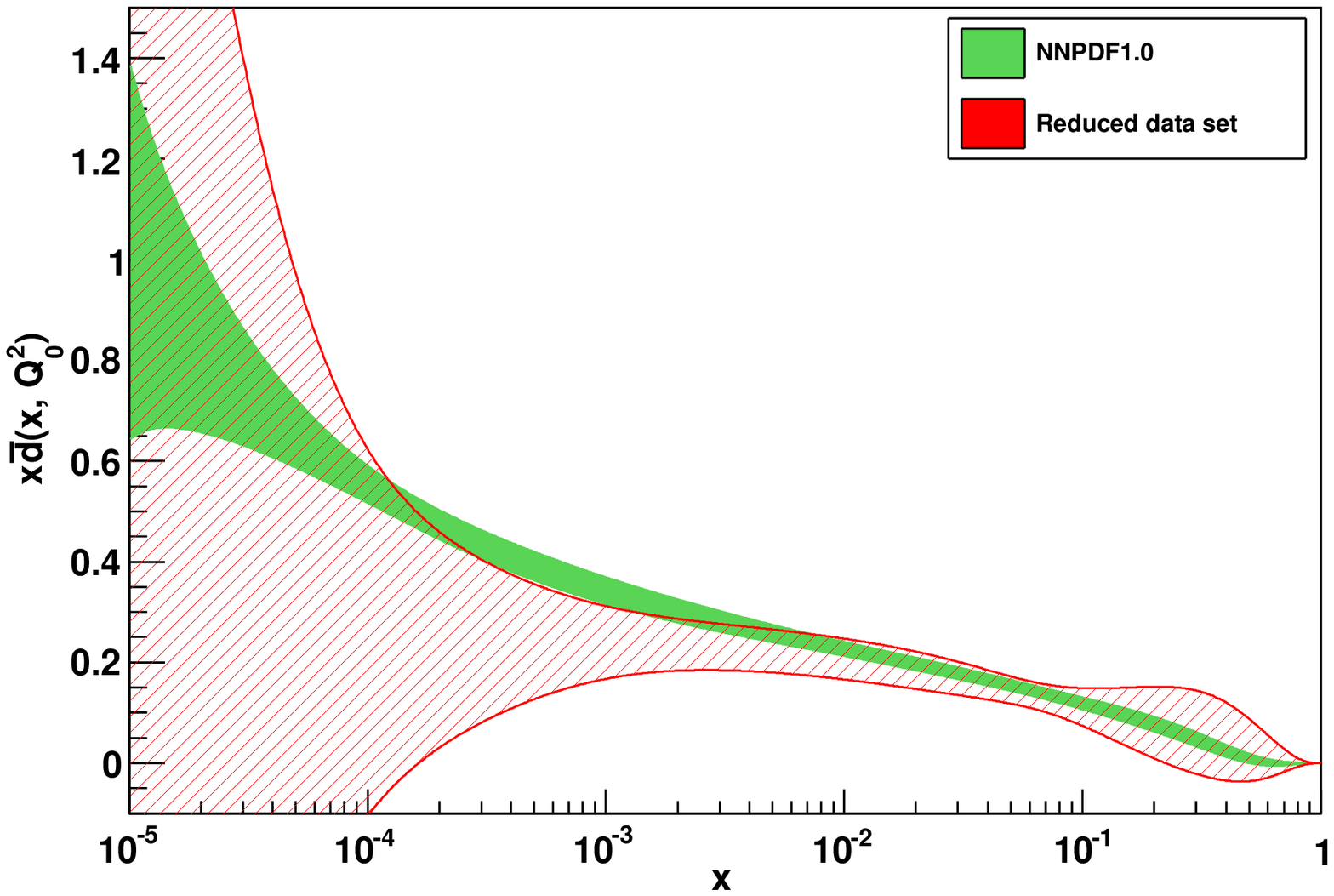}
\caption{\small 
Comparison of the NNPDF1.0 reference PDFs to those obtained from a fit
to the reduced data set of Table~\ref{tab:exps-sets-bench}. The PDFs
shown at the starting scale
$Q_0^2=2$~GeV$^2$  are
$u(x)$ (left plot)
and  $\bar{d}(x)$  (right plot).}
\label{fig:benchlike}
\end{figure}
%------------------------------------------------------------------
At first, we
simply reduce our data set to the much smaller set of
Table~\ref{tab:exps-sets-bench}, with a total of $N_{\rm dat}=733$
data points, to be compared to the  $N_{\rm dat}=3163$ of the full
fit. This is the same data set which was used for the benchmark
Ref.~\cite{heralhc}, mentioned in Sect.~\ref{sec-det}. 
A comparison of the up and
antidown PDFs obtained in this way with the ones of the reference fit is shown 
in Fig.~\ref{fig:benchlike}. Clearly, as data are removed the uncertainty
bands increase in the region where there is information
loss (such as large $x$ for the sea and small $x$ for the valence),
but central values remain compatible within uncertainties. This
shows that the parton parametrization with neural nets is flexible
enough to accommodate results from either of these data sets, and it
confirms that results are indeed independent of the parametrization.

This increase in error band  when data are removed indicates that the
data included in the fit are mostly consistent with each
other, i.e. they lead to consistent underlying PDF sets. Indeed, if
there are systematic data inconsistencies, when 
the inconsistent data are a small
subset of the full data set they will on average be fitted poorly by the
best-fit PDFs. This is due to the fact that 
when these data are included in the training set an
improvement of the quality of the fit to them does not generally lead
to an improvement of the quality of the fit to the remaining data in
the validation set. This behaviour was observed in our fits
of Ref.~\cite{f2p},
when fitting some NMC structure function data which are
known~\cite{Pumplin:2002vw}  to be
somewhat inconsistent with the rest.
If instead two sets of data are inconsistent and
roughly of the same size, the best fit PDFs will fluctuate between
those who give a good fit to either of the datasets,
according to which data are randomly
included in the training set of each replica. A mild inconsistency of
this kind was recently
observed 
in our fits when comparing ZEUS and H1 structure function
data~\cite{nnhlc}.

 In neither case will the
inclusion of inconsistent data
lead to a decrease of the uncertainty, and
in the latter case it might even lead to an increase of the
uncertainty, if the systematic discrepancy between the two datasets is
sizably larger than the uncertainty of each of them.
 In all cases, inconsistencies
will be signalled by a large value of the $\chi^2$ of the best fit. 
The behaviour displayed in Fig.~\ref{fig:benchlike}, together with 
the low value of the $\chi^2=1.34$ in Table~\ref{tab:est-fin-tot},
suggests that only minor insonsistencies are present in the NNPDF1.0
dataset, although the mild inconsistencies such as those
within the NMC dataset and
between H1 and ZEUS datasets mentioned above are likely to be present.

\begin{figure}[t!]
\centering
\epsfig{width=0.48\textwidth,figure=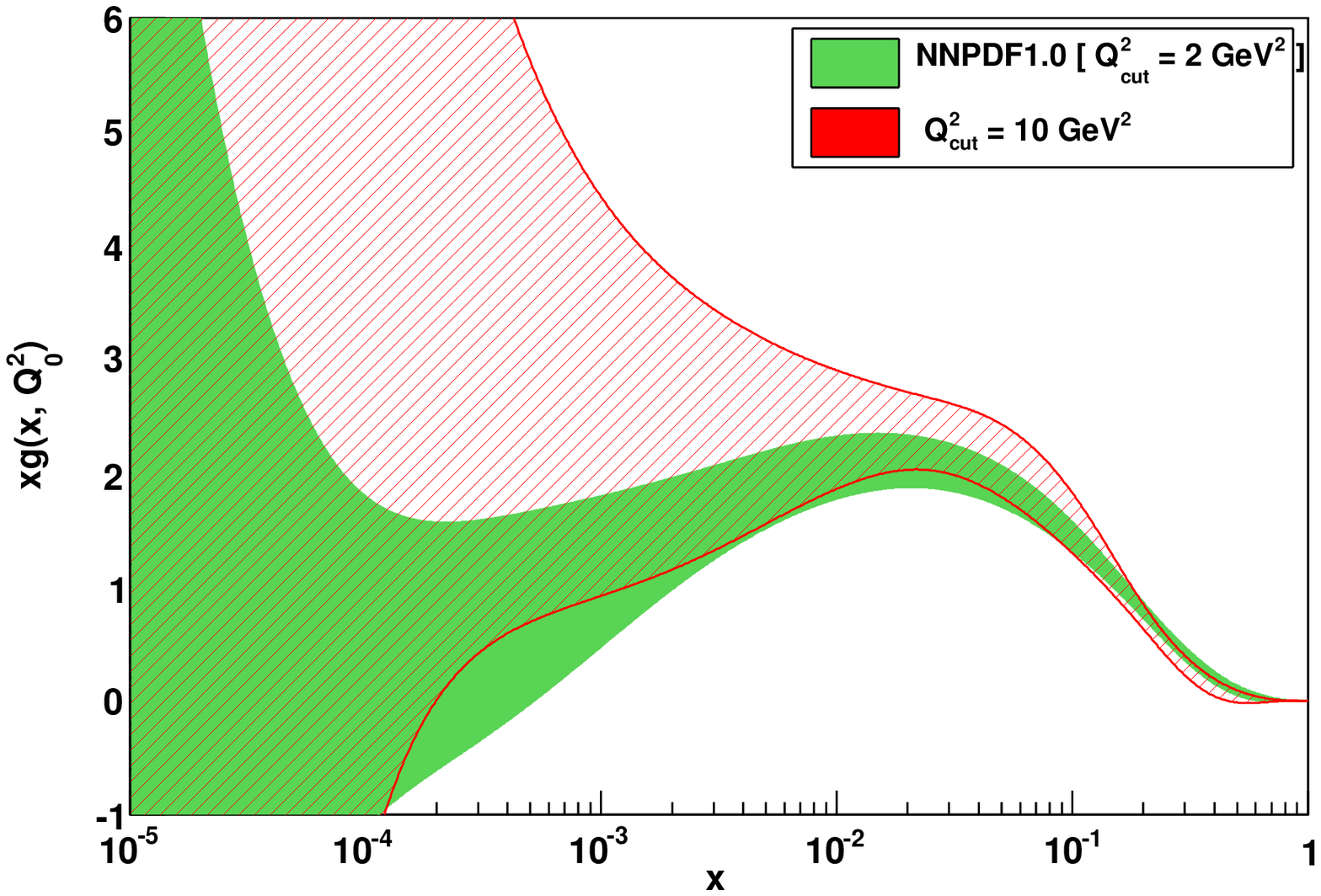}
\epsfig{width=0.48\textwidth,figure=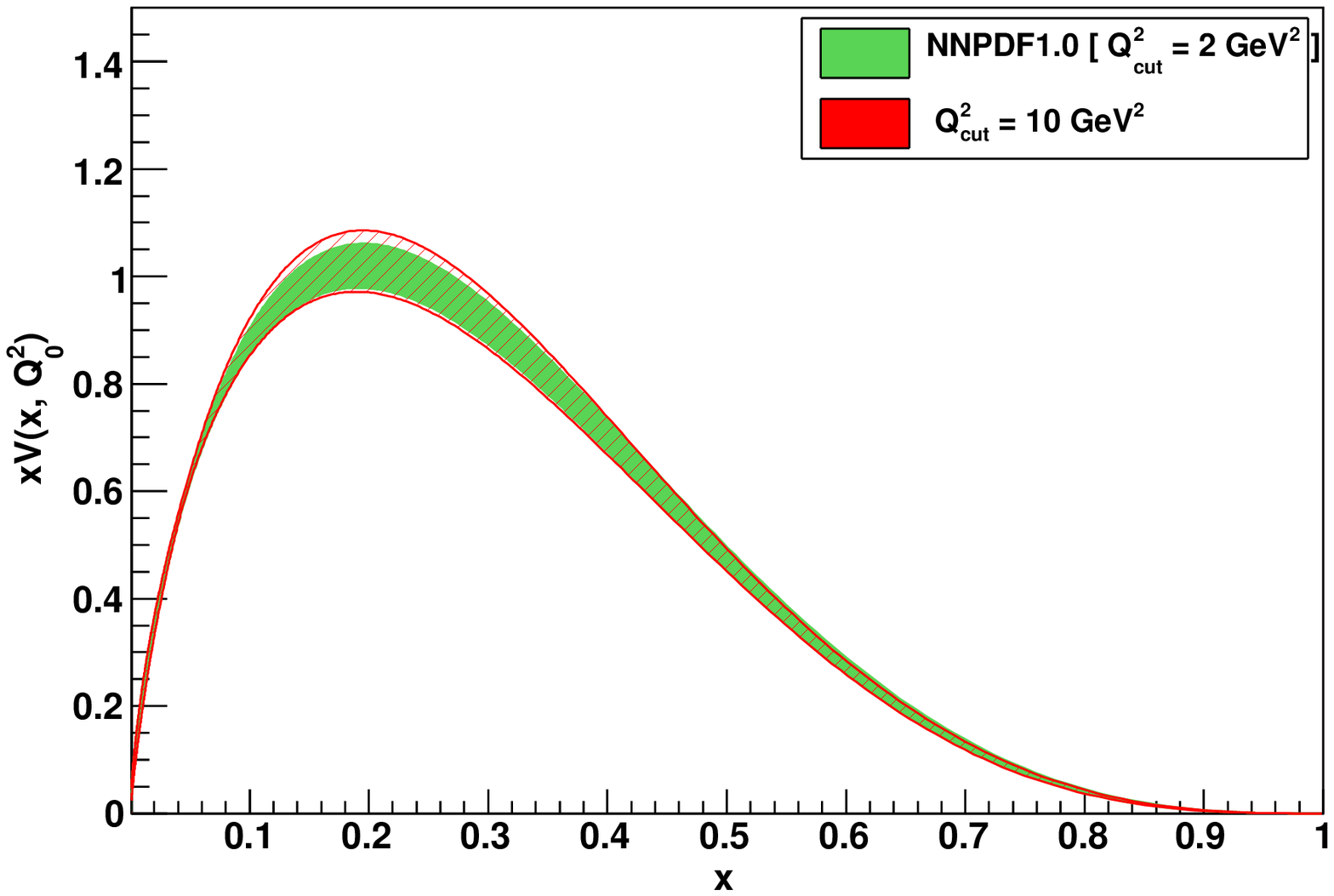}
\caption{\small 
Comparison of the reference fit
to a fit in which where the kinematical cut in $Q^2$ has
been raised from  $Q_{\rm min}^2=2$ GeV$^2$ to
$Q_{\rm min}^2=10$ GeV$^2$. The  gluon (left plot)
and the total valence PDFs (right plot) are shown.}
\label{fig:kincuts}
\end{figure}

A more detailed check of stability
can be performed by removing data in a fixed
kinematic region.
To this purpose, we repeat
the reference fit but with the cut in
$Q^2$ raised from the default
$Q_{\rm cut}^2$=2 GeV$^2$ to $Q_{\rm cut}^2$=10 GeV$^2$. 
As can be seen from Fig.~\ref{fig:dataplot}, this removes
from the analysis a sizable amount of data, leaving $N_{\rm dat}=2355$
out of the   $N_{\rm dat}=3163$ of the reference fit.
Results found in this case are displayed in Fig.~\ref{fig:kincuts},
which clearly shows the increased uncertainty in the small $x$ region
where data have been removed, and good stability in the valence region
where the data set is essentially unchanged.

\begin{table}[t!]
\begin{center}
\small
\begin{tabular}{|c|c|c|c|}
\hline 
 &  Data/Data  & Data/Extrapolation & Extrapolation/Extrapolation \\
\hline 
$\Sigma(x,Q_0^2)$ & $3~10^{-3} \le x \le 0.1$  & $5~10^{-4} \le x \le 10^{-3}$
&  $5~10^{-5} \le x \le 10^{-4}$  \\
\hline
$\la d[q]\ra$  & 2.48 &  8.04&  5.79\\
$\la d[\sigma]\ra$  & 2.85& 3.73 &  4.43\\
\hline 
\hline 
$g(x,Q_0^2)$ &$3~10^{-3} \le x \le 0.1$  & $5~10^{-4} \le x \le 10^{-3}$
&  $5~10^{-5} \le x \le 10^{-4}$  \\
\hline
$\la d[q]\ra$  & 4.27& 6.12& 5.04\\
$\la d[\sigma]\ra$  & 3.06& 3.04 & 2.03 \\
\hline 
\hline 
$T_3(x,Q_0^2)$ & $0.2 \le x \le 0.75$  & $5~10^{-1} \le x \le 0.1$ &
$10^{-3} \le x \le 10^{-2}$ \\
\hline
$\la d[q]\ra$  & 2.33& 2.00 & 0.53 \\
$\la d[\sigma]\ra$  & 1.62 &  1.34 &  1.14\\
\hline 
\hline 
$V(x,Q_0^2)$ & $0.3 \le x \le 0.6$  & $0.1 \le x \le 0.2$ &
 $3~10^{-3} \le x \le 3~10^{-2}$ \\
\hline
$\la d[q]\ra$  & 0.88 & 1.42 & 1.15 \\
$\la d[\sigma]\ra$  & 1.09 &  1.89 &  1.62\\
\hline 
\hline 
$\Delta_S(x,Q_0^2)$ & $0.3 \le x \le 0.6$  & $0.1 \le x \le 0.2$ &
 $3~10^{-3} \le x \le 3~10^{-2}$ \\
\hline
$\la d[q]\ra$  & 2.54 & 2.29 & 0.92 \\
$\la d[\sigma]\ra$  & 1.13 & 1.13 & 1.15 \\
\hline 
\end{tabular}

\end{center}
\caption{\small Distance between
results computed from  a set of 
$N_{\rm rep}=100$ replicas from the reference fit and a
set of 
$N_{\rm rep}=100$ replicas from a 
fit where the kinematic cut in $Q^2$ has
been raised from  $Q_{\rm min}^2=2$ GeV$^2$ to
$Q_{\rm min}^2=10$ GeV$^2$.
\label{tab:stabtab-kincuts}}
\end{table}
%------------------------------------------------------------
%------------------------------------------------------------------
This is assessed in a quantitative way in
Table~\ref{tab:stabtab-kincuts},
by tabulating the distance
between results in the three kinematic regions (a) where data are
available both before and
after raising the cut (Data/Data); (b) where there are data
before raising the cut but not afterwards (Data/Extrapolation); (c)
where there are no data either with the lower or higher cut
(Extrapolation/Extrapolation). In most cases we find good stability
at the 90\% confidence level: 
as shown in Fig.~\ref{fig:kincuts}
whenever central values change significantly, the uncertainties
increase correspondingly. 

An interesting exception is the case of the
Data/Extrapolation region for singlet and gluon. This is the kinematic
region of most of the HERA data: raising the cut corresponds to
excluding the HERA data from the fit. In this case, even though the 
central reference is within the uncertainty band of the fit with the
higher cut (see Fig.~\ref{fig:kincuts}), the distance between central values 
is rather larger than allowed by
statistical fluctuations. This can be understood as a consequence of
the fact that the behaviour of the gluon at small $x$
in the HERA region cannot be predicted
by simple extrapolation of the behaviour observed at larger $x$ in the
NMC region. Similar behaviour was observed in structure function
fits~\cite{f2p}, though here the problem is alleviated by the fact that
for a wide class of starting quark and gluon distributions the
shape of structure functions obtained from NLO QCD evolution is
universal~\cite{das}.

\begin{table}
\begin{center}
\small
\begin{tabular}{|c|c|c|}
\hline 
 &  Data  & Extrapolation \\
\hline 
$\Sigma(x,Q_0^2)$ & $5~10^{-4} \le x \le 0.1$  & $10^{-5} \le x \le 10^{-4}$  \\
\hline
$\la d[q]\ra$  & 2.13  &  1.95 \\
$\la d[\sigma]\ra$  & 1.17 &  1.83 \\
\hline 
\hline 
$g(x,Q_0^2)$ & $5~10^{-4} \le x \le 0.1$  & $10^{-5} \le x \le 10^{-4}$  \\
\hline
$\la d[q]\ra$  &  1.76& 1.82 \\
$\la d[\sigma]\ra$  & 1.29 & 1.87 \\
\hline 
\hline 
$T_3(x,Q_0^2)$ & $0.05 \le x \le 0.75$  & $10^{-3} \le x \le 10^{-2}$  \\
\hline
$\la d[q]\ra$  & 1.07 & 0.93 \\
$\la d[\sigma]\ra$  & 1.17 & 1.59 \\
\hline 
\hline 
$V(x,Q_0^2)$ & $0.1 \le x \le 0.6$  & $3~10^{-3} \le x \le 3~10^{-2}$  \\
\hline
$\la d[q]\ra$  & 1.42 & 1.62 \\
$\la d[\sigma]\ra$  & 1.56 & 1.64 \\
\hline 
\hline 
$\Delta_S(x,Q_0^2)$ & $0.1 \le x \le 0.6$  & $3~10^{-3} \le x \le 3~10^{-2}$  \\
\hline
$\la d[q]\ra$  & 0.99 & 1.71  \\
$\la d[\sigma]\ra$  & 1.35 & 1.73  \\
\hline 
\end{tabular}

\end{center}
\caption{\small 
Distance between
results computed from  a set of 
$N_{\rm rep}=100$ replicas from the reference fit and a
set of 
$N_{\rm rep}=100$ replicas from a 
fit where the  training fraction has been reduced from its default
$f_{\rm tr}=0.5$ to $f_{\rm tr}=0.25$.
\label{tab:stabtab-trfrac}}
\end{table}
As a final test, we reduce the number of data points by randomly
removing points from the data set. This is simply done by
reducing the fraction $f_{\rm tr}$ of
points in the training set from its default value $f_{\rm tr}=0.5$
(see  Sect.~\ref{sec-dynstop}) to $f_{\rm tr}=0.25$.
As shown in Table~\ref{tab:stabtab-trfrac}, results
move by at most two-$\sigma$ as a consequence. 
Furthermore, in this case uncertainties are unchanged:
for instance, the value of $\la \sigma^{(\net)}
\ra_{\dat}$ coincides with that of the
reference fit given in Tab.~\ref{tab:est-fin}. 
Hence, the fit behaves as one expects when the data set is changed, but
with no significant information loss. This indicates that the data set
is consistent and quite redundant. Of course, if the training fraction
were reduced further eventually one would have information loss, and
one would be led back to the situation of smaller sets of experiments
or kinematic cuts.

\subsection{Theoretical uncertainties}
\label{sec:theounc}

\begin{table}[t!]
\begin{center}
\footnotesize
\begin{tabular}{|c|c|c||c|c||c|c|}
\hline
Pert. order &  \multicolumn{2}{c||}{LO}
 & \multicolumn{2}{c||}{NLO} & 
\multicolumn{2}{c|}{NLO} \\
\hline
$\alpha_s(M_Z^2)$ &  \multicolumn{2}{c||}{0.130} & 
\multicolumn{2}{c||}{0.117}& \multicolumn{2}{c|}{0.121}  \\
\hline 
$\chi^2$  & \multicolumn{2}{c||}{1.41} & \multicolumn{2}{c||}{1.35}&
\multicolumn{2}{c|}{1.33}\\
\hline 
$\la {\rm TL}\ra$   & 
\multicolumn{2}{c||}{1153} & \multicolumn{2}{c||}{877}& 
\multicolumn{2}{c|}{736}\\
\hline 
  $\la \sigma^{(\net)}
\ra_{\dat}$  & 
\multicolumn{2}{c||}{ $1.4~10^{-2}$} & \multicolumn{2}{c||}{ $1.4~10^{-2}$}& 
\multicolumn{2}{c|}{ $1.4~10^{-2}$}\\
\hline 
  $\la \rho^{(\net)}
\ra_{\dat}$ & 
\multicolumn{2}{c||}{0.41} & \multicolumn{2}{c||}{0.40}& 
\multicolumn{2}{c|}{0.43}\\
\hline 
  $\la  {\rm cov}^{(\net)}
\ra_{\dat}$ & 
\multicolumn{2}{c||}{ $1.6~10^{-4}$} & \multicolumn{2}{c||}{$1.5~10^{-4}$}& 
\multicolumn{2}{c|}{ $2.0~10^{-4}$}\\
\hline 
\hline
 &  Data  & Extrapolation &  Data  & Extra &  Data  & Extra \\
\hline 
$\Sigma(x,Q_0^2)$ & $5~10^{-4} \le x \le 0.1$  & $10^{-5} \le x \le 10^{-4}$
  &  &  
& &    \\
\hline
$\la d[q]\ra$  &  13.42 & 3.93  & 1.72 & 1.05 & 0.73 & 0.81    \\
$\la d[\sigma]\ra$  &  1.38 & 0.93   & 1.05 &
1.03 & 1.22 & 0.95 \\
\hline 
\hline 
$g(x,Q_0^2)$ & $5~10^{-4} \le x \le 0.1$  & $10^{-5} \le x \le 10^{-4}$ 
  & &&&\\
\hline
$\la d[q]\ra$  & 18.62  & 6.42  & 4.68 &  2.29&4.12 & 0.71  \\
$\la d[\sigma]\ra$  & 1.10 & 2.10  &1.00 & 0.91 & 0.88 & 0.83  \\
\hline 
\hline 
$T_3(x,Q_0^2)$ & $0.05 \le x \le 0.75$  & $10^{-3} \le x \le 10^{-2}$ 
 & &&&  \\
\hline
$\la d[q]\ra$  &  3.72 & 1.33   & 0.71 & 0.71 & 1.55 & 0.96\\
$\la d[\sigma]\ra$  & 1.22 & 0.92  & 0.93 & 0.75 & 1.11 &
0.78  \\
\hline 
\hline 
$V(x,Q_0^2)$ & $0.1 \le x \le 0.6$  & $3~10^{-3} \le x \le 3~10^{-2}$ 
 & &&& \\
\hline
$\la d[q]\ra$  & 3.63  & 2.92  & 0.92 & 0.74 & 1.89 & 1.72  \\
$\la d[\sigma]\ra$  & 1.58 & 1.17   & 0.94 & 0.71 & 0.67 & 0.65  \\
\hline 
\hline 
$\Delta_S(x,Q_0^2)$ & $0.1 \le x \le 0.6$  & $3~10^{-3} \le x \le 3~10^{-2}$ 
  & &&&  \\
\hline
$\la d[q]\ra$  & 1.34 & 1.94  & 0.74 &  0.58& 0.86 & 1.36  \\
$\la d[\sigma]\ra$  & 0.95 & 1.06   & 0.67 & 0.83 & 0.78 &
0.76  \\
\hline 
\end{tabular}

\end{center}
\caption{\small Statistical estimators for the LO fit and the NLO fits
  with different values of $\alpha_s$ (top) and distance between
results computed from  a set of 
$N_{\rm rep}=100$ replicas from these fits and
set of 
$N_{\rm rep}=100$ replicas from the
 reference fit.
\label{tab:stabtab-theo}}
\end{table}
%------------------------------------------------------------
\begin{figure}[t!]
\centering
\epsfig{width=0.48\textwidth,figure=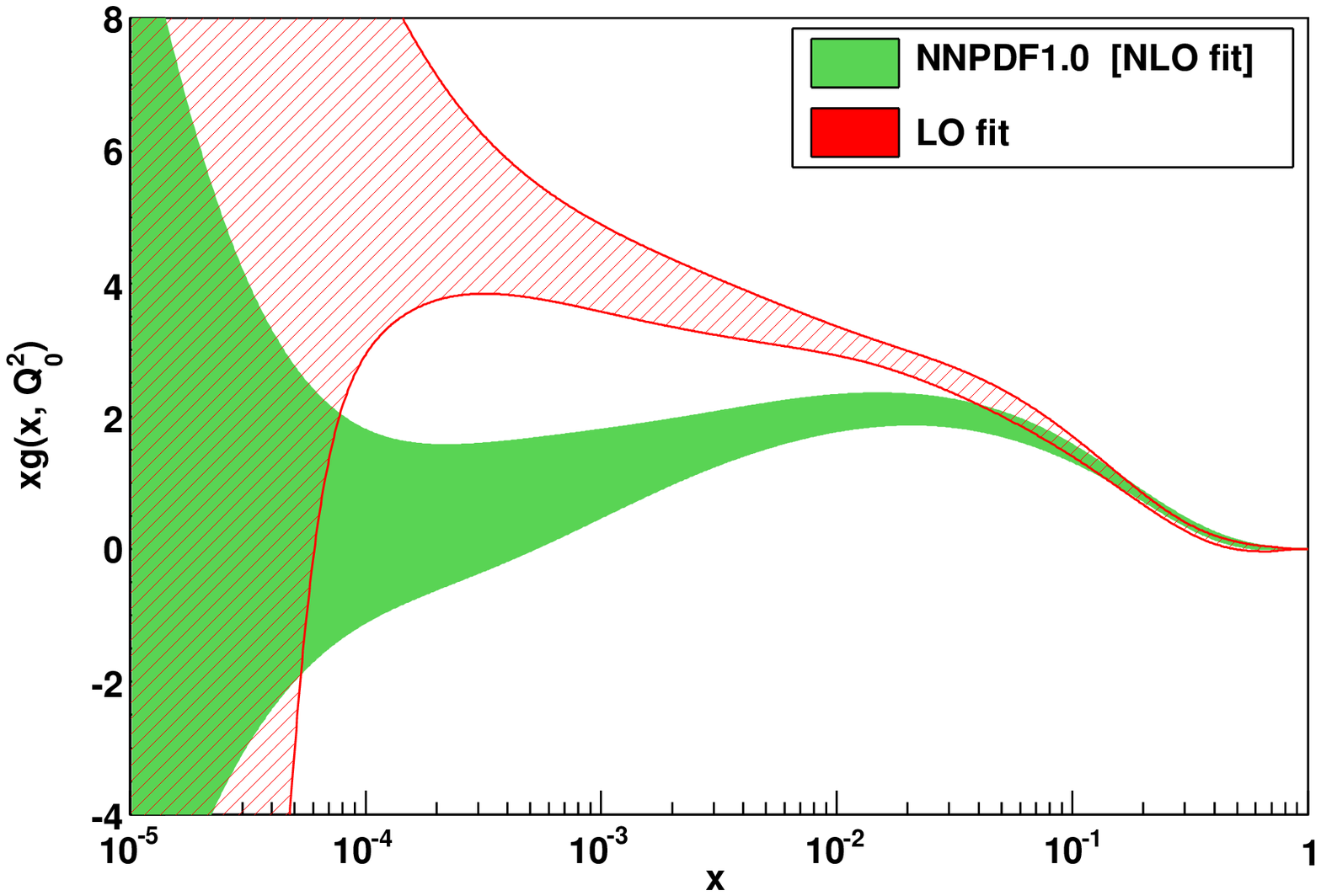}
\epsfig{width=0.48\textwidth,figure=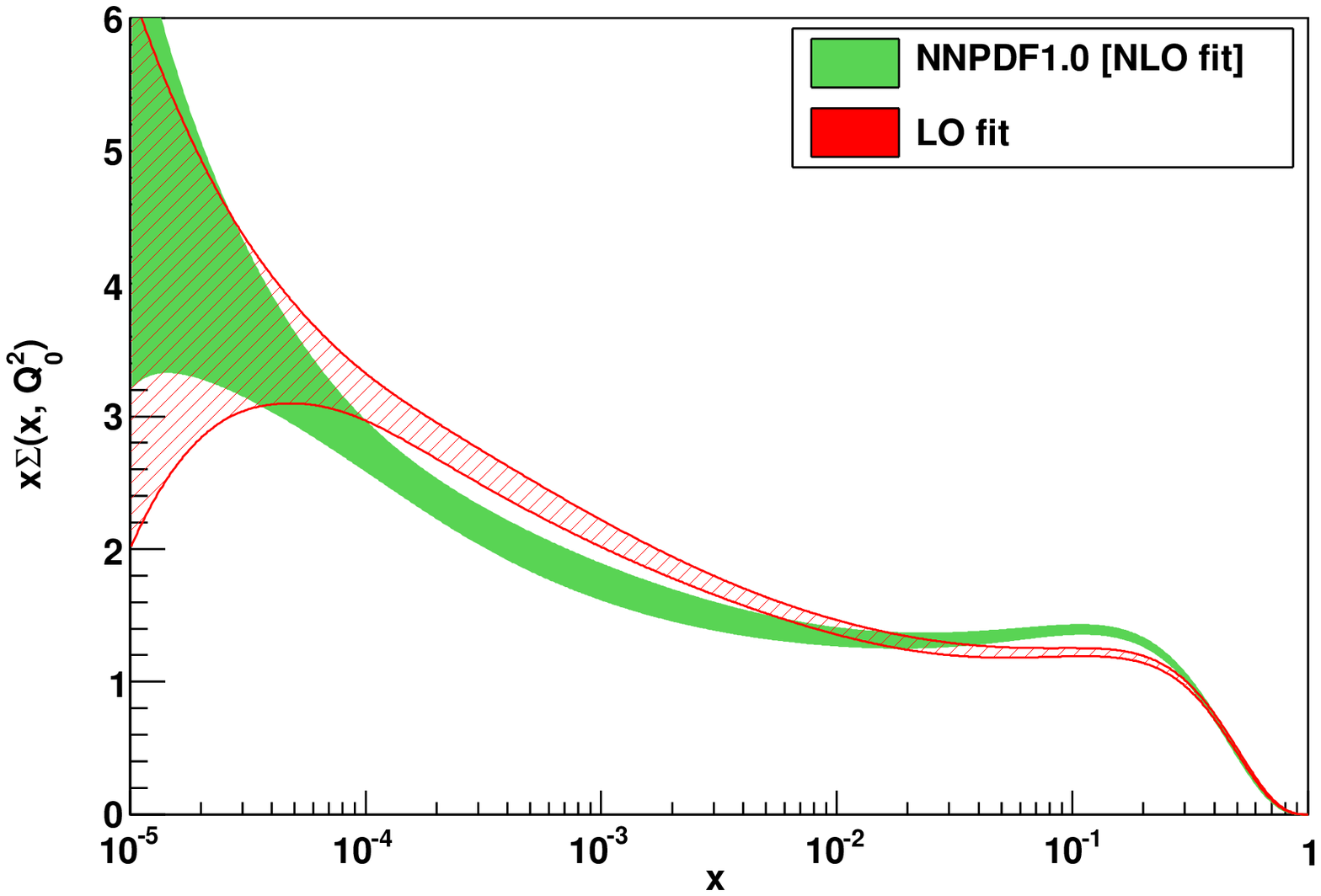}
\caption{\small 
Comparison of the NLO (reference) and LO fit results for the gluon
(left) and quark singlet PDFs (right).
\label{fig:lofit}}
\end{figure}
%------------------------------------------------------------------

In our determination of PDFs, all systematic uncertainties in the data
have been accounted for in the Monte Carlo data generation, as
discussed in Sect.~\ref{sec:expdata}: they are then propagated through
the fitting procedure onto our final results. Therefore, error
bands on the NNPDF1.0 PDFs
already include both statistical and systematic uncertainties on the
data. On top of these, however, there exists theoretical uncertainties
due to the fact that
we do not fit directly experimental observables,
but rather PDFs, which are theoretical constructs
of QCD, and are thus affected by uncertainties related to incomplete
knowledge of the theory: we shall refer to these (sometimes called
model uncertainties) as theoretical
uncertainties. 

The main theoretical uncertainties on NNPDF1.0 are  those
related to the fact that our analysis is performed at NLO, and thus
neglects effects at NNLO and beyond, and to the choice of value of the
strong coupling $\alpha_s$, which is only known to finite accuracy.
 To address these issues, we have repeated our fit
at LO, and also at NLO with two different values of the strong
coupling, $\alpha_s(M_Z^2)=0.117$ and
$\alpha_s(M_Z^2)=0.121$. The results of these studies are summarized in
Tab.~\ref{tab:stabtab-theo}, where we show the statistical features
of these fits, to be compared with those of the reference fit
Tab.~\ref{tab:est-fin}, and the distance of the result of these fits
from those of the reference.

Comparison of the LO fit to the NLO shows that the quality of the fit
deteriorates significantly when using LO theory, as one might
expect. However, the size of the uncertainties remains essentially
unchanged, as shown by the value of both $\la \sigma^{(\net)}
\ra_{\dat}$ and  of $d[\sigma]$. Furthermore, all central values of
the LO fit vary by an amount which is compatible with statistical
fluctuations, with the exception of the singlet and gluon, which vary
by about two-$\sigma$, as one expects due to the fact that the gluon
contribution to deep-inelastic coefficient functions only starts at NLO.
The change in behaviour of the singlet and gluon is displayed in
Fig.~\ref{fig:lofit}. 

The  comparison of the LO and NLO fits also indicates that the
theoretical uncertainty due to lack of inclusion of NNLO corrections
is negligible on the
scale of statistical uncertainties. This conclusion is based on the
observation that NNLO
corrections to all PDFs are known~\cite{Alekhin:2003yh,Martin:2007bv}
 to be a fraction of the typical
NLO corrections (i.e. the NLO-LO difference)
in the nonsinglet sector, which we already
find to be smaller by a factor larger
than two than statistical uncertainties.

Comparison of the NLO fits with different values of $\alpha_s$ shows a
variation in $\chi^2$ which, though small in terms of the expected statistical
fluctuations of the $\chi^2$ itself, is actually quite significant in
absolute value and thus for the determination of
$\alpha_s$ (see Ref.~\cite{Collins:2001es} for an explanation of this
distinction). The possibility of such a determination 
$\alpha_s$ will be explored further in future studies. The change in
central values of PDFs is compatible with statistical fluctuations,
except in the case of the gluon, as one expects due to the
fact that the gluon contribution to deep-inelastic structure functions
is of order $\alpha_s$. Even in this case, however, the change is a
fraction of the statistical uncertainty.

We conclude that the dominant theoretical uncertainties are small
on the scale of the statistical uncertainties in our fit. If this will no
longer be the case in future analyses, theoretical uncertainties can
be taken into account by varying the underlying parameters in a
randomized way during the fitting procedure. For example, the
uncertainty on the value of $\alpha_s$ could be accounted for by taking
it as a random variable distributed  about its central
value with a given uncertainty, and letting it fluctuate between replicas.
 For the time being, however, this does not appear to be
necessary. 

Further possible sources of 
theoretical uncertainty include effects related to large $x$
and small $x$ resummation of the perturbative expansion, 
the treatment and position of heavy quark thresholds, higher twist
corrections, and nuclear effects especially in the treatment of
neutrino data. All these corrections are expected to be  smaller than
those discussed here. They will addressed in future studies as the
precision of the analysis improves.

\subsection{Usage and delivery of the NNPDF1.0 set}
\label{sec:delivery}

The statistical features of the NNPDF1.0 parton set are those of
the full ensemble of $N_{\rm rep}=1000$ replicas. 
However, as we have explicitly verified in
Sect.~\ref{sec:stabarch}, the scaling 
with the value of $N_{\rm rep}$ of results obtained from NNPDF1.0 works 
just as one would expect on naive
statistical grounds, and therefore a smaller number of replicas may
be sufficient for the computation of quantities whose uncertainty
decreases fast enough with $N_{\rm rep}$. For example, as is 
well known~\cite{cowan},
the variance of the mean of a sample of size $N_{\mathrm{rep}}$
is $s^2/N_{\rm rep}$, where $s$ is the standard deviation of
the underlying distribution from which the sample is taken. 
Similarly, the variance of the variance itself
is (for gaussianly distributed quantities) 
$2 s^4/(N_{\rm rep}-1)$,  Therefore,
if the standard deviation corresponds to a 10\% uncertainty, 
a sample of 100 replicas will be sufficient to determine central
values with 1\% accuracy and the uncertainty itself with 5\%
accuracy. 

The computation of quantities that depend on more detailed
statistical features of the sample may be more delicate, and require
the use of the full ensemble.
An example is the computation of correlations: indeed, not only does the 
uncertainty on correlations only decrease as 
$1/\sqrt{N_{\rm rep}}$, but also the correlation computed
with the sample is only asymptotically  an unbiased estimator, since 
it overshoots the true one by an amount which only vanishes as 
$1/N_{\rm rep}$. Therefore, the stability tests of
Sects.~\ref{sec:stabarch}-\ref{sec:stabdata} were computed using
sets of $N_{\rm rep}=100$ replicas, but the correlations 
displayed in Fig.~\ref{fig:pdfcorr} were determined using the 
full set of $N_{\rm rep}=1000$ replicas.

\begin{figure}[t!]
\centering
\epsfig{width=0.48\textwidth,figure=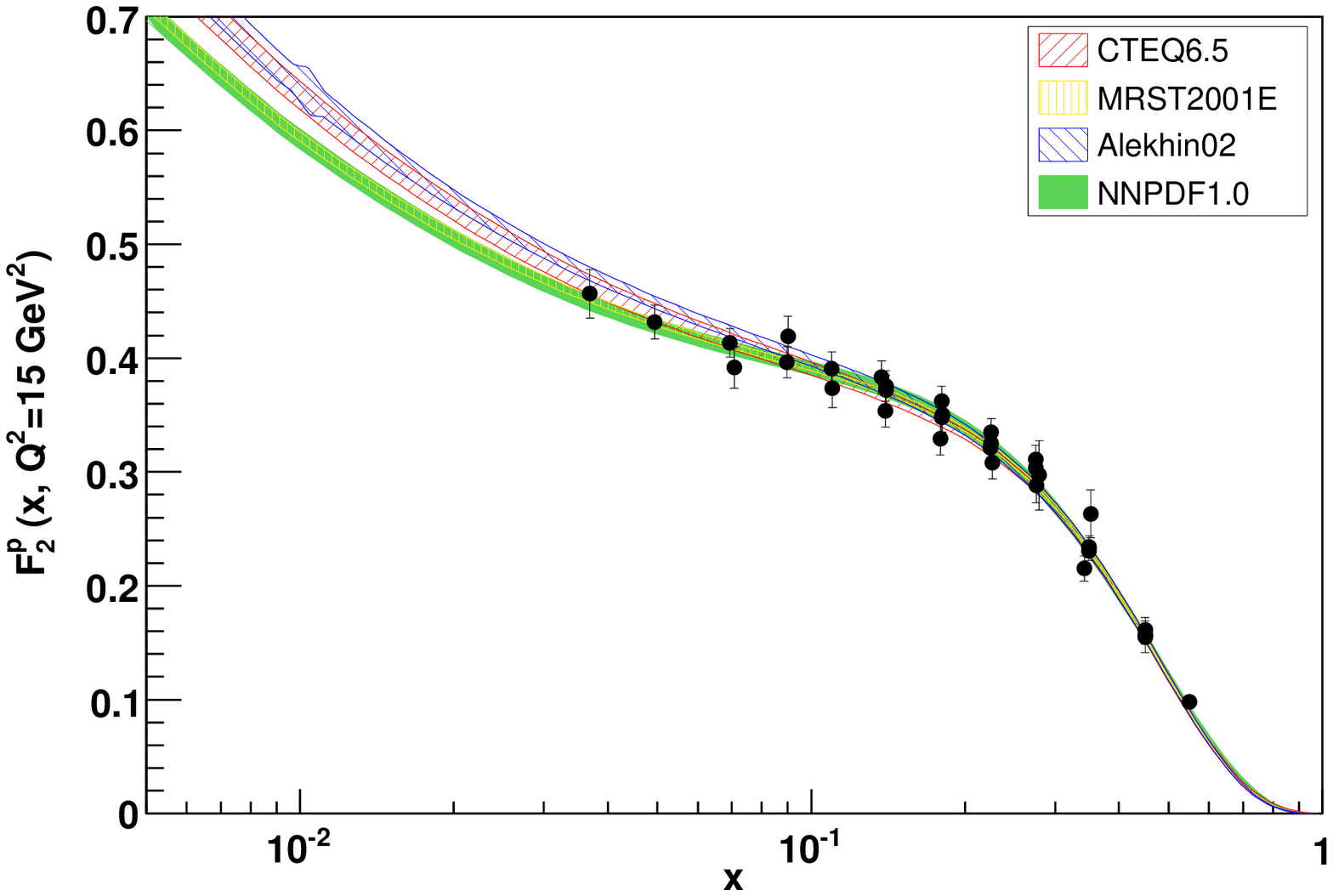} 
\epsfig{width=0.48\textwidth,figure=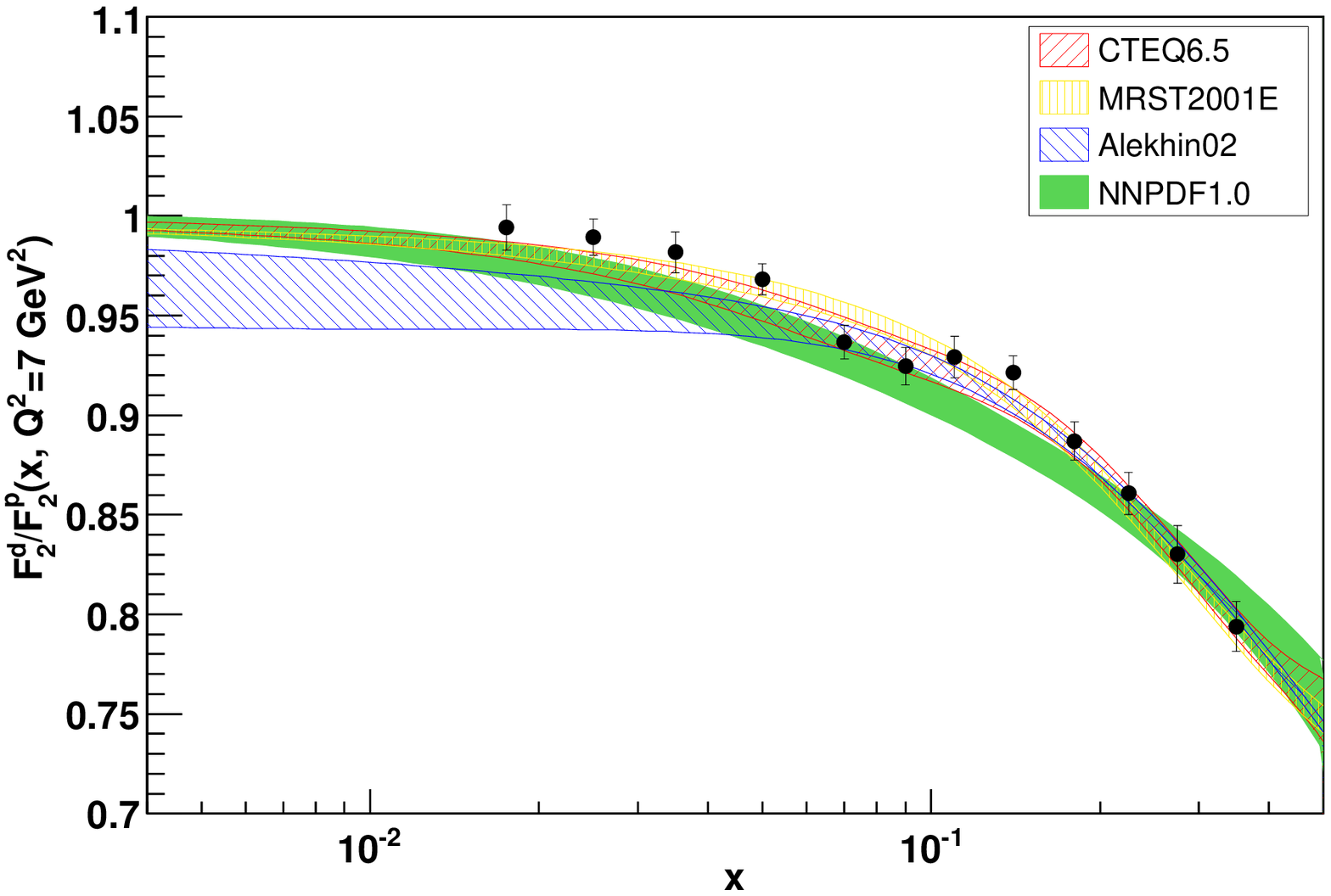} 
\epsfig{width=0.48\textwidth,figure=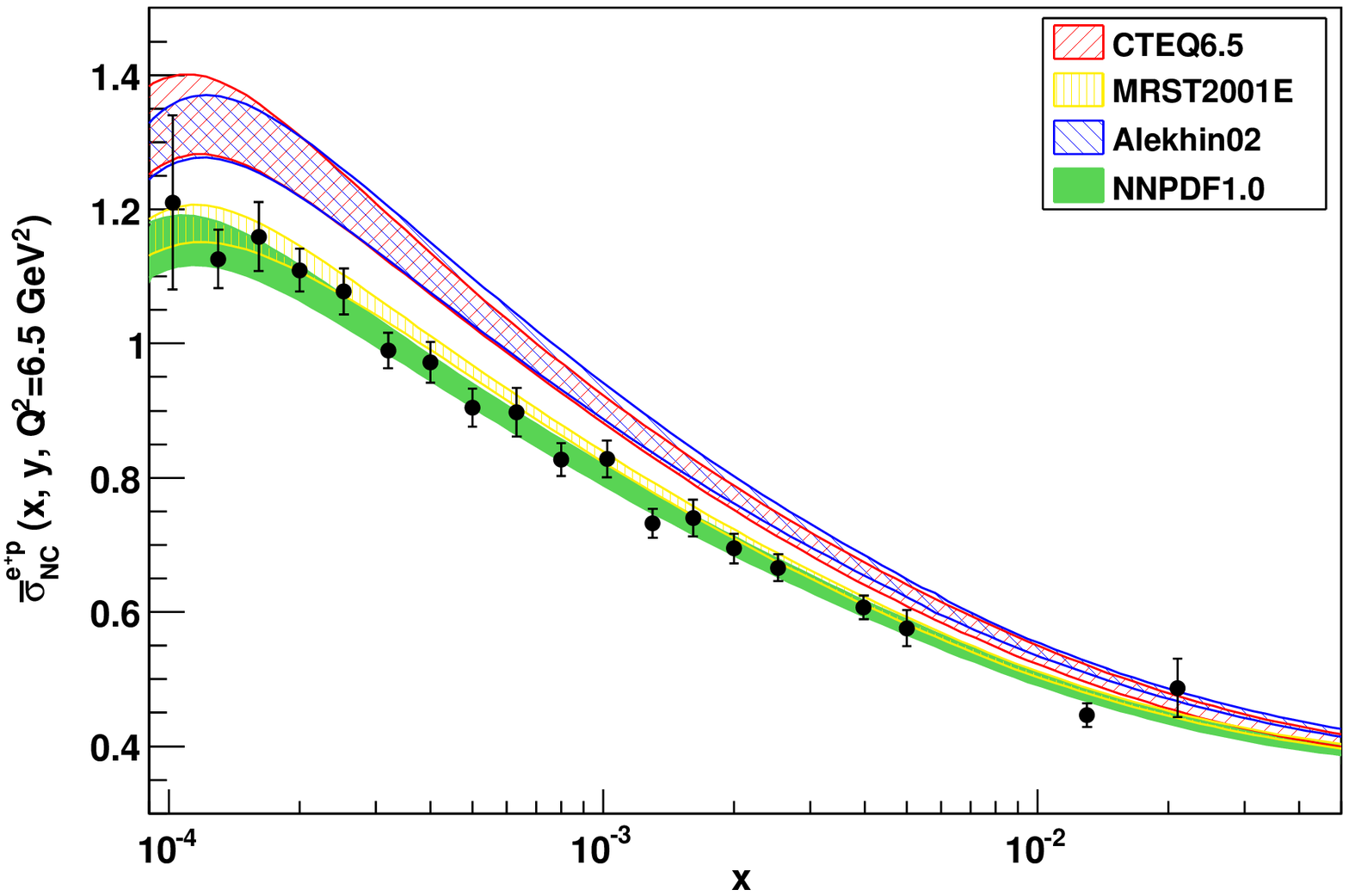} 
\epsfig{width=0.48\textwidth,figure=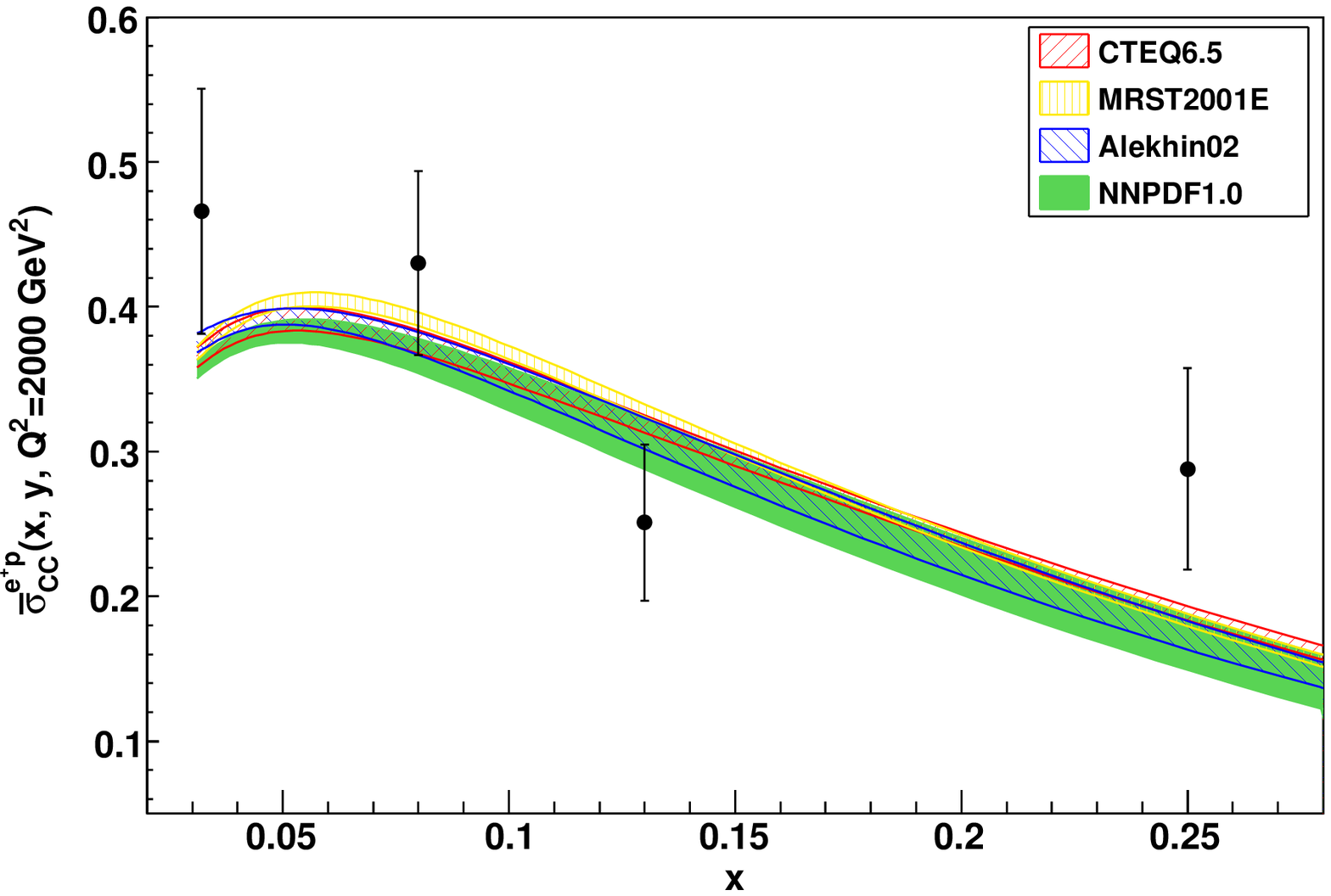}
\epsfig{width=0.48\textwidth,figure=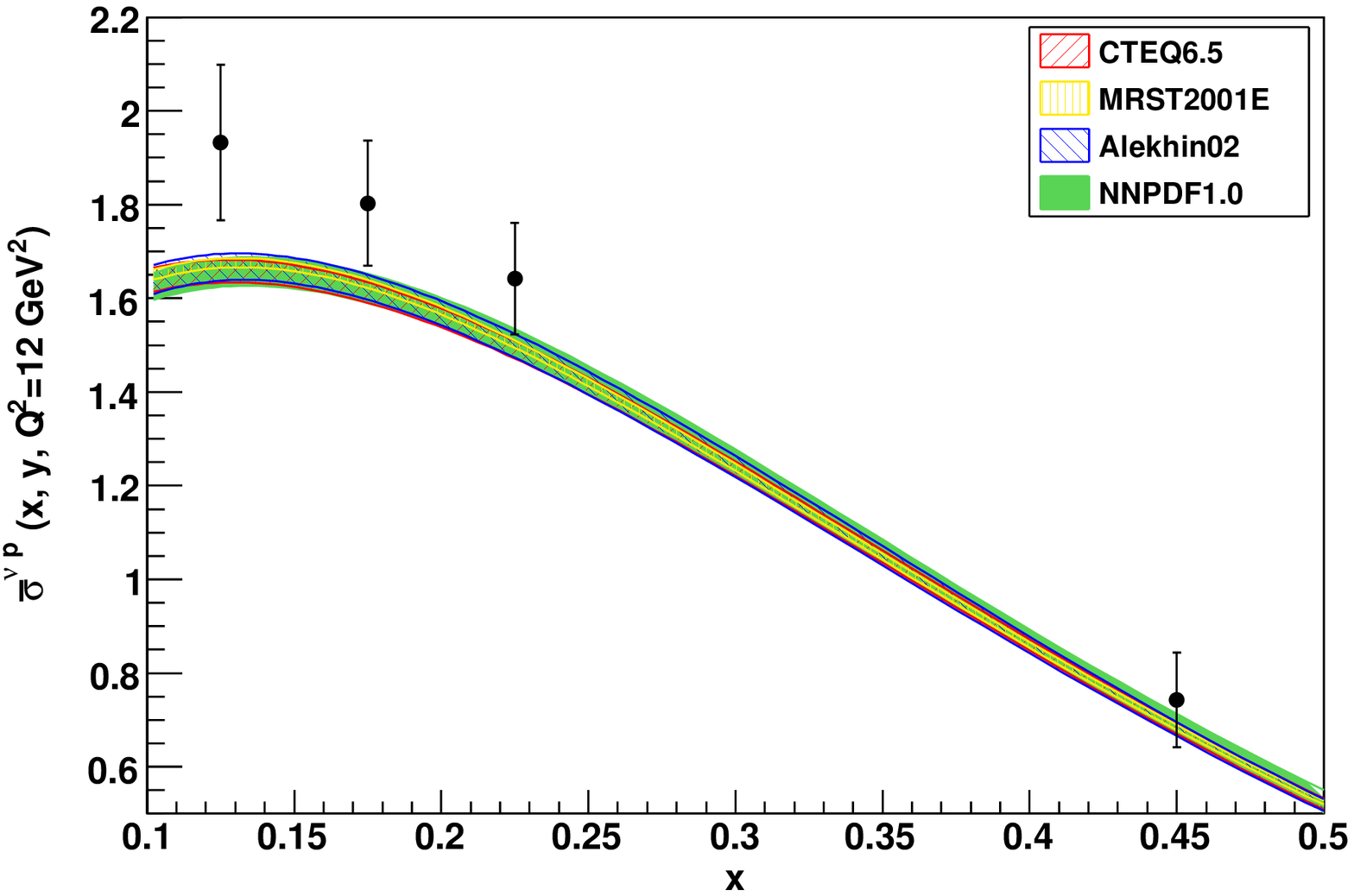}  
  \caption{Comparison of NLO theoretical predictions and data for
    various observables included in the NNPDF fits.
First row: proton $F^p_2$  at $Q^2=15$~GeV$^2$(left) and deuteron/proton ratio
$F^d_2/F^p_2$ at $Q^2=7$~GeV$^2$  (right). Second row: 
positron-proton NC (left) and CC (right) reduced cross sections at
$Q^2=6.5$~GeV$^2$. The value of $y$ is determined as a function of $x$ with
$\sqrt{s}=301$~GeV.
Third row: neutrino-proton total reduced cross-section 
 at
$Q^2=12$~GeV$^2$. The value of $y$ is determined as a function of $x$ with
$E_\nu=70$~GeV.  Theoretical uncertainties are all
one-$\sigma$ bands, while experimental error bars are total uncertainties
with statistical and systematic errors added in quadrature. 
}\label{fig:observables}
\end{figure}

For these reasons, we will provide both the full set 1000 replicas,
and a restricted set of 100 replicas. Of course, other sets can be
extracted as needed from the full set.
The NNPDF1.0 sets can be downloaded from the web site {\tt \bf
http://sophia.ecm.ub.es/nnpdf/}, together with instructions for interfacing
to commonly used codes.  They can also be accessed 
through the LHAPDF library~\cite{LHAPDFurl}. The computation of
central values and errors using Monte Carlo PDF sets is briefly summarized in 
Appendix~\ref{sec:app-pdferr}.

Given the statistical information contained in the full ensemble of PDFs, it is
possible to construct optimized sets which consist of a smaller number
of replicas, but whose statistical properties are closer to those of
the full ensemble than those of a random subset of it. This can be
done  by picking, out of the full set,
a subset that  minimizes a measure of the distance between the probability
distribution of the subset and that of the full set, such as the
relative entropy (or Kullback--Leibler divergence)~\cite{kullback}.
This construction  is the subject of current
investigation and it will be presented elsewhere.

\subsection{Comparison with present and future experimental data}
\label{sec:compdata}

A full study of the phenomenological implications of the 
NNPDF1.0 parton set is beyond the scope
of this work. However, for the sake of illustration,
in this section we present first comparisons of
theoretical predictions obtained using NNPDF1.0 with data, both for
deep-inelastic observables included in the fit and for some LHC
observables.

In Fig.~\ref{fig:observables}
the theoretical prediction obtained using NLO QCD and the NNPDF1.0 set
is compared to the data, for some
representative deep-inelastic observables included in  the data set of
Table~\ref{tab:exps-sets}. Results obtained using the most recent
NLO sets from other groups~\cite{Tung:2006tb,Martin:2002aw,Alekhin} 
included in the LHAPDF library~\cite{LHAPDFurl} are also shown.
As one might expect, the differences between predictions obtained
using different parton sets are
smaller for these observables than they are for the parton
distributions themselves (compare 
Figs.~\ref{fig:final-pdfs}-\ref{fig:final-pdfs2}). This is unsurprising
given that these data (with the exception of the CHORUS data) have
been used in the determination of all these sets.  

\begin{figure}[ht]
\centering
\epsfig{width=0.48\textwidth,figure=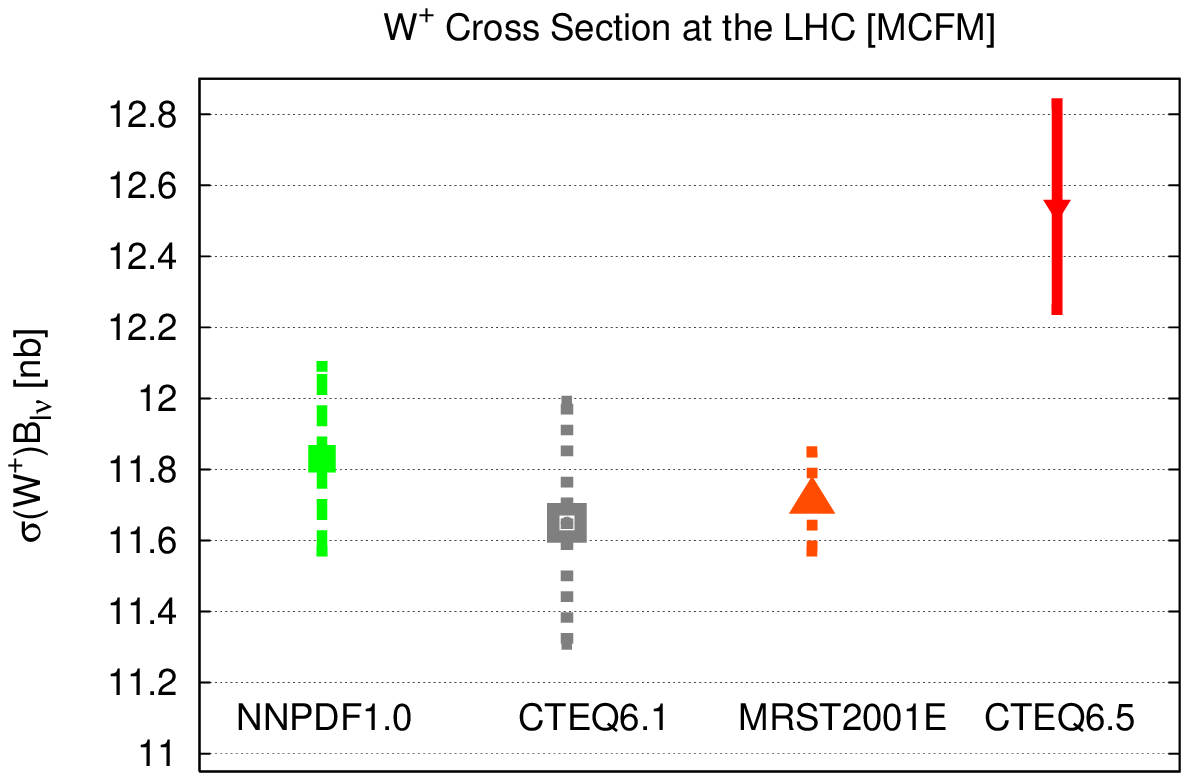} 
\epsfig{width=0.48\textwidth,figure=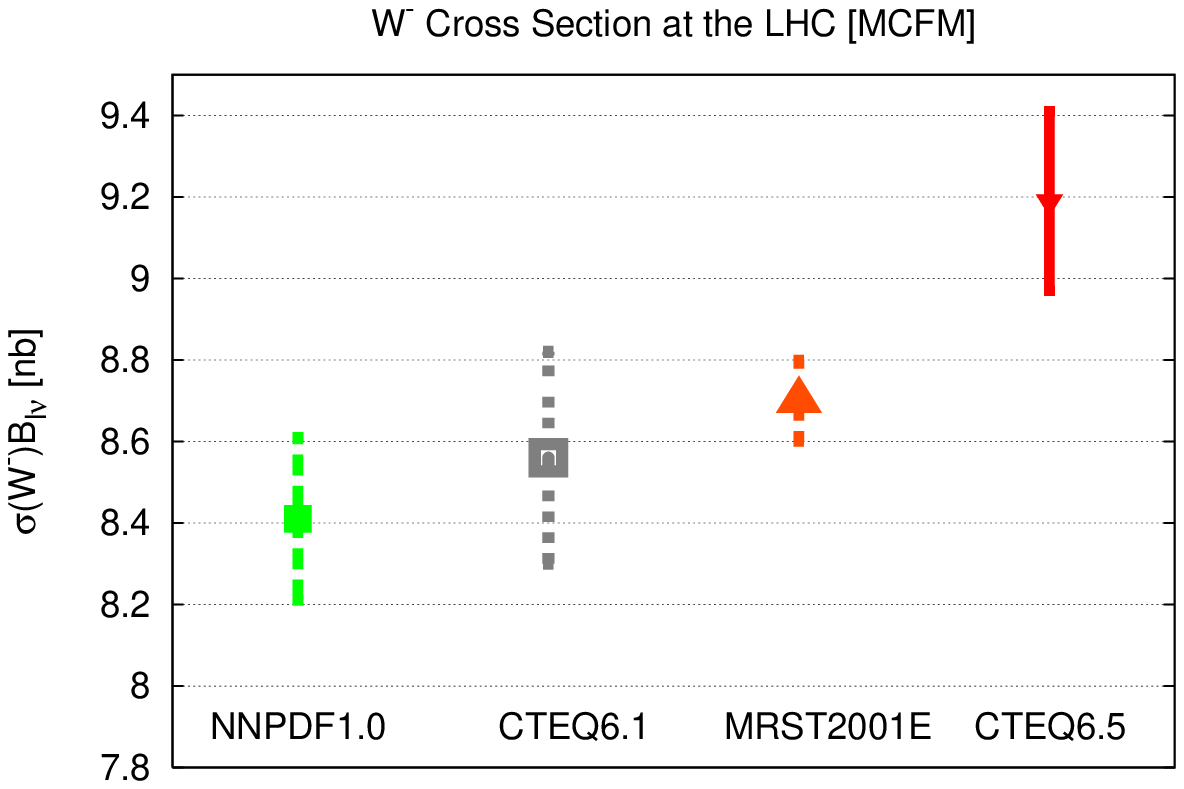} 
\epsfig{width=0.48\textwidth,figure=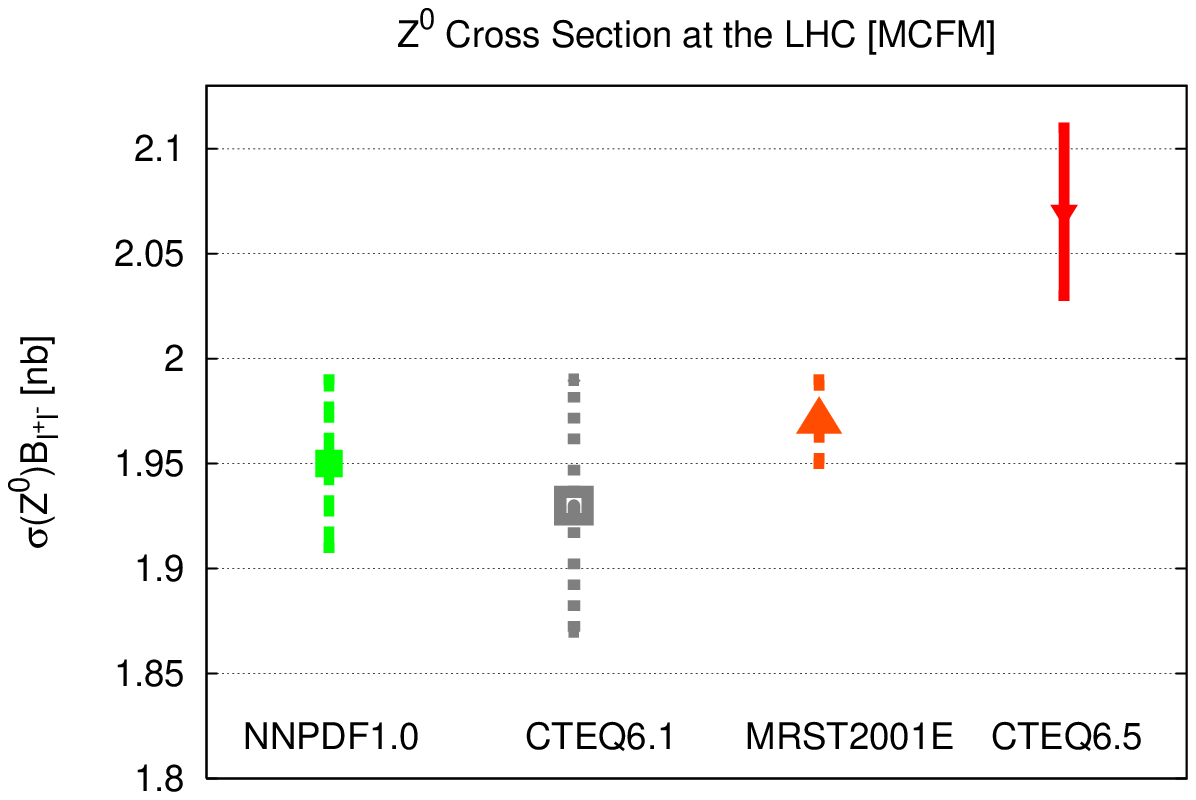} 
  \caption{The vector boson production cross 
sections of Table~\ref{tab:xsec}.}
  \label{fig:LHCobs}
\end{figure}

In Table~\ref{tab:xsec} and Fig.~\ref{fig:LHCobs} we show the
total cross sections for $W$ and $Z$ production. These
processes have been proposed as luminosity monitors at the LHC, but
they remain sensitive to various aspects of
PDFs~\cite{heralhc,Nadolsky:2008zw}.  All cross sections have been
computed at NLO using 
MCFM~\cite{Campbell:2000bg,Campbell:2002tg,Campbell:2004ch,MCFMurl},
using a sample of  $N_{\rm rep}=100$ replicas, which as discussed in
Sect.~\ref{sec:delivery} is fully adequate for this purpose. 
The results are compared to those obtained using the
MRST2001E~\cite{Martin:2002aw}, CTEQ6.1~\cite{Pumplin:2002vw} and 
CTEQ6.5~\cite{Tung:2006tb} sets. Note that the values given for
CTEQ6.5 differ somewhat from those published in
Ref.~\cite{Tung:2006tb}, which were however obtained using
WTTOT~\cite{wttot};  results for MRST2001E  and CTEQ6.1 coincide with those of
Ref.~\cite{heralhc}, also obtained using MCFM.

%---------------------------------------------------------
%
% Table with observables computed with MCFM
%
\begin{table}[ht]
\begin{center}
\small
\begin{tabular}{|c|c|c|c|c|c|c|}
\hline
&   $\sigma_{W^+}\mathcal{B}_{l^+\nu_l}$ 
& $\Delta\sigma_{W^+}/\sigma_{W^+}$
&   $\sigma_{W^-}\mathcal{B}_{l^-\nu_l}$ 
& $\Delta\sigma_{W^-}/\sigma_{W^-}$
& $\sigma_Z\mathcal{B}_{l^+l^-}$
& $\Delta\sigma_Z/\sigma_Z$
\\
\hline
NNPDF1.0  & $11.83 \pm 0.26$ & 2.2\%& $8.41 \pm 0.20$&  2.4\% 
& $1.95 \pm 0.04  $ & 2.1\%      \\
CTEQ6.1 & $11.65 \pm 0.34$ & 2.9\%& 
$8.56 \pm 0.26$& 3.0\%   &  $1.93  \pm 0.06$ & 3.1\% \\
MRST01  & $11.71 \pm 0.14$ & 1.2\%& $8.70 \pm 0.10$ & 1.1\% 
 & $1.97 \pm 0.02$ & 1.0\%   \\
CTEQ6.5 & $12.54 \pm 0.29$ & 2.3\%& $9.19 \pm 0.22$& 2.4\%  
& $2.07 \pm 0.04$ & 1.9\%  \\
\hline
\end{tabular}

\end{center}
\caption{\small Cross sections for gauge boson production at the LHC.
All quantities  have been computed at NLO with
MCFM~\cite{Campbell:2000bg,Campbell:2002tg,Campbell:2004ch,MCFMurl} 
The quoted uncertainty is the one-$\sigma$ band due to the PDF
uncertainty only. 
\label{tab:xsec} 
}
\end{table}
% -------------------------------------------------------------
%-----------------------------------------------------------

We find good agreement of central values with the CTEQ6.1
computation. This is as it should be, given that CTEQ6.1 uses a
zero-mass variable-flavour number scheme for heavy quarks as we do
(see Sect.~\ref{sec:heavy}). The difference between results obtained with
CTEQ6.1 and CTEQ6.5 is explained in Ref.~\cite{Nadolsky:2008zw} as a 
consequence of the
different treatment of heavy quark thresholds. The discrepancy between
CTEQ6.5 and MRST2001, which use a similar treatment of heavy quark
thresholds, is resolved~\cite{Nadolsky:2008zw} when
comparing to the most recent MSTW NNLO set~\cite{Martin:2007bv}.
Our estimate of the uncertainties is a little less than that of the CTEQ fits,
but is rather larger than that of MRST. Note however that unlike CTEQ and 
MRST, NNPDF1.0 does not yet include any Drell-Yan data in the determination 
of the PDFs.

% --------------------------------------------
%
%\section{Conclusions and outlook}
%
%-------------------------------------------------
%-----------------------------------------------

\section{Conclusions and outlook}
\label{sec:conclusions}

We have presented a set of parton distributions, determined using a new
method which is designed to provide results that are unbiased and
amenable to statistical analysis  for the probability distribution of PDFs.

The main features of this approach and of the NNPDF1.0 parton set
based on it are the following:

\begin{itemize}
\item The parton set is given as a Monte Carlo ensemble of replicas of the
  parton distributions which form the basis of independent PDFs. This
  ensemble provides a representation of the probability distributions
  in the space of PDFs, whence any statistical property of the parton
  distributions themselves or of any function(al) of them can be
  calculated using standard statistical methods. It is thus easy to
  compute not only central values and uncertainties, but in fact any
  desired property such as, for example, the correlation between a
  certain cross section and, say, the momentum fraction carried by a
  particular parton distribution.
\item The parton parametrization, based on neural networks,  is
  extremely redundant and independence of results on it can be verified
  explicitly. It is thus possible to verify that the reduction in the
  uncertainties of the predictions from the fit as compared to those of
  the input data is due to the combination of many consistent data
  which obey an underlying physical law, and not to parametrization bias. 
  A small residual dependence of results on the
  preprocessing in the neural networks can be accurately assessed 
  and monitored.
\item A single parton parametrization is flexible enough to
  reproduce data sets of different sizes, as demonstrated by the
  fact that as data are removed, uncertainty bands increase to
  accommodate the increasingly large fluctuations of central
  values. This in particular implies that uncertainty bands in the
  region where there are no data are determined by interpolation  or
  extrapolation of the fluctuations observed in the data region, and
  not by the form of the parametrization.
\item Results behave according to standard statistical expectations upon
  changes in the size of the Monte Carlo ensemble. This enables the
  consistent use of samples of different size according to the
  quantity one wishes to compute: for example, a small sample is often
  sufficient for the accurate determination of an average thanks to
  the fact that the variance of the average of $N$
  quantities is smaller by a factor $1/N$ than the variance
  of each of them.
\item Uncertainty bands are produced directly from the propagation
  through the Monte Carlo and fitting procedure of the uncertainty on
  the underlying data, and do not require the use of any
  tolerance criterion.
\item The best fit is not determined as the absolute minimum for a given
  functional form, but rather it is found by the cross-validation method of
  comparing to the quality of the 
  fit to a (randomly selected) control
  data sample. As a consequence, inconsistent data or underestimated
  uncertainties do not require a separate treatment, and are
  automatically accounted for and signalled by a larger than average
  value of the $\chi^2$ per degree of freedom.
 
\end{itemize}

The results discussed here provide a first full parton set based on
this novel methodology, which can be improved in many respects.
The main improvements which we envisage are the following: 
\begin{itemize}
\item A wider set of data besides deep-inelastic scattering should be
  included. This will improve the accuracy of quantities such as the
  gluon at large $x$ (from jet data) or the light antiquark sea
  asymmetry (from Drell-Yan data), and it will make possible a direct
  determination of the strange distribution which we did not determine
  here.
\item Heavy quark thresholds should be treated more accurately, by
  including terms proportional to the heavy quark mass, i.e. to
  $\frac{m^2_H}{Q^2}$. Because heavy quark distributions are generated
  radiatively by matching in the threshold region, this improves the
  accuracy in the determination of heavy quark fractions at high scale.
\item Complete independence of the preprocessing of neural networks
  should be obtained by randomizing the preprocessing functions, as
  discussed in Sect.~\ref{sec:stabarch}. Similarly contributions 
  from theoretical uncertainties such as
  the choice of factorization scale could be included by randomizing the
  corresponding parameters, as discussed in Sect.~\ref{sec:theounc}.  
\item A set of LO parton distributions should be produced in view of
  its use in Monte Carlo generators (see
  Ref.~\cite{Sherstnev:2007nd}). A set of NNLO parton distributions 
  should be produced, both 
  with the purpose of estimating uncertainties on the NLO results and
  also for some precision applications. Sets of partons including 
  large and small $x$ resummation corrections should also be considered.
\item An optimized small set of parton distributions should be
  produced which reproduces as closely as possible the statistical
  features of the full set, as briefly discussed in
  Sect.~\ref{sec:delivery}. This would then allow the efficient
  computation of detailed statistical features without having to 
  use the full ensemble of a thousand replicas.
\end{itemize}

None of these improvements (with the
possible exception of the last) requires conceptually new tools: they
can all be made by straightforward generalizations of the techniques
discussed here. Many of them will become relevant only as the quality
and accuracy of the fit improves. More importantly, 
it will be possible to monitor 
this improvement, since we now have at our disposal a statistically
reliable and accurate PDF fitting tool.

%------------------------------------------------------

{\bf Acknowledgements \\}
This work was partly supported by grants ANR-05-JCJC-0046-01 (France),
PRIN-2006 (Italy), 
MEC FIS2004-05639-C02-01
(Spain) and by the European network HEPTOOLS under contract
MRTN-CT-2006-035505. L.D.D. is funded by an STFC Advanced Fellowship and
M.U. by a SUPA graduate studentship.
We thank  S.~Alekhin, M.~Arneodo, B.~Cousins, G.~d'Agostini, A.~de~Roeck, J. Feltesse,
W.~Giele, A.~Glazov, S.~Keller, D.~Mason, P.~Nadolsky, E.~Perez, 
M.~Pierini,
K. Rabberts, G.~Salam,
 D.~Soper, M.~Spiropulu,  R.~Thorne, W.~K.~Tung for
discussions and correspondence. We thank M.~Cacciari for help in
cross-checking the implementation of NNPDF, and M.~Whalley for help
with the LHAPDF interface. 
We are especially grateful to the following people for helping us to
understand the features of various data sets: J.~Ferrando,
C.~Gwenlan and E.~Tassi for ZEUS,
 V.~Radescu, E.~Rizvi  and F.~Zomer for H1, R.~Oldeman for CHORUS and V.~Lenderman for
the H1 $F_L$ data.

%-----------------------------------

\appendix

% --------------------------------------------
%
%\section{Evolution kernels}
%
%-------------------------------------------------
%----------------------------------------------------------------

\section{Kernels for Physical Observables}
\label{sec:kernels}
\def\nn{\nonumber}
\def\GS{\Gamma_{\rm S}}
\def\GNS{\Gamma_{\rm NS}}
\def\half{\smallfrac{1}{2}}
\def\ESp{E_{\rm S}^+} 
\def\ESm{E_{\rm S}^-} 
\def\ENSp{E_{\rm NS}^+} 
\def\ENSm{E_{\rm NS}^-} 

In this appendix we expand the physical observables --- structure
functions for DIS from protons and deuterons (measured by SLAC, BCDMS
and NMC), the reduced cross-sections for neutral and charged current
DIS from protons (measured by ZEUS and H1), and the neutrino
cross-sections from a deuteron target (measured by CHORUS) --- in the
basis of PDFs given in Sec.~\ref{sec:heavy},
Eqs.~(\ref{singlet}-\ref{eq:lincombodd}). This yields explicit
expressions for the Mellin transforms of the hard kernels, in the form
Eq.~(\ref{eq:genevfactMel}), which after Mellin inversion
Eq.~(\ref{xks}) can then be used to compute the observables through
evaluation of the convolutions (\ref{kernconns},\ref{kerncons}) with
the initial PDFs determined from the neural nets
Eq.~(\ref{eq:pdfdef}).

The expressions below can be used with evolution kernels and coefficients
evaluated at any order in perturbation theory. Currently evolution kernels 
$\GNS^{\pm,v}$ and $\GS^{ij}$, $i,j = q,g$, described in Sec.~\ref{sec:eveq}, 
can be computed at NLO \cite{gNLOa,gNLOb,gNLOc,gNLOd,gNLOe,gNLOf} and 
NNLO~\cite{gnnloa,gnnlob}, while  
the DIS coefficient 
functions $C_{I,i}$, $I=2,3,L$, $i=q,g$ are known at NLO~\cite{CNLO},
NNLO~\cite{CNNLOa,CNNLOb,CNNLOc,CNNLOd,CNNLOe} and
NNNLO~\cite{Moch:2004xu,Vermaseren:2005qc,Moch:2007gx,Moch:2007rq}. 
Note that at NLO
$\GNS^v=\GNS^-=\GNS^{24}=\GNS^{35}$, 
$C^s_{I,q}=C_{I,q}$, while at LO $\GNS^+=\GNS^-=\GNS^v=\GS^{qq}$ and 
$C_{I,g}=C_{L,q}=0$, while $C_{2,q}=C_{3,q}=1$. 

The heavy quark evolution factors $\GS^{24,i}$, $\GS^{35,i}$, $i=q,g$ 
in the ZM-VFNS are given in Eqs.~(\ref{t24},\ref{t35}). Below the top 
threshold we can set $\GS^{35,q}=\GS^{qq}$, $\GS^{35,g}=\GS^{qg}$, while
below the $b$ threshold $\GS^{24,q}=\GS^{35,q}=\GS^{qq}$, 
$\GS^{24,g}=\GS^{35,g}=\GS^{qg}$, with some simplification of the results. 

Throughout this section we use a condensed notation in which the 
arguments of $f_i$, $F_I$, $\Gamma$, $C$ and $K$ are 
all suppressed: parton distributions evaluated at $Q_0^2$ are denoted by
a subscript $0$, e.g. $\Sigma_0\equiv\Sigma(x,Q_0^2)$. We will assume 
throughout that $Q_0^2=m_c^2$, as explained in Sec.~\ref{sec:heavy}.We
also assume that $b=\bar{b}$, i.e. $V_{24}=V$ at $Q^2=m_b^2$, and similarly
$t=\bar{t}$, i.e. $V_{35}=V$ at $Q^2=m_t^2$. 
However we do allow for the possibility of 
intrinsic charm, even though for the parton fits 
described in this paper this option is not exercised. The specific 
flavour assumptions described in Sec.~\ref{sec:pdfbas} can be summarised as 
$T_{8,0}=\smallfrac{2-2C_s}{2+C_s}\Sigma_0 + \smallfrac{3C_s}{2+C_s}V_0$, 
(see Eq.~({\ref{eq:strangeass})), 
$T_{15,0}=\Sigma_0$, $V_{8,0}=V_{15,0}=V_0$. 

Target mass corrections are included in the kernels using 
the procedure described in Sec.~\ref{sec:tmc}.

\subsection{Structure Functions: SLAC, BCDMS and NMC}
\label{ssec:sf}

The datasets we include in our analysis include measurements 
of $F_2^p$, $F_2^d$ and the ratio $F_2^d/F_2^p$. These observables 
can be written in terms of the
linear combinations of parton densities used in the evolution code 
(defined in Eq.~(\ref{singlet}) and Eq.~(\ref{eq:lincombeven})):
in the quark model the proton structure function
\be
\label{eq:f2p}
  F_2^p = x \sum_{i=1}^{n_f} e_i^2 q_i^+ = x\lbrace\smallfrac{5}{18} \Sigma 
             + \smallfrac{1}{6} T_3
                + \smallfrac{1}{18} (T_8 - T_{15})
        + \smallfrac{1}{30} (T_{24}-T_{35})\rbrace,
\ee
where $e_i$ are the quark charges 
($\smallfrac{2}{3}$ for $u,c,t$, $-\smallfrac{1}{3}$ for $d,s,b$), while 
the deuteron structure function
\be
  \label{eq:f2d}
  F_2^d = \half (F_2^p+F_2^n)
        =  x \lbrace\smallfrac{5}{18} \Sigma
                + \smallfrac{1}{18} (T_8 - T_{15})
        + \smallfrac{1}{30} (T_{24}-T_{35}) \rbrace\,.
\ee
In perturbative QCD, we have
\begin{eqnarray}
  \label{eq:f2pdfcomp}
  F_2^p &=& x\lbrace \smallfrac{5}{18}C^s_{2,q}\otimes \Sigma 
             + \smallfrac{1}{6}C_{2,q}\otimes( T_3
                + \smallfrac{1}{3} (T_8 - T_{15})
        + \smallfrac{1}{5} (T_{24}-T_{35}))\nn\\
 &&\qquad\qquad\qquad\qquad\qquad\qquad 
+ {\langle e_q^2\rangle}  C_{2,g}\otimes g\rbrace ,\\
  \label{eq:f2pdfcomd}
  F_2^d &=& x\lbrace \smallfrac{5}{18}C^s_{2,q}\otimes \Sigma 
             + \smallfrac{1}{6}C_{2,q}\otimes(
          \smallfrac{1}{3} (T_8 - T_{15})
        + \smallfrac{1}{5} (T_{24}-T_{35}))
        + {\langle e_q^2\rangle}  C_{2,g}\otimes g\rbrace \,.
\end{eqnarray}
where $\otimes$ denotes the Mellin convolution, 
defined in Eq.~(\ref{eq:conv}), 
and ${\langle e_q^2\rangle}$ is defined as
\be
\label{ceenf}
 {\langle e_q^2\rangle} = \smallfrac{1}{n_f}\sum_{i=1}^{n_f} e_i^2,
\ee
where $n_f$ is the number of active flavours: 
${\langle e_q^2\rangle} = \smallfrac{2}{9},
\smallfrac{5}{18},\smallfrac{11}{45},\smallfrac{5}{18}$ for $n_f = 3,4,5,6$.
We can thus write
 \bea
\label{KF2p}
  F_2^p&=&x \lbrace 
        K_{{\rm F2},\Sigma}\otimes \Sigma_0 
+K_{{\rm F2},g} \otimes g_0 
+K_{{\rm F2},+} \otimes  \left(T_{3,0}
                + \smallfrac{1}{3}(T_{8,0} - T_{15,0})\right)\rbrace,\\
\label{KF2d}
  F_2^d&=&x \lbrace 
        K_{{\rm F2},\Sigma}\otimes \Sigma_0 
+K_{{\rm F2},g} \otimes g_0 
+\smallfrac{1}{3}K_{{\rm F2},+} \otimes  
   \left(  T_{8,0} - T_{15,0}\right)
\rbrace,
 \eea
where (in Mellin space)
\begin{eqnarray}
\label{eq:Kf2S} 
K_{{\rm F2},\Sigma}&=&\smallfrac{5}{18}C^s_{2,q}\GS^{qq}
+\smallfrac{1}{30}C_{2,q}(\GS^{24,q}-\GS^{35,q})
+{\langle e_q^2\rangle}C_{2,g}\GS^{gq},\\
  \label{eq:Kf2g}
K_{{\rm F2},g}&=&\smallfrac{5}{18}C^s_{2,q}\GS^{qg}
+\smallfrac{1}{30}C_{2,q}(\GS^{24,g}-\GS^{35,g})
+{\langle e_q^2\rangle}C_{2,g}\GS^{gg},\\
  \label{eq:Kf2T}
 K_{{\rm F2},+}&=& \smallfrac{1}{6}C_{2,q}\GNS^+.
    \end{eqnarray}

\subsection{Neutral Current Reduced Cross-Sections: ZEUS and H1}
\label{ssec:nc}

For high energy DIS experiments, as is the case 
for ZEUS and H1, the contribution from weak boson exchange cannot 
be neglected. 

The neutral current deep-inelastic scattering reduced cross section 
Eq.~(\ref{eq:rednc}) is
\begin{equation}
  \label{eq:redncx}
  \widetilde{\sigma}^{\rm NC,e^{\pm}}=F_2^{NC} 
\mp \smallfrac{Y_-}{Y_+} x F_3^{NC}-\smallfrac{y^2}{Y_+} F_L^{NC},
\end{equation}
with $Y_\pm = 1\pm(1-y)^2$. In the quark model 
\be
  \label{eq:Fncq}
  F_2^{\rm NC} = F_2^{\gamma}
                       + \sum_{i=1}^{n_f}\, B_i\, q^+_i,\qquad
   F_3^{\rm NC}=\sum_{i=1}^{n_f} \,D_i \,q_i^- ,
\ee
where $F_2^\gamma$ is the purely electromagnetic 
structure function, and the charge factors are
\bea
\label{ZqB}
  B_q(Q^2)&=& -2e_qV_eV_qP_Z+(V_e^2+A_e^2)(V_q^2+A_q^2)P_Z^2,\\
\label{ZqD}
  D_q(Q^2)&=&-2e_qA_eA_qP_Z+4V_eA_eV_qA_q P_Z^2, 
\eea
\begin{table}[ht]
\vskip0.5cm
  \begin{center}
    \begin{tabular}{|c|c|c|c|}
      \hline
      fermions  & $e_f$ & $V_f$ & $A_f$  \\
      \hline
      u,c,t  & +2/3 & $(+1/2-4/3\sin^2\theta_W)$ & +1/2  \\
      d,s,b  & -1/3 & $(-1/2+2/3\sin^2\theta_W)$ & -1/2  \\
      $\nu_e,\nu_{\mu},\nu_{\tau}$  & 0 & +1/2 & +1/2  \\
      e,$\mu,\tau$  & -1 & $(-1/2+2\sin^2\theta_W)$ & -1/2  \\
      \hline
    \end{tabular}
    \caption{Coupling of fermions to the Z boson.
                                    \label{tab:coupling}} 
\end{center}
\end{table}
where $P_Z\equiv Q^2/(Q^2+M_Z^2)$, $V_f$ and $A_f$ are the vector 
and axial couplings of the
fermions to the Z boson, as given in 
Table~\ref{tab:coupling}.
%%%Table~18. 
In terms of the 
PDF evolution eigenstates Eq.~(\ref{singlet}-\ref{eq:lincombodd})
\bea
  \label{eq:F2nc}
  F_2^{\rm NC} &=& x\lbrace
\ESp\Sigma + \ENSp  (T_3+ \smallfrac{1}{3}(T_8-T_{15})
+\smallfrac{1}{5} (T_{24}-T_{35}))\rbrace,\\
  \label{eq:F3nc}
   F_3^{\rm NC}&=& \ESm V + \ENSm (V_3 
 +\smallfrac{1}{3}(V_8 -V_{15})+\smallfrac{1}{5}(V_{24} -V_{35})),
\eea
where the charge coefficients
\bea
\label{chargesE}
 \ESp &=& \smallfrac{5}{18} +\half (B_u+B_d),\qquad
 \ENSp= \smallfrac{1}{6} +\half (B_u-B_d),\nn\\
\ESm&=&\half (D_u+D_d),\qquad
 \ENSm=\half (D_u-D_d),
\eea
In perturbative QCD we can thus write the reduced cross sections 
as
\bea
\label{redxsecNC}
&&\widetilde{\sigma}^{\rm NC,e^{\pm}} = x \lbrace
( C^s_{2,q} - \smallfrac{y^2}{Y_+}C^s_{L,q})\otimes 
\ESp\Sigma + E_g ( C_{2,g} -\smallfrac{y^2}{Y_+}C_{L,g}) \otimes  g  \nn\\
&&\qquad 
+ ( C_{2,q} - \smallfrac{y^2}{Y_+}C_{L,q})
\otimes (\ENSp  (T_3+ \smallfrac{1}{3}(T_8-T_{15})
+\smallfrac{1}{5} (T_{24}-T_{35})))\nn\\
&&\qquad \mp  \smallfrac{Y_-}{Y_+}C_{3,q}\otimes ( 
\ESm V + \ENSm (V_3 + \smallfrac{1}{3}(V_8 -V_{15}))),
\rbrace
\eea
where we have set $V_{24}=V_{35}=V$, and
\be
\label{Eg}
 E_{g}= {\langle e_q^2\rangle} + {\langle B_q^2\rangle},
\ee
where
\be
\label{avB}
 {\langle B_q^2\rangle} = \smallfrac{1}{n_f}\sum_{i=1}^{n_f} B_i  
= \begin{cases}
\smallfrac{1}{3}\lp B_u+2B_d\rp, & {\rm if}\quad n_f=3,\\
\half \lp B_u+B_d\rp, & {\rm if}\quad n_f=4,\\
\smallfrac{1}{5}\lp 2B_u+3B_d\rp, & {\rm if}\quad n_f=5,\\
\half \lp B_u+B_d \rp, & {\rm if}\quad n_f=6.
\end{cases}
\ee
Note that the gluon coefficient function $C_{2,g}$ is the same
quantity which appears in Eq.~(\ref{eq:f2pdfcomp}).
In terms of hard kernels, we thus have
\bea
\label{KNC}
\widetilde{\sigma}^{\rm NC,e^{\pm}} &=& x \lbrace K_{{\rm NC},\Sigma}
\otimes\Sigma_0
+ K_{{\rm NC},g}\otimes g_0 + K_{{\rm NC},+}
\otimes (T_{3,0}+\smallfrac{1}{3}(T_{8,0}-T_{15,0}))\nn\\
&& \qquad \mp K_{{\rm NC},V} \otimes V_0 \mp K_{{\rm NC},-} 
\otimes (V_{3,0}+ \smallfrac{1}{3}(V_{8,0}-V_{15,0}))\rbrace,
\eea  
where in Mellin space
\bea
\label{KNCS}
K_{{\rm NC},\Sigma} &=& ( C^s_{2,q} - \smallfrac{y^2}{Y_+}C^s_{L,q})
\ESp \GS^{qq}+ E_g ( C_{2,g} -\smallfrac{y^2}{Y_+}C_{L,g})\GS^{gq}\nn\\
&&\qquad\qquad +\smallfrac{1}{5}(C_{2,q} - \smallfrac{y^2}{Y_+}C_{L,q})
\ENSp(\GS^{24,q}-\GS^{35,q}),\\
\label{KNCg}
K_{{\rm NC},g} &=& (C^s_{2,q} - \smallfrac{y^2}{Y_+}C^s_{L,q}) 
\ESp \GS^{qg}+ E_g ( C_{2,g} -\smallfrac{y^2}{Y_+}C_{L,g})\GS^{gg}  \nn\\
&&\qquad\qquad +\smallfrac{1}{5}(C_{2,q} - \smallfrac{y^2}{Y_+}C_{L,q})
\ENSp(\GS^{24,g}-\GS^{35,g}),\\
\label{KNC+}
K_{{\rm NC},+} &=& \ENSp( C_{2,q} - \smallfrac{y^2}{Y_+}C_{L,q}) 
\GNS^{+},\\
\label{KNCV}
K_{{\rm NC},V} &=& \ESm\smallfrac{Y_-}{Y_+}C_{3,q}\GNS^{v},\qquad
K_{{\rm NC},-} = \ENSm\smallfrac{Y_-}{Y_+}C_{3,q}\GNS^{-}. 
\eea
  
It is also convenient to write down here the expression for 
$F_L^{\rm NC}$ which we use to fix the positivity constraint and fit
the H1 $F_L$ data:
\begin{equation}
\label{FLNC}
  F_L^{\rm NC}= x \lbrace 
        K_{{\rm FL},\Sigma}\otimes \Sigma_0
        +K_{{\rm FL},g}\otimes g_0 
        + K_{{\rm FL},+} \otimes  \left(T_{3,0}
                + \smallfrac{1}{3}(T_{8,0}-T_{15,0})\right)
\rbrace,
\end{equation}
where
\bea
\label{KFLS}
K_{{\rm FL},\Sigma} &=& 
C^s_{L,q}\ESp \GS^{qq}+\smallfrac{1}{5}C_{L,q}\ENSp(\GS^{24,q}-\GS^{35,q}) 
+ E_g C_{L,g}\GS^{gq},\\
\label{KFLg}
K_{{\rm FL},g} &=& C^s_{L,q} 
\ESp\GS^{qg}+\smallfrac{1}{5}C_{L,q}\ESm(\GS^{24,g}-\GS^{35,g})   
+ E_g C_{L,g}\GS^{gg},\\
\label{KFL+}
K_{{\rm FL},+} &=& \ENSp C_{L,q} 
\GNS^{+}.
\eea

\subsection{Charged-current scattering: ZEUS and H1}
\label{ssec:cc}

For charged current deep-inelastic scattering, the reduced cross-section
Eq.~(\ref{eq:redcc})
\be
\label{sigCC}
  \widetilde{\sigma}^{{\rm CC},e^\pm}=\half (
  Y_+ F_2^{CC,e^{\pm}}\mp Y_- x F_3^{CC,e^{\pm}}
  -y^2 F_L^{CC,e^{\pm}}) \ ,
\ee
where the factor of a half comes from the average over the helicity of 
the incoming leptons. 

For charged current and neutrino scattering the inclusion of the third 
generation is more subtle than for neutral current scattering, since the 
scattering of $W^\pm$ involves a transition between $b$ and $t$. 
In the ZM-VFNS this means that below the top threshold neither 
$b$ nor $t$ can contribute
to the cross-section, even above $b$ threshold \cite{CSB:1997xy}. 
We must thus consider the
linear decomposition of the structure functions below and above the
top threshold as two separate cases. Of course, all the current 
data are effectively below top threshold, so the second possibility is at 
present rather academic.

Below top threshold (but above charm threshold) in the quark model
\bea
\label{f2ccpm}
  F_2^{CC,e^+}&=& 2x\lbrace \bar{u}+d+s
  +\bar{c} \rbrace,\qquad 
  F_2^{CC,e^-}= 2x\lbrace u+\bar{d}
+\bar{s}+c \rbrace.\\
\label{f3ccpm}
  F_3^{CC,e^+}&=& 2\lbrace -\bar{u}+d+s-\bar{c}\rbrace,\qquad\quad
F_3^{CC,e^-}= 2\lbrace u-\bar{d}-\bar{s}+c\rbrace.
\eea
Below charm threshold both the charm and the strange
contributions would be absent.
In terms of the PDF evolution eigenstates 
Eq.~(\ref{singlet}-\ref{eq:lincombodd})
we thus have
\bea
\label{f2ccpmcomb}
F_2^{CC,e^\pm}&=& x\lbrace  
(\smallfrac{4}{5}\Sigma + \smallfrac{1}{5}T_{24}) 
\mp (V_3 + \smallfrac{1}{3}(V_8-V_{15})\rbrace,\\
\label{f3ccpmcomb}
F_3^{CC,e^\pm}&=&  
(\smallfrac{4}{5}V + \smallfrac{1}{5}V_{24})
 \mp ( T_3 + \smallfrac{1}{3}(T_8-T_{15}).
\eea
The first term in Eq.~(\ref{f2ccpmcomb}) is subtle:
$(\smallfrac{4}{5}\Sigma + \smallfrac{1}{5}T_{24}) = u^++d^++s^++c^+$
both above and below the $b$ threshold, so the $b$ quarks do not couple to the 
$W^\pm$, consistent with Eq.~(\ref{f2ccpm}). Similar considerations apply to
the first term in Eq.~(\ref{f3ccpmcomb}).
In perturbative QCD we thus have
\bea
\label{f2ccpmpqcd}
F_2^{CC,e^\pm}&=& x\lbrace  
C^s_{2,q}\otimes (\smallfrac{4}{5}\Sigma + \smallfrac{1}{5}T_{24}) 
\mp C_{2,q}\otimes(V_3 + \smallfrac{1}{3}(V_8-V_{15}))\nn\\
&&\qquad\qquad\qquad\qquad\qquad\qquad 
+ r_f C_{2,g}\otimes g  \rbrace,\\
\label{fLccpmpqcd}
F_L^{CC,e^\pm}&=& x\lbrace  
C^s_{L,q}\otimes (\smallfrac{4}{5}\Sigma + \smallfrac{1}{5}T_{24}) 
\mp  C_{L,q}\otimes( V_3 + \smallfrac{1}{3}(V_8-V_{15}))\nn\\
&&\qquad\qquad\qquad\qquad\qquad\qquad 
+ r_f C_{L,g}\otimes g  \rbrace,\\
\label{f3ccpmpqcd}
F_3^{CC,e^\pm}&=&   
C^s_{3,q}\otimes (\smallfrac{4}{5} V + \smallfrac{1}{5}V_{24}) 
\mp C_{3,q}\otimes\left( T_3 + \smallfrac{1}{3}(T_8-T_{15})\right),
\eea
where the factor $r_f$ counts the proportion of active flavors:
denoting with $[n]$ the integer part of $n$,
\begin{equation}
\label{rfdef}
		r_f = [n_f/2]/(n_f/2) = \begin{cases} 1~~~~~~~\hbox{when}~n_f~\hbox{is even}\\
 1-\frac{1}{n_f}~\hbox{when}~n_f~\hbox{  is odd.}
\end{cases}
\end{equation}
 The gluon coefficient functions are the same quantities which appear
in Eqs.~(\ref{eq:f2pdfcomp},\ref{redxsecNC}).
We can thus write the reduced cross sections as
\bea
\label{KCC}
\widetilde{\sigma}^{\rm CC,e^{\pm}} &=& \half x \lbrace K_{{\rm CC},\Sigma}
\otimes\Sigma_0
+ K_{{\rm CC},g}\otimes g_0 + K_{{\rm CC},+}
\otimes (T_{3,0}+\smallfrac{1}{3}(T_{8,0}-T_{15,0}))\nn\\
&& \qquad \mp K_{{\rm CC},V} \otimes V_0 
\mp K_{{\rm CC},-} 
\otimes (V_{3,0}+ \smallfrac{1}{3}(V_{8,0} -V_{15,0}))\rbrace,
\eea  
where in Mellin space
\bea
\label{KCCS}
K_{{\rm CC},\Sigma} &=& 
\left(Y_+C^s_{2,q}-y^2C^s_{L,q}\right)(\smallfrac{4}{5}\GS^{qq}
 +\smallfrac{1}{5}\GS^{24,q})\nn\\
&&\qquad\qquad\qquad 
+r_f\left(Y_+C_{2,g}-y^2C_{L,g}\right)\GS^{gq},\\
\label{KCCg}
K_{{\rm CC},g} &=& 
\left(Y_+C^s_{2,q}-y^2C^s_{L,q}\right)(\smallfrac{4}{5}\GS^{qg}
 +\smallfrac{1}{5}\GS^{24,g})\nn\\
 &&\qquad\qquad\qquad +r_f\left(Y_+C_{2,g}-y^2C_{L,g}\right)\GS^{gg},\\
\label{KCC+}
K_{{\rm CC},+} &=& Y_-C_{3,q}\GNS^{+},\\
\label{KCCV}
K_{{\rm CC},V} &=& Y_-C^s_{3,q}(\smallfrac{4}{5}\GNS^{v} 
+ \smallfrac{1}{5}\GNS^{24}),\\
\label{KCC-}
K_{{\rm CC},-} &=& \left(Y_+C_{2,q}-y^2C_{L,q}\right)\GNS^{-}.
\eea

Above top threshold, $K_{{\rm CC},\Sigma}$, $K_{{\rm CC},g}$ and 
$K_{{\rm CC},V}$ become instead
\bea
\label{KCCSt}
K^t_{{\rm CC},\Sigma} &=& \left(Y_+C^s_{2,q}-y^2C^s_{L,q}\right)\GS^{qq}
  +r_f\left(Y_+C_{2,g}-y^2C_{L,g}\right)\GS^{gq}\nn\\
&&\qquad\qquad\qquad +\smallfrac{1}{5} Y_- C_{3,q}(\GS^{24,q}-\GS^{35,q}),\\
\label{KCCgt}
K^t_{{\rm CC},g} &=& \left(Y_+C^s_{2,q}-y^2C^s_{L,q}\right)\GS^{qg}
  +
r_f\left(Y_+C_{2,g}-y^2C_{L,g}\right)\GS^{gg}\nn\\
&&\qquad\qquad\qquad +\smallfrac{1}{5} Y_- C_{3,q}(\GS^{24,g}-\GS^{35,g}),\\
\label{KCCVt}
K^t_{{\rm CC},V} &=& Y_-C^s_{3,q}\GNS^{v}+\smallfrac{1}{5}
\left(Y_+C_{2,q}-y^2C_{L,q}\right)(\GNS^{24}-\GNS^{35}).
\eea

\subsection{Neutrino Scattering: CHORUS}
\label{ssec:nu}

The cross-section for high energy neutrino scattering off an isoscalar nucleon
is given by Eq.~(\ref{eq:nuxsec}), which we write as
\be
\label{nuxsecx}
\widetilde
\sigma^{\nu(\bar{\nu})}
=\kappa[\widetilde{Y}_+ F_2^{\nu(\bar{\nu})}
-y^2 F_L^{\nu(\bar{\nu})}\pm \,Y_-\,xF_3^{\nu(\bar{\nu})}],
\ee
where $\kappa= G_F^2M_N/(2\pi(1+Q^2/M_W^2)^2)$, $\widetilde{Y}_+ 
= Y_+ - 2M^2_Nx^2y^2/Q^2$.

Remembering that the target is isoscalar, 
in the quark model below the top threshold, but above charm
threshold, 
\begin{eqnarray}
\label{f2nu}
  F_2^{\nu}&=& x\lbrace  u+\bar{u}+d+\bar{d}+2s
  +2\bar{c} \rbrace,\qquad
  F_2^{\bar{\nu}}= x\lbrace  u+\bar{u}+d+\bar{d}
+2\bar{s}+2c\rbrace,\\
\label{f3nu}
  F_3^{\nu}&=&  u-\bar{u}+d-\bar{d}+2s-2\bar{c},\qquad
F_3^{\bar{\nu}}= u-\bar{u}+ d-\bar{d}-2\bar{s}+2c.
\end{eqnarray}
Below charm threshold both the charm and the strange
contributions would be absent.
Of course these expressions can be easily deduced from 
Eqs.~(\ref{f2ccpm},\ref{f3ccpm}), since 
\be\label{crossing}
F^{\nu,p}_{2,L,3}
=F^{CC, e^+}_{2,L,3}, \qquad F^{\bar{\nu},p}_{2,L,3}=F^{CC, e^-}_{2,L,3}.
\ee
In terms of the PDF evolution eigenstates 
Eq.~(\ref{singlet}-\ref{eq:lincombodd})  
we now have in perturbative QCD
\bea
\label{f2nupqcd}
F_2^{\nu(\bar{\nu})}&=& x\lbrace  
C^s_{2,q}\otimes (\smallfrac{4}{5}\Sigma 
+ \smallfrac{1}{5}T_{24}) 
\mp \smallfrac{1}{3}C_{2,q}\otimes(V_8-V_{15})
+ r_f C_{2,g}\otimes g  \rbrace,\\
\label{fLnupqcd}
F_L^{\nu(\bar{\nu})}&=& x\lbrace  
C^s_{L,q}\otimes (\smallfrac{4}{5}\Sigma + 
\smallfrac{1}{5}T_{24}) 
\mp  \smallfrac{1}{3}C_{L,q}\otimes(V_8-V_{15}) 
+ r_f C_{L,g}\otimes g  \rbrace,\\
\label{f3nupqcd}
F_3^{\nu(\bar{\nu})}&=& \, C^s_{3,q}\otimes 
(\smallfrac{4}{5}V + \smallfrac{1}{5}V_{24}) 
\mp \smallfrac{1}{3} C_{3,q}\otimes(T_8-T_{15}),
\eea
the $b$ threshold being accommodated in the same way as in 
Eq.~(\ref{f2ccpmpqcd}), and with $r_f$ defined as in Eq.~(\ref{rfdef})
We can thus write the neutrino cross sections as
\bea
\label{Knu}
\widetilde{\sigma}^{\nu(\bar{\nu})} &=& \kappa\, x \lbrace K_{\nu,\Sigma}
\otimes\Sigma_0
+ K_{\nu,g}\otimes g_0 + K_{\nu,+}
\otimes \smallfrac{1}{3}(T_{8,0}-T_{15,0})\nn\\
&& \qquad \mp K_{\nu,V} \otimes V_0 
\mp K_{\nu,-} 
\otimes \smallfrac{1}{3}(V_{8,0} -V_{15,0})\rbrace,
\eea  
where in Mellin space
\bea
\label{KnuS}
K_{\nu,\Sigma} &=& 
\left(\widetilde{Y}_+C^s_{2,q}-y^2C^s_{L,q}\right)(\smallfrac{4}{5}\GS^{qq}
 +\smallfrac{1}{5}\GS^{24,q})\nn\\
&&\qquad\qquad\qquad 
+r_f\left(\widetilde{Y}_+C_{2,g}-y^2C_{L,g}\right)\GS^{gq},\\
\label{Knug}
K_{\nu,g} &=& 
\left(\widetilde{Y}_+C^s_{2,q}-y^2C^s_{L,q}\right)(\smallfrac{4}{5}\GS^{qg}
 +\smallfrac{1}{5}\GS^{24,g})
\nn\\
 &&\qquad\qquad\qquad 
+r_f\left(\widetilde{Y}_+C_{2,g}-y^2C_{L,g}\right)\GS^{gg},\\
\label{Knu+}
K_{\nu,+} &=& Y_-C_{3,q}\GNS^{+},\\
\label{KnuV}
K_{\nu,V} &=& Y_-C^s_{3,q}(\smallfrac{4}{5}\GNS^{v} 
+ \smallfrac{1}{5}\GNS^{24}),\\
\label{Knu-}
K_{\nu,-} &=&  (\widetilde{Y}_+
C_{2,q}-y^2C_{L,q})\GNS^{-}.
\eea

%----------------------------------------------------

% --------------------------------------------
%
%\section{PDF uncertainty determination}
%
%-------------------------------------------------
%----------------------------------------------------------------

\section{Computation  of PDF uncertainties}
\label{sec:app-pdferr}

Within our approach, 
the expectation value of any function 
$ \mathcal{F} [ \{  q \}]$ which depends
on the PDFs is computed as an average over the ensemble of PDFs, using
the master formula
\be
\label{masterave}
\la \mathcal{F} [ \{  q \}] \ra
= \frac{1}{N_{\rm set}} \sum_{k=1}^{N_{\rm set}}
\mathcal{F} [ \{  q^{(k)} \}],
\ee
where $N_{\rm set}=N_{\rm rep}$ is the number of sets of PDFs in the ensemble,
equal to the number of replicas. 
The associated uncertainty is found as the standard deviation of the
sample, according to the usual formula
\bea
\sigma_{\mathcal{F}} 
&=& \left( \frac{N_{\rm set}}{N_{\rm set}-1}   
\lp \la \mathcal{F} [ \{  q \}]^2\ra 
-   \la \mathcal{F} [ \{  q \}] \ra^2 
\rp \right)^{1/2}\nn\\
&=& \left( \frac{1}{N_{\rm set}-1}
\sum_{k=1}^{N_{\rm set}}   
\lp \mathcal{F} [ \{  q^{(k)} \}] 
-   \la \mathcal{F} [ \{  q \}] \ra\rp^2 
 \right)^{1/2}.
\label{mastersig}
\eea
These formulae may also be used for the determination of central values and
uncertainties of the parton distribution themselves, in which case the
functional $\mathcal{F}$ is identified with the parton distribution $q$ :  
$\mathcal{F}[ \{ q\}]\equiv q$.

A full discussion of statistical estimators for our PDF set, including 
correlators and covariances, is given
in Appendix B of Ref.~\cite{DelDebbio:2007ee}. Here we briefly compare 
the procedures we use to determine central values and errors with those used 
for the various other PDFs available through HEPDATA.

Available methods for the determination of PDF uncertainties 
fall broadly into two distinct categories, which we shall refer to as
the HEPDATA method (used as a  default 
in the PDF server at the HEPDATA database~\cite{HEPDATAurl})
and the { Monte Carlo method}.
In both methods sets of PDFs with uncertainties are given as
an ensemble of $N_{\rm set}$ sets of PDFs, 
\be\{ q^{(k)} \} \ , \qquad k=0,\ldots,N_{\rm set}.
\ee 
Conventionally the PDF set $q^{(0)}$ corresponds to a ``central'' set. 

In the HEPDATA method, the central set is a best fit set
of PDFs, which thus provides the central value for PDFs
themselves.
 The central value of any quantity $\mathcal{F}[ \{  q
\}]$ is obtained in this method
by evaluating it as a function of the central set: 
\be
\label{naiveav}
\mathcal{F}^{(0)}=
\mathcal{F}[ \{  q^{(0)} \}].
\ee

In the Monte Carlo method, 
the central values of  any quantity  $\mathcal{F}[ \{  q
\}]$ is instead given by  Eq.~(\ref{masterave}). Therefore, 
the central
value for PDFs themselves is given by 
 \be
\label{mcav}
q^{(0)} \equiv \la  q \ra = \frac{1}{N_{\rm set}}
\sum_{k=1}^{N_{\rm set}} q^{(k)} \ .
\ee
This set is provided as set $q^{(0)}$ in the
NNPDF1.0 PDFs. Hence, in the Monte Carlo method the central (best
fit) PDF is obtained as an average of the replica  best fits.
However, for any quantity 
$\mathcal{F}[ \{  q \}]$ which 
depends nonlinearly on the PDFs
\be
\label{linearav}
\la \mathcal{F} [ \{  q \}] \ra 
\not= 
 \mathcal{F} [\{  q^{(0)}\}] .
\ee
Hence,
Eq.~(\ref{masterave}) must be used  for the determination of the
central value, and  use of the  set  $q^{(0)}$ is not 
recommended. However, for a quantity that does depend linearly on the
PDFs, such as a DIS structure function, Eq.~(\ref{naiveav}) with the
central PDFs Eq.~(\ref{mcav}) gives the
same result as Eq.~(\ref{masterave}), and thus it may be used
also with the Monte Carlo method.
Note that set $q^{(0)}$ should not be included when
computing an average with Eq.~(\ref{masterave}), because it is 
itself already an average.

The determination of uncertainties with  the HEPDATA method is
based on the idea that sets $ q^{(k)}$ with $k>0$ provide upper and
lower variations (for even and odd values of $k$)
away from the central set  $ q^{(0)}$ which correspond to eigenvectors
in parameter space. 
The one-$\sigma$ uncertainty is then found by adding in quadrature 
these variations:
\be
\label{hepdataerr}
\sigma_{\mathcal{F} }^{\rm hepdata} = \frac{1}{2C_{90}}\lp 
\sum_{k=1}^{N_{\rm set}/2} 
\lp \mathcal{F} [ \{  q^{(2k-1)} \}]
- \mathcal{F} [ \{  q^{(2k)} \}] \rp^2\rp^{1/2},
\ee
where the factor
\be
C_{90} \equiv \sqrt{2}{\rm Erf}^{-1}[ 0.90 ] = 1.64485
\ee 
accounts for the fact that the upper and lower parton sets correspond
to 90\% confidence levels rather than to one-$\sigma$ uncertainties.
This method should be used with the CTEQ and MRST/MSTW sets
Refs.~\cite{Pumplin:2002vw,Huston:2005jm,Tung:2006tb,Owens:2007kp,
Lai:2007dq,Martin:2002aw,Martin:2003sk,Martin:2004ir,Martin:2007bv}.

A slightly different application of the HEPDATA method is required for the
Alekhin PDF sets 
Ref.~\cite{Alekhin:2002fv,Alekhin:2003yh,Alekhin:2005gq,Alekhin:2006zm}.  
With these PDFs, sets $ q^{(k)}$ with $k>0$ each provide the
uncertainty limits from
the central set, with upper and lower PDFs symmetrical by
construction and already corresponding to one-$\sigma$ uncertainties.
So for these PDFs
\be
\label{alekerr}
\sigma_{\mathcal{F} }^{\rm hepdata} = \lp \sum_{k=1}^{N_{\rm set}} 
\lp \mathcal{F} [ \{  q^{(k)} \}]
- \mathcal{F} [ \{  q^{(0)} \}] \rp^2\rp^{1/2}.
\ee

In the Monte Carlo method, instead, the one-$\sigma$ uncertainty is found using
the variance formula Eq.~(\ref{mastersig}). 
This formula should be used not only for the NNPDF1.0 sets, but also
for the parton sets of Refs.~\cite{Giele:2001mr,Botje:1999dj,Alekhin:2000ch}.

\bibliography{nnsinglet}

\end{document}